%% file: version4.tex
\documentclass[12pt]{article}
\usepackage{latexsym}
\usepackage{amsmath}
\usepackage{epsfig,graphics}

\newcommand{\be}{\begin{equation}}
\newcommand{\ee}{\end{equation}}

\newlength{\figsize}
\begin{document}

\begin{titlepage}

\vspace*{0.60in}

\begin{center}
{\large\bf Matching of Wilson loop eigenvalue densities in 1+1, 2+1 and 3+1 dimensions\\ }
\vspace*{0.65in}
{Francis Bursa$^a$\\
\vspace*{.25in}
$^a$\emph{Institut f\"ur Theoretische Physik, 
 Universit\"at Regensburg, 93040 Regensburg, Germany}\\
}
\end{center}

\vspace*{0.50in}

\begin{center}
{\bf Abstract}
\end{center}

We investigate the matching of eigenvalue densities of Wilson loops in $SU(N)$ lattice gauge theory: the eigenvalue densities in 1+1, 2+1 and 3+1 dimensions are nearly identical when the traces of the loops are equal. We show that the matching is present to at least second order in the strong--coupling expansion, and also to second order in perturbation theory. We find that in the continuum limit there is matching at all values of the trace for bare Wilson loops. We confirm numerically that there is matching in these limits and find there are small violations away from them. We discuss the implications for the bulk transitions and for non--analytic gap formation at $N = \infty$ in 2+1 and 3+1 dimensions.

\end{titlepage}

\setcounter{page}{1}
\newpage
\pagestyle{plain}

\section{Introduction}
\label{section_intro}
The eigenvalues of an $SU(N)$ matrix are pure phases, $e^{i \alpha}$, and are gauge invariant for a closed Wilson loop. In~\cite{BurTep06} the eigenvalue distributions of Wilson loops in $D=1+1$ and $D=2+1$ $SU(N)$ lattice gauge theories were studied, and a remarkable result was found: if the eigenvalue distributions of Wilson loops $U_w^{n\times n}$ of size $n\times n$ were calculated in $D$ and $D^{'}$ dimensions, with the couplings $\beta_D$ and $\beta_{D^\prime}$ chosen so that the expectation values of the traces in the fundamental representation $u_w^{n\times n}=\frac{1}{N}\mathrm{Re}\mathrm{Tr}_F\{U_w^{n\times n}\}$ were equal,
\begin{equation}
\langle u_w^{n\times n}(\beta_D)\rangle_D
=
\langle u_w^{n\times n}(\beta_{D^\prime})\rangle_{D^\prime},
\label{eqn_matching traces}
\end{equation}
then the two eigenvalue distributions would match. This was found to apply over a wide range of 
couplings, loop sizes, and values of $N$. Preliminary results showed that the matching also applied 
to 3+1 dimensions.
Matching between eigenvalue distributions in 1+1 and 3+1 dimensions was also found in~\cite{Belova83} for $SU(2)$.
The eigenvalue distribution of \emph{smeared} Wilson loops in 3+1 
dimensions in the large--$N$ limit appears to approach the $D=1+1$ $SU(\infty)$ 
distribution~\cite{NN2006}, suggesting the matching may also work for smeared loops, at least at 
$N=\infty$.
An approximate matching of smeared Wilson loops between 1+1 and 3+1 dimensions has also been observed in $SU(2)$~\cite{Brzoska04}.

If the matching found in~\cite{BurTep06} is exact it will have implications for the 2+1 dimensional and 3+1 dimensional theories. Firstly, in 1+1 dimensional lattice gauge theory there is the Gross--Witten transition~\cite{GW}
at $N=\infty$. This is a third--order phase transition which occurs when the trace of the plaquette is 0.5. At the transition a gap opens in the plaquette eigenvalue density. That is,
on the strong--coupling side of the transition, the eigenvalue density
is non--zero for all $\alpha$, but on the weak--coupling side it 
is only non--zero for some range of angles $-\alpha_c < \alpha < \alpha_c$ with $\alpha_c < \pi$.
If the matching is exact this gap opening, and hence the Gross--Witten transition, would be reproduced in 2+1 and 3+1 dimensions. In 2+1 dimensions there is a bulk transition at this value of the trace which is however different to the Gross--Witten transition, having a specific heat peak which is not present in 1+1 dimensions~\cite{BurTep06}, so if the Gross--Witten transition is present it would have hidden in some way by this transition. Similarly in 3+1 dimensions there is a strong first--order bulk transition at $N=\infty$ (in fact for all $N\geq 5$)~\cite{Campo98,OxG01,OxT05}, which would have to hide the Gross--Witten transition.

Secondly, in 1+1 dimensional $SU(\infty)$ gauge theory in the continuum there is
a non--analyticity, the Durhuus--Olesen transition,
where a gap opens in the Wilson loop eigenvalue distribution at a critical Wilson loop area $A_\mathrm{crit}$~\cite{DurOle,BasGriVian}. If the matching is exact such a non--analyticity would also have to be present in the $D=2+1$ and $D=3+1$ cases. In the latter case, and if it occurs at a finite physical lengthscale, it could provide an explanation for the rapid crossover between perturbative and non--perturbative physics that is observed in the strong interactions~\cite{Teper:2004pk,Jaffe:2004,Narayanan:2005en}. It has also been suggested~\cite{NN2006} that the non--analyticity at $N=\infty$ could be used to match perturbative and non--perturbative physics and thus calculate the string tension in terms of $\Lambda_\mathrm{QCD}$.

However, exact matching would also imply that Casimir Scaling~\cite{APO84},
which is exact in 1+1 dimensions, 
must also be exact in 2+1 and 3+1 dimensions. Casimir Scaling has been found to be a good approximation for tensions between sources in various representations of $SU(3)$ in 3+1 dimensions~\cite{Deldar00,Bali00}, but it is known not to be exact. It is also not exact for $k$--string tensions in both 2+1 and 3+1 dimensions~\cite{LucTep01}. Hence the matching cannot be exact, and must presumably be violated at some level.

The results in~\cite{BurTep06} were not the main purpose of that work and hence were somewhat limited; for example, the matching of traces in~eq. (\ref{eqn_matching traces}) was not exact and there was no error analysis. The purpose of the present work is to carry out a more focused, quantitative study of the matching and of any violations that may be present, and to assess the implications for gap formation.

The structure of the paper is as follows: in the next section we show analytically that matching is present in both the strong--coupling and weak--coupling limits, and we consider the continuum limit. In Section~\ref{section_results} we present our numerical results. We conclude in Section~\ref{section_conclusions}.

\section{Strong--coupling and weak--coupling limits}
\label{section_strong and weak}

\subsection{Strong coupling limit}
\label{section_strong}
The eigenvalue density of a Wilson loop $\rho(\alpha)$ can be expressed as a sum of Fourier components $F_k\ \mathrm{cos}(k\alpha)$. The Fourier coefficients $F_k$ are just the traces of powers of the Wilson loop:
\begin{equation}
F_k=\frac{1}{N}\mathrm{Re}\ \mathrm{Tr}(W^k).
\end{equation}
So if the traces of all powers of two Wilson loops are equal their eigenvalue densities will necessarily be equal too.

Consider the evaluation of such traces in the strong--coupling expansion for a planar Wilson loop of area $A$ in lattice units. Apart from a possible constant term, the leading order contribution comes from the diagram in which we tile the Wilson loop with $k$ sheets of plaquettes, giving a contribution proportional to $\langle u_p\rangle^{kA}$, where $\langle u_p\rangle$ is the trace of the plaquette. This is independent
of the number of dimensions $D$, as are the group--theoretical prefactors.
The trace in the fundamental representation is just $\langle u_p\rangle^{A}$, so the condition~(\ref{eqn_matching traces}) is equivalent to matching the values of $\langle u_p\rangle$. Thus to leading order in the strong--coupling expansion we expect that the traces of all powers of the Wilson loops, and thus their eigenvalue distributions, will indeed match.

There is a possible loophole in this argument. The traces of powers of a Wilson loop can be 
reexpressed as combinations of traces of the Wilson loop in different irreducible 
representations. Some of these representations (for example the adjoint) can be screened in 2+1 
and 3+1 dimensions, leading to a perimeter law instead of an area law. This should lead to 
violations of the matching. However, in practice what happens is that for small loops, the tiling with fewest plaquettes, which will give the leading term in the strong coupling expansion, is the one in which there are sheets of plaquettes tiling the plane of the loop, giving an area law. Tilings with sheets of plaquettes forming ``tubes'' around the outside of the loop, giving a perimeter law, only start to have fewer plaquettes for rather large loops. In the case of a square $L\times L$ loop, this will happen approximately when $L^2=16(L-1)$, or at about $L=15$. Such large loops are difficult to study numerically, and in any case will have extremely small traces: if we are in strong coupling, so $\langle u_p\rangle < 0.4$, say, the trace in the fundamental representation of a $15\times15$ loop will be less than approximately $0.4^{15^2}\simeq10^{-90}$. Traces in other representations will be even smaller. So the eigenvalue density will be indistinguishable from the Haar measure in practice, and there will trivially be matching between different dimensions.

Higher--order terms in the strong--coupling expansion do depend on the dimensionality of space--time, so we expect these to cause violations of the matching. The first such term is one in which one of the sheets of plaquettes tiling the Wilson loop has a ``bump'' in it. This involves an additional four plaquettes compared to tilings in which all the sheets are flat, so is suppressed by a factor $\langle u_p\rangle^4$. However, we show in Appendix~\ref{appendix_bump} that in fact the contributions from the bump are identical in all dimensions when the traces are matched as in eq.~(\ref{eqn_matching traces}).

So the first possible violations come in at the next order in the strong--coupling expansion. In $SU(2)$, the next order is tilings where one sheet has a ``bump'' in it, all the other sheets are flat, and there is an extra pair of plaquettes somewhere in one of the sheets. This is order $\langle u_p\rangle^{kA+6}$. In $SU(3)$ we must add three plaquettes instead of two, so the tiling is of order $\langle u_p\rangle^{kA+7}$, and for higher $N$ we can add another ``bump'', requiring an extra 4 plaquettes and giving tilings of order $\langle u_p\rangle^{kA+8}$. 

These contributions are very small for all but the smallest loops. The largest contributions are those for $k=2$ and for $SU(2)$. Assuming the prefactors to be of order one, and taking the largest $\langle u_p\rangle$ for which the strong--coupling expansion is valid to be approximately 0.4, these will be of order 0.001 for the plaquette, which should be easily observable, but only of order $10^{-6}$ for the $2\times2$ loop, which is unlikely to be detectable, and even smaller and hence almost certainly undetectable for larger loops. Thus we expect that when strong coupling applies, the eigenvalue densities of Wilson loops will be observed to match, except for the very smallest loops.

\subsection{Weak coupling limit}
\label{section_weak}
We now consider the eigenvalue densities of Wilson loops in perturbation theory. As discussed in Section~\ref{section_strong}, if the traces of all powers of two Wilson loops are equal their eigenvalue densities will match. Furthermore, traces of powers of a Wilson loop can be expressed as combinations of traces of the Wilson loop in various representations. Thus to show that two Wilson loops will have matching eigenvalue densities it is sufficient to show that their traces in all representations are the same.

Wilson loops satisfy ``Casimir Scaling'' to second order in perturbation theory; that is, 
the trace of a Wilson loop in a representation $R$ is given by
\begin{equation}
\frac{1}{N}\mathrm{Tr}_RW=e^{-C_R f_D(g^2)},
\label{Casimir scaling}
\end{equation}
where $C_R$ is the quadratic Casimir for the representation $R$, and $f_D(g^2)$ is a 
representation--independent function (which can however depend on the dimension of space--time $D$).

If eq.~(\ref{eqn_matching traces}) is satisfied it is clear the \emph{values} of the functions $f_D$ and $f_{D^\prime}$ must match. Then it follows that the traces
in all other representations will match and hence the eigenvalue spectra
will match also.

Thus it seems that we expect matching when the coupling constant $g^2$ is sufficiently small 
that second--order perturbation theory is a good approximation. However, there is a subtlety 
involving representations with 
numerically large Casimirs $C_R$. These can be important when $g^2$ is small
since then the trace of the Wilson loop approaches 1 so the eigenvalue
distribution must be narrow. Then Fourier components with large $k$,
that is traces of large powers of the Wilson loop, are important, and these can in general involve representations with large $C_R$. Now consider the trace of the Wilson loop in such a representation to leading order in perturbation theory:
\begin{equation}
\frac{1}{N}\mathrm{Tr}_RW=1-W_1C_Rg^2.
\label{Tree level Wilson loop}
\end{equation}
Here $W_1$ is a representation--independent constant which depends on the Wilson loop size and 
the dimension $D$. If $C_R$ is large the quadratic term in~(\ref{Tree level Wilson loop}) can 
also be large even though $g^2$ is small. Then we expect that higher--order terms in 
perturbation theory 
could also be large, including those which violate Casimir Scaling at third order and higher.

However, consider the next term in the expansion of $\mathrm{Tr}_RW$. This is of the form:
\begin{equation}
\frac{1}{N}\mathrm{Tr}_RW=1-W_1C_Rg^2+(\frac{1}{2}W_1^2C_R^2-W_2C_R)g^4.
\label{One loop Wilson loop}
\end{equation}
In the limit of very large $C_R$ the quartic piece is dominated by the $\frac{1}{2}W_1^2C_R^2g^4$ term. This term comes from the exponentiation of the tree--level term $W_1C_Rg^2$:
\begin{equation}
e^{-W_1C_Rg^2}=1-W_1C_Rg^2+\frac{1}{2}W_1^2C_R^2g^4+\mathcal{O}(g^6).
\end{equation}
The exponential will contribute a piece proportional to $C_R^n$ to the $g^{2n}$ term in the 
perturbation expansion. If the remaining parts of the $g^{2n}$ term are lower order in $C_R$, as 
is the case for the $g^2$ and $g^4$ terms, the pieces coming from the exponential will dominate. 
The higher--order terms are not known, but if this is the case to all 
orders in 
$g^2$, which seems plausible, the trace of the Wilson loop will be dominated by a term of the 
form~(\ref{Casimir scaling}) and there will be Casimir Scaling, and hence matching of the eigenvalue densities.

In the particular case of $N=\infty$ there is another argument for matching
in perturbation theory which does not depend on the behaviour
of representations with large Casimirs. For sufficiently weak coupling
the eigenvalues are all near unity and then they
will plausibly be in the same universality class as the Gaussian Unitary
Ensemble, and so the density will be a ``Wigner semicircle''.
This will be true in all dimensions, again leading to matching.

\subsection{Continuum limit}
\label{section_continuum}
The strong--coupling analysis in Section~\ref{section_strong} applies when the lattice coupling 
constant, $\beta=\frac{2N}{g^2}$, is sufficiently small. This is the exact opposite of the 
continuum limit, for which we take $\beta$ to infinity, so we do not expect the analysis to be 
useful in 
the continuum limit. Specifically, we expect that strong--coupling will apply below some critical coupling $\beta_c$. The trace of a Wilson loop of area $A$ is approximately $\langle u_p\rangle^{A}$, where $\langle u_p\rangle\propto \beta$ is the trace of the plaquette. Hence we expect the strong--coupling analysis to apply only for traces less than $\beta_c^A$, which indeed rapidly goes to zero in the continuum limit.

\vspace{2ex}

In 1+1 dimensions in the continuum Wilson loops obey an exact area law, $\langle 
u_w\rangle=e^{-\sigma a^2L^2}$, where $\sigma$ is the string tension, $a$ is the lattice spacing, and $L$ is the size of the loop in lattice units. So the boundary of the regime in which the weak--coupling 
analysis 
in Section~\ref{section_weak} applies corresponds to a finite value of the trace. However, this is not the case in either 2+1 or 3+1 dimensions. In both these cases there is a ``perimeter term'', due to the self--energy of the sources which propagate around the perimeter of the loop. The leading correction to $\ln\langle u_w\rangle$ is $aL V(a)$, where $V(r)$ is the Coulomb potential, which is proportional to $g^2N\ln(r)$ and $g^2N/r$ in $D=2+1$ and $D=3+1$ respectively. Combining this correction with the area term, we get
\begin{equation}
\langle u_w\rangle\propto
\begin{cases}
\mathrm{exp}(-caLg^2N\ln(ag^2N)-\sigma a^2L^2)&  D=2+1 \\ 
\mathrm{exp}(-cg^2NL-\sigma a^2L^2)&  D=3+1 
\end{cases}
\end{equation}
We see that in both cases if we take the continuum limit, keeping $aL$ fixed, at a constant value of the coupling $g^2N$, the trace will decrease. So the boundary of the weak--coupling regimes will shift to lower traces in the continuum limit. This will happen only logarithmically in $D=2+1$, but linearly in $D=3+1$. 

On the weak--coupling side of this boundary the traces of the Wilson loop in different representations will obey Casimir Scaling. But in $D=1+1$ there is Casimir Scaling at all couplings, not just in weak--coupling. And by the arguments of Section~\ref{section_weak}, Casimir Scaling implies matching, so as long as we are on the weak--coupling side of the boundary there will be matching with loops in 1+1 dimensions. For sufficiently large $L$ the minimum trace for which the coupling is weak will become arbitrarily small, and there should be matching at \emph{all} values of the trace.

\section{Results}
\label{section_results}
\subsection{Preliminaries}
\label{results_preliminaries}

We discretise $D$--dimensional Euclidean space--time to a periodic cubic lattice
with lattice spacing $a$ and size $L^D$ in 
lattice units. We assign $SU(N)$ matrices, $U_l$, to the links 
$l$ of the lattice. We use the standard Wilson plaquette action
\begin{equation}
S=\beta \sum_{p} \{1-\frac{1}{N}\mathrm{Re}\mathrm{Tr}U_p\}
\label{eqn_action}
\end{equation}
where $U_p$ is the ordered product of the $SU(N)$ link matrices 
around the boundary of the plaquette $p$. We simulate the theory
using a conventional 
mixture of heat bath and over--relaxation steps applied to the 
$SU(2)$ subgroups of the $SU(N)$ link matrices.

As discussed above, we expect that matching of eigenvalue spectra will
only occur in certain limits. Specifically we expect to see matching in the 
strong--coupling limit, and when the physical size of the loops is small enough that second--order  perturbation theory is valid. As discussed in Section~\ref{section_continuum}, due to the ``perimeter term'' this will be the case for loops with any finite trace in the continuum limit in both $D=2+1$ and $D=3+1$. So we expect to see matching for any value of the trace in the continuum limit also.

Thus we expect that, for each $N$, the part of parameter space in which matching does not occur 
will be constrained on three sides: in the direction of decreasing trace by strong coupling, in the direction of increasing trace by perturbation theory, and in the direction of increasing Wilson loop size (in lattice units) by the perimeter term. A fourth constraint is given by the loop size, which of course cannot become negative. Our strategy will be to determine for each Wilson loop size whether there is indeed a range of traces where matching does not occur, bounded by the strong--coupling regime on one side and the weak--coupling regime on the other, and then to increase the Wilson loop size to check whether this region moves to lower values of the trace, as expected due to the perimeter term.

We begin our investigation in $SU(2)$ since it is easiest to obtain the large statistics necessary to detect small violations there. We then turn to $SU(3)$ and $SU(6)$, to look at how things change with increasing $N$ and, in the latter case, to investigate possible effects due to the first--order bulk transition which is present in $D=3+1$~\cite{Campo98,OxG01,OxT05}. These calculations will allow us to draw some conclusions about what happens at $N=\infty$, and in particular
whether the Gross--Witten and Durhuus--Olesen transitions will be reproduced
there.

To perform a comparison for a particular loop size $n$ at a particular value of the trace 
$\langle u_w\rangle$ we first generate configurations at couplings $\beta_\mathrm{run}$ in $D=1+1$, $D=2+1$, and $D=3+1$. We then reweight to couplings $\beta_\mathrm{rew}$ where the traces match the target value of the trace exactly. On every sweep we calculate Wilson loops on every 1 to 2 lattice sites, and extract their eigenvalues. We store these in a histogram with a large number $n_\mathrm{bin}$ (typically 4096) of bins.

From these histograms we estimate the first $\frac{n_\mathrm{bin}}{2}$ Fourier components $F_k$, 
defined by $F_k=\frac{1}{N}\sum_{i=1}^{N} \cos(k\alpha_i)$ where $\lambda_i=e^{i \alpha_i}$ is the 
$i$'th eigenvalue, and the first $\frac{n_\mathrm{bin}}{2}$ moments $\mu_k$ of the eigenvalue 
distribution. The finite width of the histogram bins means that these estimates will have errors 
of $\mathcal{O}\left(\frac{k^2}{n_{bin}^2}\right)$. As we shall see below, the relevant values 
of $k$ are usually in the range 2 to 20, for which this is $\sim 10^{-5}$ and hence 
negligible. We compare the Fourier components and the moments of the eigenvalue distributions, searching for any statistically significant differences between the two distributions. We also compare the two histograms directly. To reduce the noise we also construct coarser histograms with $\frac{n_\mathrm{bin}}{2}, \frac{n_\mathrm{bin}}{4}, \hdots, 4$ bins, and compare these.

Of course, these channels in which violations can show up are not independent; a generic violation will give a signal in all of them. What we find in practice is that the Fourier components are the most sensitive. That is, as we increase the statistics (at a loop size and trace for which we expect to see a violation), we first see a statistically significant difference in the Fourier components, and only for higher levels of statistics do the differences in the moments and histograms become significant. Thus from here on we will concentrate on our results for the Fourier components.

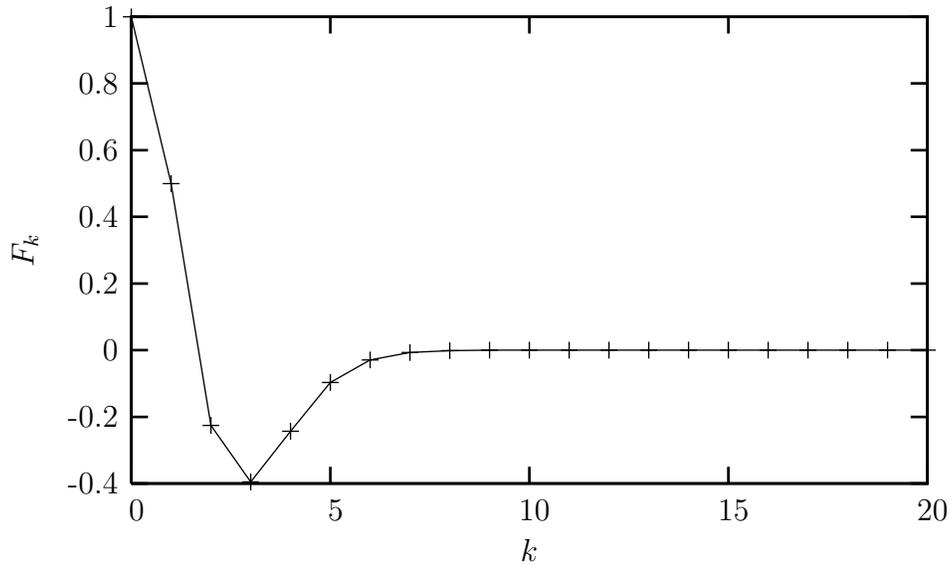
\begin{figure}[p]
\begin{center}
\leavevmode
\input{figures/typicalfourier}
\caption{Fourier components $F_k$ for the plaquette in $SU(2)$ with $\langle u_p\rangle=0.5$ on a $4^2$ lattice.
\label{Typical Fourier}}
\end{center}
\end{figure}

Furthermore, there is also a pattern in which Fourier components show significant differences first. We find that in $SU(2)$, a graph of the Fourier coefficients $F_k$ typically looks like that shown in Fig.~\ref{Typical Fourier}: $F_0$ is always 1 by definition, $F_1$ is the trace and is always positive, and for larger $k$ there is a rapid decrease in $F_k$ to some maximally negative value, followed by an asymptotic approach to zero. We find that the Fourier components where significant differences first show up are those around the minimum. This is perhaps not surprising; is is easier to observe a small absolute difference if the quantity being observed is large. If the Fourier component is large and positive it is not far from 1 and thus largely perturbative. In this case we expect that we will not see any violations, as we discussed in Section~\ref{section_weak}. So indeed we expect that violations should be easiest to see for the Fourier components which are large and negative.

For larger $N$ the graph of Fourier components $F_k$ has more maxima and minima; we then find significant differences first appear in the Fourier components around the first minimum. We will refer to the value of $k$ at which the first minimum in $F_k$ occurs as $k^\prime$. The corresponding Fourier component is then $F_{k^\prime}$. Since the significant differences first appear in the Fourier components around the first minimum $F_{k^\prime}$ is a useful quantity to use as a measure of the size of the violations, if any.

\subsection{$SU(2)$}
We now turn to our results, starting with those for $SU(2)$. Presumably violations of eigenvalue spectrum matching will be largest when we are furthest from the limits analysed in Section~\ref{section_strong and weak} where we expect matching. This leads us to look at small loops at moderate values of the trace. We begin with the smallest loop, the plaquette.

\begin{figure}[p]
\begin{center}
\leavevmode
\input{figures/su2plaq_0.7densities}
\caption{Eigenvalue densities for the plaquette in $SU(2)$ with $\langle u_p\rangle=0.7$ in $D=1+1$~(solid line), $D=2+1$~(dashed) and $D=3+1$~(dotted).
\label{Biggest action diffs}}
\end{center}
\end{figure}
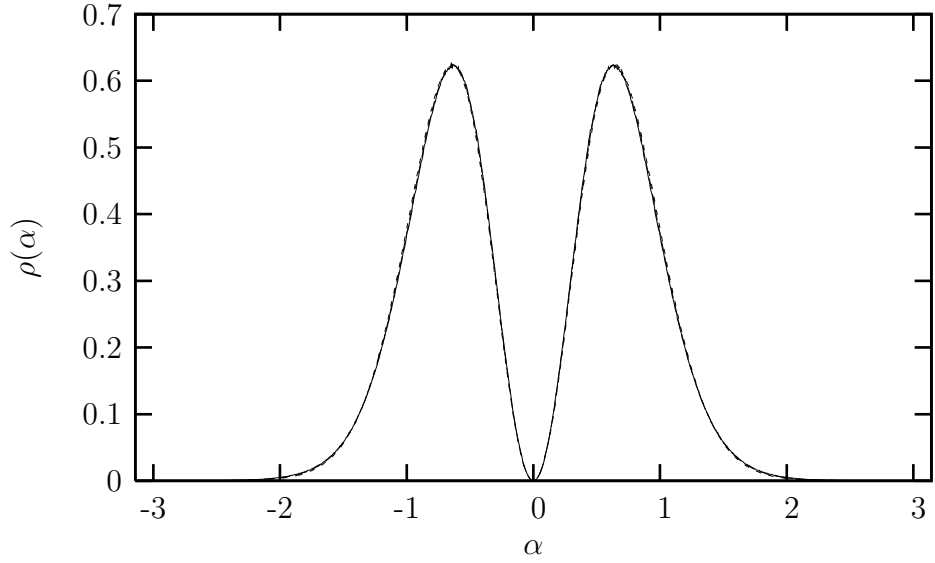

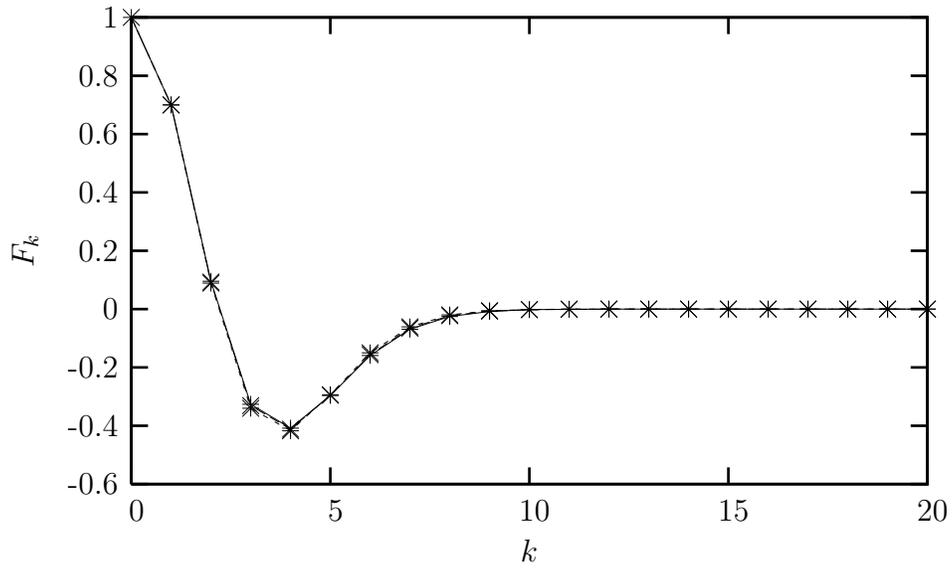
\begin{figure}[p]
\begin{center}
\leavevmode
\input{figures/su2plaq_0.7fouriers}
\caption{Fourier components $F_k$ for the plaquette in $SU(2)$ with $\langle u_p\rangle=0.7$ in $D=1+1$~($+$), $D=2+1$~($\times$) and $D=3+1$~($\ast$).
\label{Biggest Fourier diffs}}
\end{center}
\end{figure}

Our results for the plaquette are summarised in Table~\ref{table_plaquettes SU(2)}.
We see that, as found in~\cite{BurTep06}, the eigenvalue distributions in different numbers of dimensions when the traces are matched are indeed very similar. We illustrate this in Fig.~\ref{Biggest action diffs}, where we plot the plaquette eigenvalue densities at $\langle u_p\rangle=0.7$, where the differences are largest. Despite this, we see that the differences are barely visible by eye. Turning to the Fourier components, which we plot in Fig.~\ref{Biggest Fourier diffs}, we see that the differences are slightly more clear, but still extremely small. Note that these are the largest differences we see; at most values of the trace the differences would be impossible to see on such plots.

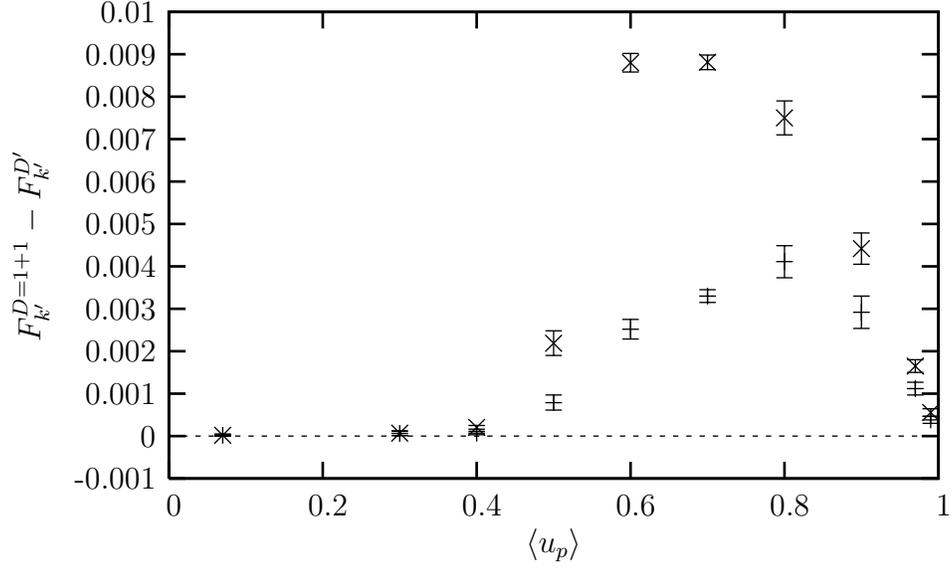
\begin{figure}[p]
\begin{center}
\leavevmode
\input{figures/su2plaqdiffs}
\caption{Difference between $F_{k^\prime}$ in 1+1 dimensions and $D^\prime$ dimensions, for $D^\prime=2+1$~($+$) and $D^\prime=3+1$~($\times$), as a function of the trace $\langle u_p\rangle$. All for the plaquette in $SU(2)$.
\label{SU(2) plaquette differences}}
\end{center}
\end{figure}

We see that, as discussed at the end of Section~\ref{results_preliminaries}, the largest differences in Fig.~\ref{Biggest Fourier diffs} are around the Fourier component at the minimum, $F_{k^\prime}$. To investigate how the differences vary with the trace $\langle u_p\rangle$, we plot in Fig.~\ref{SU(2) plaquette differences} the differences in $F_{k^\prime}$ between 1+1 and 2+1 dimensions and between 1+1 and 3+1 dimensions, as a function of the trace.

Proceeding from small to large values of the trace, we see first of all that
there are no detectable violations at $\langle u_p\rangle=0.07$, where the coupling is very
strong. Violations become visible, possibly at $\langle u_p\rangle=0.3$, and certainly
at $\langle u_p\rangle=0.4$, still within the strong--coupling region. After this they
rapidly increase, reaching a maximum when the trace is around 0.8 for the
difference between 1+1 and 2+1 dimensions and 0.6--0.7 for the difference
between 1+1 and 3+1 dimensions. Finally they decrease again as we approach
$\langle u_p\rangle=1$. We see the same features for other values of $k$ near $k^\prime$, and
also (though with relatively larger statistical errors) for the moments
and histograms of the distributions.

All this is as we would expect. There is a large range of traces in strong
coupling for which there are no observable violations of the matching.
Violations start to appear around the upper limit of strong coupling,
due to the higher--order terms in the strong--coupling expansion 
(though they are smaller than we estimated; presumably the prefactors are small). Then there is an intermediate--coupling region where there is no reason to expect matching and where we indeed see violations.  Finally the coupling becomes weak and the matching reappears due to Casimir Scaling.

For the larger values of the trace, the volumes used in $D=2+1$ and especially in $D=3+1$ are extremely small. Indeed, our calculations for $\langle u_p\rangle=0.99$ in 3+1 dimensions were carried out at a coupling $\beta=75.2$, for which the lattice spacing is $\sim 10^{-80}$ fm! Thus one might worry that the small differences we see there are not due to true infinite--volume differences between the theories but are instead due to finite--volume effects. To check for this, we have repeated our calculations on different sized lattices. We find no difference in the violations in matching as we change the lattice size, making it unlikely that these are finite--volume effects.

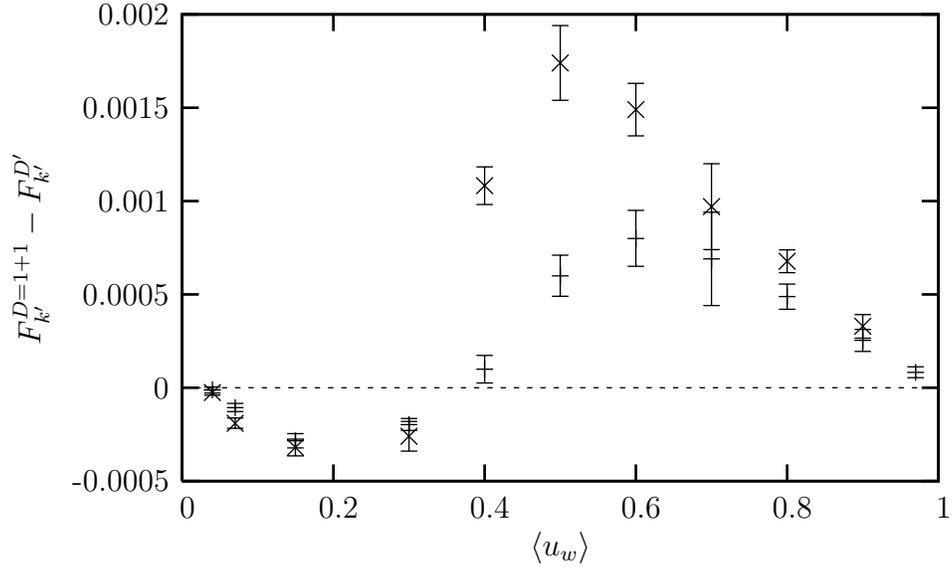
\begin{figure}[p]
\begin{center}
\leavevmode
\input{figures/su2_2x2diffs}
\caption{Difference between $F_{k^\prime}$ in 1+1 dimensions and $D^\prime$ dimensions, for $D^\prime=2+1$~($+$) and $D^\prime=3+1$~($\times$), as a function of the trace $\langle u_w\rangle$. All for the $2 \times 2$ Wilson loop in $SU(2)$.
\label{SU(2) 2x2 differences}}
\end{center}
\end{figure}

We now turn to our results for $2\times2$ Wilson loops. These are listed in Table~\ref{table_2x2 
SU(2)}, and the corresponding differences in $F_{k^\prime}$ are plotted in Fig.~\ref{SU(2) 2x2 differences}. We can compare these to our results for the plaquette in Table~\ref{table_plaquettes SU(2)} and Fig.~\ref{SU(2) plaquette differences}. Proceeding from small to large traces, we see first of all that there are violations for $2\times2$ loops at much smaller values of the trace than for the plaquette. This is because the boundary of strong--coupling has shifted to a much smaller trace --- indeed at $\langle u_w\rangle=0.04$ the trace of the plaquette is 0.447, 0.444 and 0.438 in 1+1, 2+1 and 3+1 dimensions respectively, so we are already near
the edge of strong coupling. The violations of matching are just becoming detectable here. They then become larger, as for the plaquette, but reach maxima and start decreasing at lower values of the trace than for the plaquette. This is presumably the effect of the perimeter term lowering the value of the trace at which weak--coupling begins. We see that the peak is shifted to a lower value of the trace
in 3+1 dimensions compared to 2+1 dimensions, as expected since the
perimeter term is stronger in the former case.

We also see that the magnitude of the violations is smaller by a factor of
5 or so compared to the plaquette. This is not just because of the perimeter
term, which causes larger loops to be at a weaker coupling at the same
value of the trace. For example, comparing the plaquette at a trace of 0.8
to the $2\times2$ Wilson loop at a trace of 0.5, we see the violations are much
larger for the plaquette, despite the fact that the couplings are roughly
the same. Presumably what is going on is that there are relatively large
lattice corrections to Casimir Scaling for the plaquette, which spoil
the matching, but these corrections are smaller for the $2\times2$ loop.

Finally we look at our results for $4\times4$ loops, which are listed in
Table~\ref{table_4x4 SU(2)} and plotted in Fig.~\ref{SU(2) 4x4 differences}.
We see that the difference between 1+1 and 2+1 dimensions is non--zero
for much smaller values of the trace than for the smaller loops, as expected. 
For the difference between 1+1 and 3+1 dimensions the situation is less clear
--- the peak appears to be around $\langle u_w\rangle=0.5$, as for $2\times 2$ loops ---
but it is clear the violations are now extremely small.

Above we have described the behaviour of the Fourier component $F_{k^\prime}$
for the $2\times 2$ and $4\times4$ loops, but as for the plaquettes we see
essentially the same behaviour for other Fourier components with $k$
near $k^\prime$, and for the moments and histograms of the eigenvalue densities.

\subsection{$SU(3)$ and $SU(6)$}
We now turn to our results for $SU(3)$ and $SU(6)$. Our results here are less extensive and are aimed mainly at establishing that the features we saw for $SU(2)$ carry over to larger $N$.

\begin{figure}[p]
\begin{center}
\leavevmode
\input{figures/su3plaqdiffs}
\caption{Difference between $F_{k^\prime}$ in 1+1 dimensions and $D^\prime$ dimensions, for $D^\prime=2+1$~($+$) and $D^\prime=3+1$~($\times$), as a function of the trace $\langle u_p\rangle$. All for the plaquette in $SU(3)$.
\label{SU(3) plaquette differences}}
\end{center}
\end{figure}
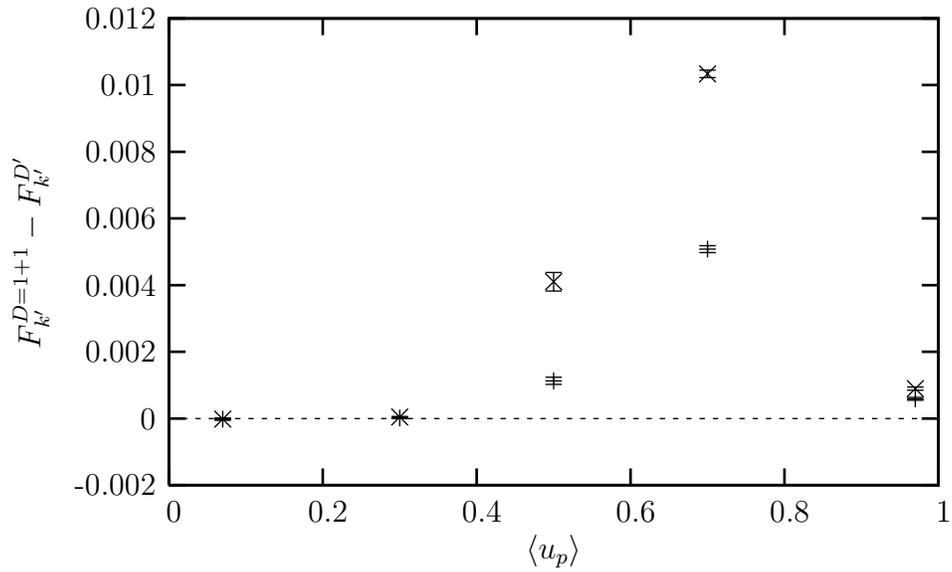

Our results for the plaquette in $SU(3)$ are summarised in
Table~\ref{table_plaquettes SU(3)} and are plotted in
Fig.~\ref{SU(3) plaquette differences}. We see a similar pattern to the
corresponding results in $SU(2)$: there is very good matching at small
values of the trace, followed by violations of the matching at intermediate
values, which decrease again as we approach $\langle u_p\rangle=1$. The violations
appear to be somewhat larger than for $SU(2)$.

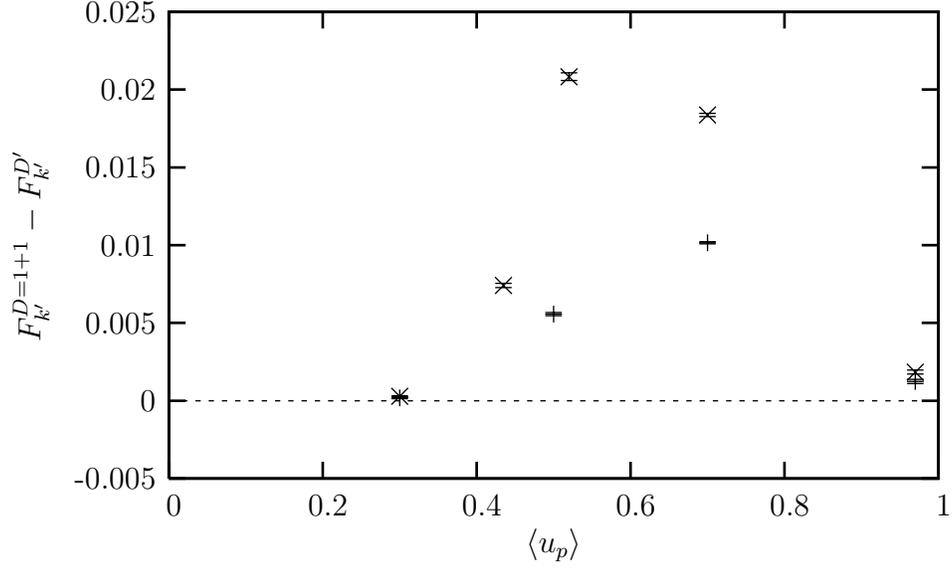
\begin{figure}[p]
\begin{center}
\leavevmode
\input{figures/su6plaqdiffs}
\caption{Difference between $F_{k^\prime}$ in 1+1 dimensions and $D^\prime$ dimensions, for $D^\prime=2+1$~($+$) and $D^\prime=3+1$~($\times$), as a function of the trace $\langle u_p\rangle$. All for the plaquette in $SU(6)$.
\label{SU(6) plaquette differences}}
\end{center}
\end{figure}

Our results for the plaquette in $SU(6)$ are summarised in Table~\ref{table_plaquettes SU(6)} and are plotted in Fig.~\ref{SU(6) plaquette differences}. Note that due to the bulk transition there is a range of $\langle u_p\rangle$ around 0.5 which is inaccessible in 3+1 dimensions; the closest we can get is $\langle u_p\rangle=0.435$ and $\langle u_p\rangle=0.52$ on the strongly--coupled and weakly--coupled sides respectively. The overall pattern of violations is similar to that for $SU(2)$ and $SU(3)$: again we have good matching at small and large values of the trace and violations in between.There is no sign of anything exceptional happening around $\langle u_p\rangle=0.5$,
the region where the bulk transitions occur.

The maximum violations now appear quite large ($\sim0.02$). 
To judge whether this is in fact large, we can compare
the values of $F_{k^\prime}$ at a trace of 0.5 
in Tables~\ref{table_plaquettes SU(6)} and~\ref{table_2x2 SU(6)}.
We see that they are around $-0.04$ for the plaquettes and $-0.105$ for the
$2\times2$ loops, so distributions in which the $F_{k^\prime}$ differ by $\sim0.06$
are certainly possible. So the violations observed are still significantly
smaller than they could be.

\begin{figure}[p]
\begin{center}
\leavevmode
\input{figures/su3_2x2diffs}
\caption{Difference between $F_{k^\prime}$ in 1+1 dimensions and $D^\prime$ dimensions, for $D^\prime=2+1$~($+$) and $D^\prime=3+1$~($\times$), as a function of the trace $\langle u_w\rangle$. All for the $2 \times 2$ Wilson loop in $SU(3)$.
\label{SU(3) 2x2 differences}}
\end{center}
\end{figure}
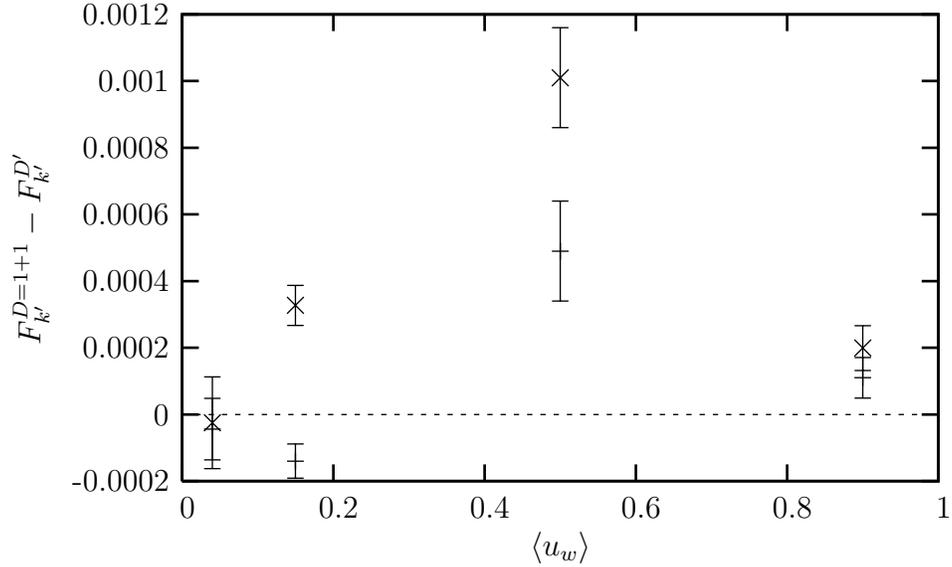

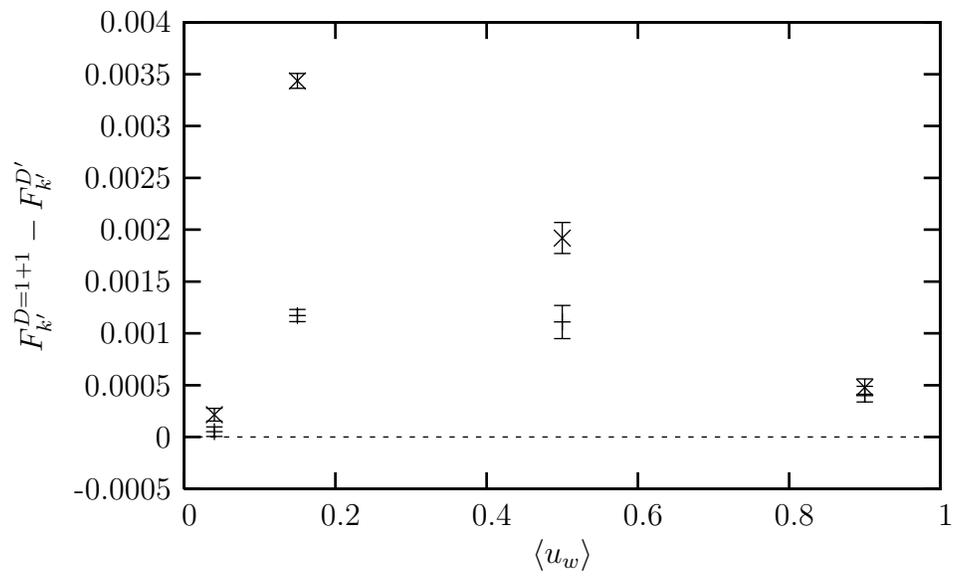
\begin{figure}[p]
\begin{center}
\leavevmode
\input{figures/su6_2x2diffs}
\caption{Difference between $F_{k^\prime}$ in 1+1 dimensions and $D^\prime$ dimensions, for $D^\prime=2+1$~($+$) and $D^\prime=3+1$~($\times$), as a function of the trace $\langle u_w\rangle$. All for the $2 \times 2$ Wilson loop in $SU(6)$.
\label{SU(6) 2x2 differences}}
\end{center}
\end{figure}

Our results for $2\times2$ loops are summarised in Tables~\ref{table_2x2 SU(3)} and~\ref{table_2x2 SU(6)} for $SU(3)$ and $SU(6)$ respectively. They are plotted in Figs.~\ref{SU(3) 2x2 differences} and~\ref{SU(6) 2x2 differences}. Again, the pattern of violations is broadly similar to that seen in $SU(2)$, shown in Fig.~\ref{SU(2) 2x2 differences}. The biggest change is that the peak violations are larger and occur at a lower value of the trace in $SU(6)$. Why this should be is unclear; it occurs in the intermediate--coupling region where we do not have an analytic understanding of the source of the violations. In any case the violations are still small, and they decrease in both the strong--coupling and weak--coupling regions, as they should.

The differences described above are for the Fourier component $F_{k^\prime}$,
but we see the same patterns for nearby Fourier components and for the
moments and histograms.

In summary, we see that despite some differences in the details, the violations in $SU(3)$ and $SU(6)$ follow the same broad patterns as in $SU(2)$, and are consistent with our expectations based on strong--coupling and weak--coupling analysis.

\section{Conclusions}
\label{section_conclusions}
We have shown that the remarkable matching between eigenvalue spectra of Wilson loops observed in~\cite{BurTep06,Belova83,NN2006,Brzoska04} can be understood analytically in both the strong--coupling and weak--coupling limits. In the strong--coupling limit, it follows from the fact that the leading--order contributions come from tilings that are flat, and hence the same, in all dimensions. In the weak--coupling limit it follows from the fact that Wilson loops satisfy Casimir Scaling up to second order in perturbation theory in all dimensions. Due to the ``perimeter term'', perturbation theory applies at all values of the trace in the continuum limit, so the matching will apply there also.

However, our numerical results have shown that away from these limits the matching of eigenvalue densities is not exact. These differences remain small, of order 0.02 at most. Why this should be when we are well away from both the strong--coupling and weak--coupling regimes is unclear.

Apart from their small size, the overall pattern of violations follows our expectations: it approaches zero in both the strong--coupling and weak--coupling limits, and has a peak at intermediate couplings. The trace at which this peak occurs decreases as the size of the loop increases, as expected due to the effect of the perimeter term.

In the large--$N$ limit in 1+1 dimensions, the Gross--Witten transition occurs when the trace of the plaquette is 0.5~\cite{GW}. At this value of the trace we observe finite differences between the eigenvalue densities in 1+1 and 2+1 dimensions which show no sign of decreasing (indeed, they appear to increase) with $N$. This implies that the Gross--Witten transition will not be reproduced in $D=2+1$, which is consistent with the results found in~\cite{BurTep06}, where a peak was observed in the specific heat at the bulk transition in 2+1 dimensions which is not present at the Gross--Witten transition.

In 3+1 dimensions a jump in the trace of the plaquette occurs at the bulk transition making values around 0.5 inaccessible. 
However, we see differences in the eigenvalue densities on either side of this jump, suggesting that even if the bulk transition was removed there would still be significant differences at a trace of 0.5 and thus any ``underlying'' transition would again be different to the Gross--Witten transition.

The situation is different for the Durhuus--Olesen non--analyticity in
Wilson loop eigenvalue densities which occurs at $N=\infty$ in the continuum in 1+1 dimensions~\cite{DurOle,BasGriVian}. This occurs when the trace of the Wilson loop is $e^{-2}\approx 0.135$. Due to the perimeter term this will be in the weak--coupling limit in the continuum limit, where we expect matching. Thus we expect that the non--analyticity will be reproduced in both 2+1 and 3+1 dimensions. However, since it occurs in the weak--coupling limit it is in the far ultraviolet, so it seems unlikely to be physically significant.

This problem could be avoided by using smeared, instead of bare, Wilson loops, as was done in~\cite{NN2006,Brzoska04}. If the perimeter term was eliminated completely by smearing the Wilson loops would have a trace of $e^{-2}$ at a constant lengthscale in physical units in the continuum limit (although the lengthscale would be determined by the details of the smearing procedure). However, the Wilson loops would then be given completely by the area term, which does not satisfy Casimir Scaling exactly in 2+1 or 3+1 dimensions. Then the traces of Wilson loops in different representations will not match and so neither will the eigenvalue densities, and so the non--analyticity in $D=1+1$ will \emph{not} be reproduced for smeared Wilson loops in 2+1 or 3+1 dimensions.

We still expect gap formation to occur, since the eigenvalue density becomes flat as $u_w\to0$ on the one hand, and forms a Wigner semicircle as $u_w\to1$ on the other. In between a gap must form.
However, since the lengthscale it will occur at will depend on the details of the smearing procedure, it is unclear whether the gap formation will have any physical significance. Also, since the eigenvalue density will be different from that at the Durhuus--Olesen non--analyticity, it is not clear what universality class the gap formation will fall into. This is a question that needs to be resolved in order to implement the idea of matching perturbative and non--perturbative physics through the transition, as set out in~\cite{NN2006}.

In summary, we have understood the matching of Wilson loop eigenvalue densities in both the strong--coupling and weak--coupling limits, and shown it is violated away from these limits. These violations imply that the Gross--Witten transition will not be matched in 2+1 and 3+1 dimensions. The Durhuus--Olesen non--analyticity will be reproduced for bare loops in 2+1 and 3+1 dimensions, but at an ultraviolet length--scale, and it will not be reproduced for smeared loops.

\vspace*{0.20in}

\section*{Acknowledgments}
We are grateful to Mike Teper for many useful discussions throughout this work.
This work was supported by the EC Hadron Physics I3 Contract RII3-CT-2004-506078
and by the BMBF.

\vspace*{0.20in}

\appendix

\section{Matching of eigenvalue spectra in the strong--coupling expansion}
\label{appendix_bump}

As shown in Section~\ref{section_strong}, matching the traces of two Wilson loops according to eq.~(\ref{eqn_matching traces}) leads to matching of the traces of all powers of the loops, and hence to matching of the spectra, to leading order in the strong--coupling expansion. In this Appendix we consider the next order, described by tilings with one ``bump'', and show that the matching also works there.

We consider first the plaquette in $SU(2)$. The strong--coupling expansion for the trace in $D=1+1$ dimensions is
\begin{equation}
\mathrm{Tr}(P)=a_1\beta_1+a_3\beta_1^3+a_5\beta_1^5+\mathcal{O}(\beta_1^7),
\label{Plaquette 1+1}
\end{equation}
and in $D^{\prime}=2+1,3+1$ dimensions it is
\begin{equation}
\mathrm{Tr}(P)=a_1\beta_{D^{\prime}}+a_3\beta_{D^{\prime}}^3+b_5\beta_{D^{\prime}}^5+\mathcal{O}(\beta_{D^{\prime}}^7).
\label{Plaquette 3+1}
\end{equation}
The $\beta$ and $\beta^3$ terms are the same in both cases since they correspond to only planar tilings, but the $\beta^5$ terms are different since they can include a ``bump'' in 2+1 or 3+1 dimensions.

Similarly, the strong--coupling expansion of the trace of the plaquette squared is
\begin{equation}
\mathrm{Tr}(P^2)=c_2\beta_1^2+c_4\beta_1^4+c_6\beta_1^6+\mathcal{O}(\beta_1^8)
\label{P^2 1+1}
\end{equation}
in 1+1 dimensions, and
\begin{equation}
\mathrm{Tr}(P^2)=c_2\beta_{D^{\prime}}^2+c_4\beta_{D^{\prime}}^4+d_6\beta_{D^{\prime}}^6+\mathcal{O}(\beta_{D^{\prime}}^8)
\label{P^2 3+1}
\end{equation}
in $D^{\prime}$ dimensions. The first term is $\mathcal{O}(\beta^2)$, rather than $\mathcal{O}(\beta)$, since we need to tile the doubly-winding loop with two plaquettes, the $\beta^4$ term has 4 plaquettes, the $ \beta^6$ term has either 6 plaquettes in the plane or 1 in the plane and 5 in a bump, and so on.

The matching condition~(\ref{eqn_matching traces}) is that~(\ref{Plaquette 1+1}) and~(\ref{Plaquette 3+1}) are equal. We can expand $\beta_{D^{\prime}}$ in $\beta_1$; the condition gives
\begin{equation}
\beta_{D^{\prime}}=\beta_1+\frac{a_5-b_5}{a_1}\beta_1^5+\mathcal{O}(\beta_1^7).
\label{matching}
\end{equation}
Substituting this into~(\ref{P^2 3+1}) we find the values of $\mathrm{Tr}(P^2)$ will match in 1+1 and $D^{\prime}$ dimensions if the coefficients in the expansion are related as follows:
\begin{equation}
d_6=c_6+2\frac{c_2}{a_1}(b_5-a_5).
\label{condition}
\end{equation}
The first term on the right--hand side is the $\mathcal{O}(\beta_1^6)$ term in~(\ref{P^2 1+1}), which is the same as the contribution from flat tilings to $d_6$. The second term should then give the contribution from tilings with a bump if~(\ref{condition}) is to be satisfied. Now $b_5-a_5$ is the difference between the $\mathcal{O}(\beta^5)$ contributions to~(\ref{Plaquette 3+1}) and~(\ref{Plaquette 1+1}) i.e. it is the contribution from a ``bump'' in a single sheet. The factor $\frac{c_2}{a_1}$ accounts for the different multiplicative factors we require when tiling a doubly--winding loop with two sheets rather than a singly--winding loop with one sheet. Finally the factor of 2 accounts for the fact that the ``bump'' can occur in either of the two sheets. Putting all these together gives exactly the correct expression for $d_6$, so~(\ref{condition}) is indeed satisfied.

Generalising this to the trace of the $k$'th power of the Wilson loop $\mathrm{Tr}(P^k)$ is straightforward. It is also straightforward to generalise to Wilson loops of area $A$. Finally, changing $N$ changes the number of plaquettes we can add to a tiling at one site from 2 to $N$; this will change the $\mathcal{O}(\beta^3)$ terms in~(\ref{Plaquette 1+1}) and~(\ref{Plaquette 3+1}) to $\mathcal{O}(\beta^{1+N})$, but it will not affect the ``bump'' contribution at $\mathcal{O}(\beta^5)$ since this is singly--tiled everywhere. Hence the proof above will also work for all values of $N$.

So the first term in the strong--coupling expansion corresponding to tilings in which one of the sheets has a ``bump'' will indeed give matching of traces of powers of the Wilson loop if the matching condition~(\ref{eqn_matching traces}) is satisfied. Hence the eigenvalue densities will also match up to that order in the strong--coupling expansion.

\vfill\eject

\begin{table}[p]
\begin{center}
\begin{tabular}{|c|c|r@{.}l|r@{.}l|c|r@{.}l|} \hline
\multicolumn{9}{|c|}{Plaquette in $SU(2)$} \\ \hline
$\langle u_p\rangle$ & Lattice size & \multicolumn{2}{|c|}{$\beta_\mathrm{run}$} & \multicolumn{2}{|c|}{$\beta_\mathrm{rew}$} & $k^\prime$ & \multicolumn{2}{|c|}{$F_{k^\prime}$} \\ \hline
0.07 & $4^2$ & 0&281 & 0&28095 & 2 & -0&495066(20) \\
& $4^3$ & 0&2807 & 0&28078 & & -0&495087(13) \\
& $4^4$ & 0&2808 & 0&28082 & & -0&495087(8) \\ \hline
0.3 & $4^2$ & 1&279 & 1&27893 & 2 & -0&407127(32) \\
& $4^3$ & 1&2605 & 1&26049 & & -0&407189(37) \\
& $4^4$ & 1&242 & 1&24140 & & -0&407195(45) \\ \hline
0.4 & $4^2$ & 1&803 & 1&80340 & 3 & -0&351050(35) \\
& $4^3$ & 1&729 & 1&72878 & & -0&351121(18) \\
& $4^4$ & 1&648 & 1&64803 & & -0&351256(25) \\ \hline
0.5 & $4^2$ & 2&444 & 2&44677 & 3 & -0&39550(17) \\
& $4^3$ & 2&235 & 2&23511 & & -0&39629(7) \\
& $4^4$ & 1&996 & 1&99071 & & -0&39769(24) \\ \hline
0.6 & $8^2$ & 3&31 & 3&3114 & 3 & -0&39758(18) \\
& $8^3$ & 2&83 & 2&8279 & & -0&40010(14) \\
& $8^4$ & 2&29 & 2&2933 & & -0&40638(13) \\ \hline
0.7 & $8^2$ & 4&64 & 4&6440 & 4 & -0&40855(14) \\
& $8^3$ & 3&67 & 3&6734 & & -0&41185(6) \\
& $8^4$ & 2&804 & 2&8033 & & -0&41736(9) \\ \hline
0.8 & $8^2$ & 7&17 & 7&1940 & 5 & -0&41510(35) \\
& $8^3$ & 5&3 & 5&3006 & & -0&41921(14) \\
& $8^4$ & 4&0 & 4&0062 & & -0&42260(20) \\ \hline
0.9 & $8^2$ & 14&7 & 14&719 & 7 & -0&43006(34) \\
& $8^3$ & 10&27 & 10&253 & & -0&43298(17) \\
& $8^4$ & 7&73 & 7&729 & & -0&43448(15) \\ \hline
0.97 & $8^2$ & 49&4 & 49&598 & 12 & -0&44220(14) \\
& $8^3$ & 33&5 & 33&545 & & -0&44332(5) \\
& $8^4$ & 25&2 & 25&218 & & -0&44385(5) \\ \hline
0.99 & $12^2$ & 149&01 & 149&534 & 21 & -0&444903(84) \\
& $12^3$ & 100&15 & 100&211 & & -0&445286(19) \\
& $12^4$ & 75&2 & 75&194 & & -0&445457(33) \\  \hline
\end{tabular}
\caption{Fourier components $F_{k^\prime}$ at the minimum $k^\prime$ of the Fourier spectrum of the eigenvalue distribution, for plaquettes in $SU(2)$. 
\label{table_plaquettes SU(2)}}
\end{center}
\end{table}

\begin{table}[p]
\begin{center}
\begin{tabular}{|c|c|r@{.}l|r@{.}l|c|r@{.}l|} \hline
\multicolumn{9}{|c|}{$2 \times 2$ Wilson loops in $SU(2)$} \\ \hline
$\langle u_w\rangle$ & Lattice size & \multicolumn{2}{|c|}{$\beta_\mathrm{run}$} & \multicolumn{2}{|c|}{$\beta_\mathrm{rew}$} & $k^\prime$ & \multicolumn{2}{|c|}{$F_{k^\prime}$} \\ \hline
0.04 & $8^2$ & 2&0883 & 2&08746 & 2 & -0&499370(10) \\
& $8^3$ & 1&9434 & 1&94312 & & -0&499358(13) \\
& $8^4$ & 1&7915 & 1&79090 & & -0&499344(10) \\ \hline
0.07 & $8^2$ & 2&555 & 2&55399 & 2 & -0&497887(13) \\
& $8^3$ & 2&273 & 2&27472 & & -0&497781(18) \\
& $8^4$ & 1&985 & 1&98421 & & -0&497697(25) \\ \hline
0.15 & $12^2$ & 3&552 & 3&55501 & 2 & -0&487888(29) \\
& $12^3$ & 2&877 & 2&87655 & & -0&487604(24) \\
& $12^4$ & 2&233 & 2&23233 & & -0&487568(34) \\ \hline
0.3 & $12^2$ & 5&438 & 5&4385 & 2 & -0&435366(23) \\
& $12^3$ & 3&85 & 3&8501 & & -0&435169(22) \\
& $12^4$ & 2&646 & 2&6459 & & -0&436626(77) \\ \hline
0.4 & $12^2$ & 7&019 & 7&0179 & 3 & -0&376907(63) \\
& $12^3$ & 4&642 & 4&6417 & & -0&377007(39) \\
& $12^4$ & 3&119 & 3&1167 & & -0&377989(79) \\ \hline
0.5 & $12^2$ & 9&13 & 9&1355 & 3 & -0&43410(7) \\
& $12^3$ & 5&7 & 5&7104 & & -0&43470(8) \\
& $12^4$ & 3&8 & 3&8015 & & -0&43584(19) \\ \hline
0.6 & $12^2$ & 12&22 & 12&2357 & 3 & -0&44130(13) \\
& $12^3$ & 7&285 & 7&2886 & & -0&44210(8) \\
& $12^4$ & 4&82 & 4&8230 & & -0&44279(5) \\ \hline
0.7 & $12^2$ & 17&3 & 17&3138 & 4 & -0&43348(21) \\
& $12^3$ & 9&87 & 9&8833 & & -0&43417(14) \\
& $12^4$ & 6&54 & 6&5368 & & -0&43445(10) \\ \hline
0.8 & $12^2$ & 27&39 & 27&3759 & 5 & -0&432074(48) \\
& $12^3$ & 15&08 & 15&0791 & & -0&432562(47) \\
& $12^4$ & 9&97 & 9&9604 & & -0&432752(37) \\ \hline
0.9 & $16^2$ & 57&36 & 57&4346 & 7 & -0&439275(47) \\
& $16^3$ & 30&66 & 30&6838 & & -0&439529(36) \\
& $16^4$ & 20&23 & 20&2416 & & -0&439604(42) \\ \hline
0.97 & $24^2$ & 197&5 & 197&411 & 12 & -0&445541(25) \\
& $24^3$ & 103&5 & 103&483 & & -0&445624(14) \\ \hline
\end{tabular}
\caption{Fourier components $F_{k^\prime}$ at the minimum $k^\prime$ of the Fourier spectrum of the eigenvalue distribution, for $2 \times 2$ Wilson loops in $SU(2)$. 
\label{table_2x2 SU(2)}}
\end{center}
\end{table}

\begin{table}[p]
\begin{center}
\begin{tabular}{|c|c|r@{.}l|r@{.}l|c|r@{.}l|} \hline
\multicolumn{9}{|c|}{$4 \times 4$ Wilson loops in $SU(2)$} \\ \hline
$\langle u_w\rangle$ & Lattice size & \multicolumn{2}{|c|}{$\beta_\mathrm{run}$} & \multicolumn{2}{|c|}{$\beta_\mathrm{rew}$} & $k^\prime$ & \multicolumn{2}{|c|}{$F_{k^\prime}$} \\ \hline
0.04 & $16^2$ & 7&933 & 7&93321 & 2 & -0&499699(10) \\
& $16^3$ & 4&497 & 4&49635 & & -0&499642(12) \\
& $16^4$ & 2&636 & 2&63619 & & -0&499688(9) \\ \hline
0.07 & $16^2$ & 9&506 & 9&50617 & 2 & -0&498705(8) \\
& $16^3$ & 5&067 & 5&06701 & & -0&498632(10) \\
& $16^4$ & 2&8908 & 2&89075 & & -0&498708(9) \\ \hline
0.2 & $16^2$ & 15&4 & 15&4000 & 2 & -0&479298(11) \\
& $16^3$ & 7&1843 & 7&18433 & & -0&479154(8) \\
& $16^4$ & 3&9905 & 3&99028 & & -0&479322(11) \\ \hline
0.4 & $16^2$ & 26&69 & 26&6860 & 3 & -0&379362(15) \\
& $16^3$ & 11&219 & 11&2183 & & -0&379351(18) \\
& $16^4$ & 6&231 & 6&23144 & & -0&379401(20) \\ \hline
0.5 & $16^2$ & 35&124 & 35&1209 & 3 & -0&437299(16) \\
& $16^3$ & 14&224 & 14&2245 & & -0&437275(17) \\
& $24^4$ & 7&9327 & 7&93339 & & -0&437364(12) \\ \hline
0.7 & $24^2$ & 67&74 & 67&7909 & 4 & -0&435227(19) \\
& $24^3$ & 25&974 & 26&0474 & & -0&435269(21) \\
& $24^4$ & 14&563 & 14&5656 & & -0&435259(16) \\ \hline
\end{tabular}
\caption{Fourier components $F_{k^\prime}$ at the minimum $k^\prime$ of the Fourier spectrum of the eigenvalue distribution, for $4 \times 4$ Wilson loops in $SU(2)$. 
\label{table_4x4 SU(2)}}
\end{center}
\end{table}

\begin{table}[p]
\begin{center}
\begin{tabular}{|c|c|r@{.}l|r@{.}l|c|r@{.}l|} \hline
\multicolumn{9}{|c|}{Plaquette in $SU(3)$} \\ \hline
$\langle u_p\rangle$ & Lattice size & \multicolumn{2}{|c|}{$\beta_\mathrm{run}$} & \multicolumn{2}{|c|}{$\beta_\mathrm{rew}$} & $k^\prime$ & \multicolumn{2}{|c|}{$F_{k^\prime}$} \\ \hline
0.07 & $4^2$ & 1&15 & 1&15022 & 2 & -0&062686(18) \\
& $4^3$ & 1&15 & 1&15038 & & -0&062663(25) \\
& $4^4$ & 1&152 & 1&15112 & & -0&062670(20) \\ \hline
0.3 & $4^2$ & 4&268 & 4&2690 & 2 & -0&165392(13) \\
& $4^3$ & 4&187 & 4&1864 & & -0&165408(13) \\
& $4^4$ & 4&102 & 4&1017 & & -0&165438(13) \\ \hline
0.5 & $4^2$ & 7&3 & 7&2996 & 2 & -0&11133(6) \\
& $4^3$ & 6&484 & 6&4920 & & -0&11246(9) \\
& $4^4$ & 5&51 & 5&4985 & & -0&11543(27) \\ \hline
0.7 & $8^2$ & 12&94 & 12&9408 & 3 & -0&16766(8) \\
& $8^3$ & 9&91 & 9&9328 & & -0&17274(6) \\
& $8^4$ & 7&54 & 7&5356 & & -0&17800(8) \\ \hline
0.97 & $8^2$ & 132&8 & 132&798 & 11 & -0&193903(31) \\
& $8^3$ & 89&6 & 89&597 & & -0&194490(31) \\
& $8^4$ & 67&4 & 67&338 & & -0&194797(34) \\ \hline
\end{tabular}
\caption{Fourier components $F_{k^\prime}$ at the minimum $k^\prime$ of the Fourier spectrum of the eigenvalue distribution, for plaquettes in $SU(3)$. 
\label{table_plaquettes SU(3)}}
\end{center}
\end{table}

\begin{table}[p]
\begin{center}
\begin{tabular}{|c|c|r@{.}l|r@{.}l|c|r@{.}l|} \hline
\multicolumn{9}{|c|}{Plaquette in $SU(6)$} \\ \hline
$\langle u_p\rangle$ & Lattice size & \multicolumn{2}{|c|}{$\beta_\mathrm{run}$} & \multicolumn{2}{|c|}{$\beta_\mathrm{rew}$} & $k^\prime$ & \multicolumn{2}{|c|}{$F_{k^\prime}$} \\ \hline
0.3 & $4^2$ & 20&78 & 20&7839 & 2 & -0&031469(28) \\
& $4^3$ & 20&42 & 20&4343 & & -0&031658(51) \\
& $4^4$ & 20&09 & 20&0838 & & -0&031747(26) \\ \hline
0.435 & $4^2$ & 29&7 & 29&355 & 2 & -0&04909(4) \\
& $4^4$ & 24&45 & 24&451 & & -0&05650(12) \\ \hline
0.5 & $4^2$ & 34&0 & 33&970 & 2 & -0&03759(8) \\
& $4^3$ & 29&7 & 29&882 & & -0&04316(6) \\ \hline
0.52 & $4^2$ & 35&5 & 35&575 & 2 & -0&03020(8) \\
& $4^4$ & 24&15 & 24&129 & & -0&05103(24) \\ \hline
0.7 & $8^2$ & 58&0 & 57&920 & 3 & -0&09947(5) \\
& $8^3$ & 43&8 & 43&788 & & -0&10963(5) \\
& $8^4$ & 33&1 & 33&160 & & -0&11784(9) \\ \hline
0.97 & $8^2$ & 582&3 & 582&73 & 10 & -0&13828(11) \\
& $8^3$ & 391&0 & 391&28 & & -0&13952(7) \\
& $8^4$ & 294&2 & 294&59 & & -0&14012(7) \\ \hline
\end{tabular}
\caption{Fourier components $F_{k^\prime}$ at the first minimum $k^\prime$ of the Fourier spectrum of the eigenvalue distribution, for plaquettes in $SU(6)$. 
\label{table_plaquettes SU(6)}}
\end{center}
\end{table}

\begin{table}[p]
\begin{center}
\begin{tabular}{|c|c|r@{.}l|r@{.}l|c|r@{.}l|} \hline
\multicolumn{9}{|c|}{$2 \times 2$ Wilson loops in $SU(3)$} \\ \hline
$\langle u_w\rangle$ & Lattice size & \multicolumn{2}{|c|}{$\beta_\mathrm{run}$} & \multicolumn{2}{|c|}{$\beta_\mathrm{rew}$} & $k^\prime$ & \multicolumn{2}{|c|}{$F_{k^\prime}$} \\ \hline
0.04 & $8^2$ & 6&39 & 6&3875 & 2 & -0&03891(5) \\
& $8^3$ & 5&83 & 5&8241 & & -0&03887(7) \\
& $8^4$ & 5&22 & 5&2101 & & -0&03889(13) \\ \hline
0.15 & $12^2$ & 10&12 & 10&1213 & 2 & -0&130071(21) \\
& $12^3$ & 7&857 & 7&8574 & & -0&129931(46) \\
& $12^4$ & 5&774 & 5&7742 & & -0&130398(56) \\ \hline
0.5 & $12^2$ & 24&86 & 24&8285 & 2 & -0&14448(8) \\
& $12^3$ & 15&17 & 15&1684 & & -0&14497(13) \\
& $12^4$ & 10&1 & 10&1042 & & -0&14548(13) \\ \hline
0.9 & $12^2$ & 153&5 & 153&412 & 6 & -0&194108(25) \\
& $12^3$ & 81&7 & 81&767 & & -0&194218(56) \\
& $12^4$ & 53&97 & 53&959 & & -0&194307(62) \\ \hline
\end{tabular}
\caption{Fourier components $F_{k^\prime}$ at the minimum $k^\prime$ of the Fourier spectrum of the eigenvalue distribution, for $2 \times 2$ Wilson loops in $SU(3)$. 
\label{table_2x2 SU(3)}}
\end{center}
\end{table}

\begin{table}[p]
\begin{center}
\begin{tabular}{|c|c|r@{.}l|r@{.}l|c|r@{.}l|} \hline
\multicolumn{9}{|c|}{$2 \times 2$ Wilson loops in $SU(6)$} \\ \hline
$\langle u_w\rangle$ & Lattice size & \multicolumn{2}{|c|}{$\beta_\mathrm{run}$} & \multicolumn{2}{|c|}{$\beta_\mathrm{rew}$} & $k^\prime$ & \multicolumn{2}{|c|}{$F_{k^\prime}$} \\ \hline
0.04 & $8^2$ & 30&18 & 30&1791 & 2 & -0&008415(31) \\
& $8^3$ & 27&47 & 27&4617 & & -0&008467(34) \\
& $8^4$ & 24&42 & 24&4228 & & -0&008630(52) \\ \hline
0.15 & $12^2$ & 45&8 & 45&8021 & 2 & -0&066150(48) \\
& $12^3$ & 34&8 & 34&8034 & & -0&067323(34) \\
& $12^4$ & 25&035 & 25&0292 & & -0&069585(53) \\ \hline
0.5 & $12^2$ & 109&5 & 109&584 & 2 & -0&10342(6) \\
& $12^3$ & 66&4 & 66&259 & & -0&10453(15) \\
& $12^4$ & 44&25 & 44&2464 & & -0&10534(13) \\ \hline
0.9 & $12^2$ & 672&4 & 672&859 & 6 & -0&137916(64) \\
& $12^3$ & 358&0 & 357&767 & & -0&138330(41) \\
& $12^4$ & 236&0 & 236&098 & & -0&138396(50) \\ \hline
\end{tabular}
\caption{Fourier components $F_{k^\prime}$ at the first minimum $k^\prime$ of the Fourier spectrum of the eigenvalue distribution, for $2 \times 2$ Wilson loops in $SU(6)$. 
\label{table_2x2 SU(6)}}
\end{center}
\end{table}

\begin{figure}[p]
\begin{center}
\leavevmode
\input{figures/su2_4x4diffs}
\caption{Difference between $F_{k^\prime}$ in 1+1 dimensions and $D^\prime$ dimensions, for $D^\prime=2+1$~($+$) and $D^\prime=3+1$~($\times$), as a function of the trace $\langle u_w\rangle$. All for the $4 \times 4$ Wilson loop in $SU(2)$.
\label{SU(2) 4x4 differences}}
\end{center}
\end{figure}
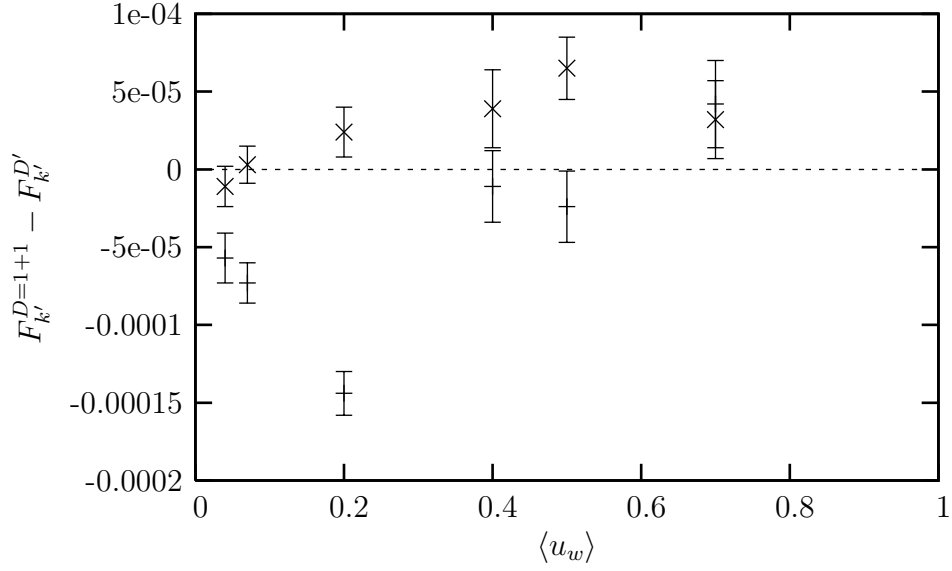

\end{document}

%% file: figures/typicalfourier.tex
\begingroup%
  \makeatletter%
  \newcommand{\GNUPLOTspecial}{%
    \@sanitize\catcode`\%=14\relax\special}%
  \setlength{\unitlength}{0.1bp}%
\begin{picture}(3600,2160)(0,0)%
{\GNUPLOTspecial{"
/gnudict 256 dict def
gnudict begin
/Color false def
/Solid false def
/gnulinewidth 5.000 def
/userlinewidth gnulinewidth def
/vshift -33 def
/dl {10.0 mul} def
/hpt_ 31.5 def
/vpt_ 31.5 def
/hpt hpt_ def
/vpt vpt_ def
/Rounded false def
/M {moveto} bind def
/L {lineto} bind def
/R {rmoveto} bind def
/V {rlineto} bind def
/N {newpath moveto} bind def
/C {setrgbcolor} bind def
/f {rlineto fill} bind def
/vpt2 vpt 2 mul def
/hpt2 hpt 2 mul def
/Lshow { currentpoint stroke M
  0 vshift R show } def
/Rshow { currentpoint stroke M
  dup stringwidth pop neg vshift R show } def
/Cshow { currentpoint stroke M
  dup stringwidth pop -2 div vshift R show } def
/UP { dup vpt_ mul /vpt exch def hpt_ mul /hpt exch def
  /hpt2 hpt 2 mul def /vpt2 vpt 2 mul def } def
/DL { Color {setrgbcolor Solid {pop []} if 0 setdash }
 {pop pop pop 0 setgray Solid {pop []} if 0 setdash} ifelse } def
/BL { stroke userlinewidth 2 mul setlinewidth
      Rounded { 1 setlinejoin 1 setlinecap } if } def
/AL { stroke userlinewidth 2 div setlinewidth
      Rounded { 1 setlinejoin 1 setlinecap } if } def
/UL { dup gnulinewidth mul /userlinewidth exch def
      dup 1 lt {pop 1} if 10 mul /udl exch def } def
/PL { stroke userlinewidth setlinewidth
      Rounded { 1 setlinejoin 1 setlinecap } if } def
/LTw { PL [] 1 setgray } def
/LTb { BL [] 0 0 0 DL } def
/LTa { AL [1 udl mul 2 udl mul] 0 setdash 0 0 0 setrgbcolor } def
/LT0 { PL [] 1 0 0 DL } def
/LT1 { PL [4 dl 2 dl] 0 1 0 DL } def
/LT2 { PL [2 dl 3 dl] 0 0 1 DL } def
/LT3 { PL [1 dl 1.5 dl] 1 0 1 DL } def
/LT4 { PL [5 dl 2 dl 1 dl 2 dl] 0 1 1 DL } def
/LT5 { PL [4 dl 3 dl 1 dl 3 dl] 1 1 0 DL } def
/LT6 { PL [2 dl 2 dl 2 dl 4 dl] 0 0 0 DL } def
/LT7 { PL [2 dl 2 dl 2 dl 2 dl 2 dl 4 dl] 1 0.3 0 DL } def
/LT8 { PL [2 dl 2 dl 2 dl 2 dl 2 dl 2 dl 2 dl 4 dl] 0.5 0.5 0.5 DL } def
/Pnt { stroke [] 0 setdash
   gsave 1 setlinecap M 0 0 V stroke grestore } def
/Dia { stroke [] 0 setdash 2 copy vpt add M
  hpt neg vpt neg V hpt vpt neg V
  hpt vpt V hpt neg vpt V closepath stroke
  Pnt } def
/Pls { stroke [] 0 setdash vpt sub M 0 vpt2 V
  currentpoint stroke M
  hpt neg vpt neg R hpt2 0 V stroke
  } def
/Box { stroke [] 0 setdash 2 copy exch hpt sub exch vpt add M
  0 vpt2 neg V hpt2 0 V 0 vpt2 V
  hpt2 neg 0 V closepath stroke
  Pnt } def
/Crs { stroke [] 0 setdash exch hpt sub exch vpt add M
  hpt2 vpt2 neg V currentpoint stroke M
  hpt2 neg 0 R hpt2 vpt2 V stroke } def
/TriU { stroke [] 0 setdash 2 copy vpt 1.12 mul add M
  hpt neg vpt -1.62 mul V
  hpt 2 mul 0 V
  hpt neg vpt 1.62 mul V closepath stroke
  Pnt  } def
/Star { 2 copy Pls Crs } def
/BoxF { stroke [] 0 setdash exch hpt sub exch vpt add M
  0 vpt2 neg V  hpt2 0 V  0 vpt2 V
  hpt2 neg 0 V  closepath fill } def
/TriUF { stroke [] 0 setdash vpt 1.12 mul add M
  hpt neg vpt -1.62 mul V
  hpt 2 mul 0 V
  hpt neg vpt 1.62 mul V closepath fill } def
/TriD { stroke [] 0 setdash 2 copy vpt 1.12 mul sub M
  hpt neg vpt 1.62 mul V
  hpt 2 mul 0 V
  hpt neg vpt -1.62 mul V closepath stroke
  Pnt  } def
/TriDF { stroke [] 0 setdash vpt 1.12 mul sub M
  hpt neg vpt 1.62 mul V
  hpt 2 mul 0 V
  hpt neg vpt -1.62 mul V closepath fill} def
/DiaF { stroke [] 0 setdash vpt add M
  hpt neg vpt neg V hpt vpt neg V
  hpt vpt V hpt neg vpt V closepath fill } def
/Pent { stroke [] 0 setdash 2 copy gsave
  translate 0 hpt M 4 {72 rotate 0 hpt L} repeat
  closepath stroke grestore Pnt } def
/PentF { stroke [] 0 setdash gsave
  translate 0 hpt M 4 {72 rotate 0 hpt L} repeat
  closepath fill grestore } def
/Circle { stroke [] 0 setdash 2 copy
  hpt 0 360 arc stroke Pnt } def
/CircleF { stroke [] 0 setdash hpt 0 360 arc fill } def
/C0 { BL [] 0 setdash 2 copy moveto vpt 90 450  arc } bind def
/C1 { BL [] 0 setdash 2 copy        moveto
       2 copy  vpt 0 90 arc closepath fill
               vpt 0 360 arc closepath } bind def
/C2 { BL [] 0 setdash 2 copy moveto
       2 copy  vpt 90 180 arc closepath fill
               vpt 0 360 arc closepath } bind def
/C3 { BL [] 0 setdash 2 copy moveto
       2 copy  vpt 0 180 arc closepath fill
               vpt 0 360 arc closepath } bind def
/C4 { BL [] 0 setdash 2 copy moveto
       2 copy  vpt 180 270 arc closepath fill
               vpt 0 360 arc closepath } bind def
/C5 { BL [] 0 setdash 2 copy moveto
       2 copy  vpt 0 90 arc
       2 copy moveto
       2 copy  vpt 180 270 arc closepath fill
               vpt 0 360 arc } bind def
/C6 { BL [] 0 setdash 2 copy moveto
      2 copy  vpt 90 270 arc closepath fill
              vpt 0 360 arc closepath } bind def
/C7 { BL [] 0 setdash 2 copy moveto
      2 copy  vpt 0 270 arc closepath fill
              vpt 0 360 arc closepath } bind def
/C8 { BL [] 0 setdash 2 copy moveto
      2 copy vpt 270 360 arc closepath fill
              vpt 0 360 arc closepath } bind def
/C9 { BL [] 0 setdash 2 copy moveto
      2 copy  vpt 270 450 arc closepath fill
              vpt 0 360 arc closepath } bind def
/C10 { BL [] 0 setdash 2 copy 2 copy moveto vpt 270 360 arc closepath fill
       2 copy moveto
       2 copy vpt 90 180 arc closepath fill
               vpt 0 360 arc closepath } bind def
/C11 { BL [] 0 setdash 2 copy moveto
       2 copy  vpt 0 180 arc closepath fill
       2 copy moveto
       2 copy  vpt 270 360 arc closepath fill
               vpt 0 360 arc closepath } bind def
/C12 { BL [] 0 setdash 2 copy moveto
       2 copy  vpt 180 360 arc closepath fill
               vpt 0 360 arc closepath } bind def
/C13 { BL [] 0 setdash  2 copy moveto
       2 copy  vpt 0 90 arc closepath fill
       2 copy moveto
       2 copy  vpt 180 360 arc closepath fill
               vpt 0 360 arc closepath } bind def
/C14 { BL [] 0 setdash 2 copy moveto
       2 copy  vpt 90 360 arc closepath fill
               vpt 0 360 arc } bind def
/C15 { BL [] 0 setdash 2 copy vpt 0 360 arc closepath fill
               vpt 0 360 arc closepath } bind def
/Rec   { newpath 4 2 roll moveto 1 index 0 rlineto 0 exch rlineto
       neg 0 rlineto closepath } bind def
/Square { dup Rec } bind def
/Bsquare { vpt sub exch vpt sub exch vpt2 Square } bind def
/S0 { BL [] 0 setdash 2 copy moveto 0 vpt rlineto BL Bsquare } bind def
/S1 { BL [] 0 setdash 2 copy vpt Square fill Bsquare } bind def
/S2 { BL [] 0 setdash 2 copy exch vpt sub exch vpt Square fill Bsquare } bind def
/S3 { BL [] 0 setdash 2 copy exch vpt sub exch vpt2 vpt Rec fill Bsquare } bind def
/S4 { BL [] 0 setdash 2 copy exch vpt sub exch vpt sub vpt Square fill Bsquare } bind def
/S5 { BL [] 0 setdash 2 copy 2 copy vpt Square fill
       exch vpt sub exch vpt sub vpt Square fill Bsquare } bind def
/S6 { BL [] 0 setdash 2 copy exch vpt sub exch vpt sub vpt vpt2 Rec fill Bsquare } bind def
/S7 { BL [] 0 setdash 2 copy exch vpt sub exch vpt sub vpt vpt2 Rec fill
       2 copy vpt Square fill
       Bsquare } bind def
/S8 { BL [] 0 setdash 2 copy vpt sub vpt Square fill Bsquare } bind def
/S9 { BL [] 0 setdash 2 copy vpt sub vpt vpt2 Rec fill Bsquare } bind def
/S10 { BL [] 0 setdash 2 copy vpt sub vpt Square fill 2 copy exch vpt sub exch vpt Square fill
       Bsquare } bind def
/S11 { BL [] 0 setdash 2 copy vpt sub vpt Square fill 2 copy exch vpt sub exch vpt2 vpt Rec fill
       Bsquare } bind def
/S12 { BL [] 0 setdash 2 copy exch vpt sub exch vpt sub vpt2 vpt Rec fill Bsquare } bind def
/S13 { BL [] 0 setdash 2 copy exch vpt sub exch vpt sub vpt2 vpt Rec fill
       2 copy vpt Square fill Bsquare } bind def
/S14 { BL [] 0 setdash 2 copy exch vpt sub exch vpt sub vpt2 vpt Rec fill
       2 copy exch vpt sub exch vpt Square fill Bsquare } bind def
/S15 { BL [] 0 setdash 2 copy Bsquare fill Bsquare } bind def
/D0 { gsave translate 45 rotate 0 0 S0 stroke grestore } bind def
/D1 { gsave translate 45 rotate 0 0 S1 stroke grestore } bind def
/D2 { gsave translate 45 rotate 0 0 S2 stroke grestore } bind def
/D3 { gsave translate 45 rotate 0 0 S3 stroke grestore } bind def
/D4 { gsave translate 45 rotate 0 0 S4 stroke grestore } bind def
/D5 { gsave translate 45 rotate 0 0 S5 stroke grestore } bind def
/D6 { gsave translate 45 rotate 0 0 S6 stroke grestore } bind def
/D7 { gsave translate 45 rotate 0 0 S7 stroke grestore } bind def
/D8 { gsave translate 45 rotate 0 0 S8 stroke grestore } bind def
/D9 { gsave translate 45 rotate 0 0 S9 stroke grestore } bind def
/D10 { gsave translate 45 rotate 0 0 S10 stroke grestore } bind def
/D11 { gsave translate 45 rotate 0 0 S11 stroke grestore } bind def
/D12 { gsave translate 45 rotate 0 0 S12 stroke grestore } bind def
/D13 { gsave translate 45 rotate 0 0 S13 stroke grestore } bind def
/D14 { gsave translate 45 rotate 0 0 S14 stroke grestore } bind def
/D15 { gsave translate 45 rotate 0 0 S15 stroke grestore } bind def
/DiaE { stroke [] 0 setdash vpt add M
  hpt neg vpt neg V hpt vpt neg V
  hpt vpt V hpt neg vpt V closepath stroke } def
/BoxE { stroke [] 0 setdash exch hpt sub exch vpt add M
  0 vpt2 neg V hpt2 0 V 0 vpt2 V
  hpt2 neg 0 V closepath stroke } def
/TriUE { stroke [] 0 setdash vpt 1.12 mul add M
  hpt neg vpt -1.62 mul V
  hpt 2 mul 0 V
  hpt neg vpt 1.62 mul V closepath stroke } def
/TriDE { stroke [] 0 setdash vpt 1.12 mul sub M
  hpt neg vpt 1.62 mul V
  hpt 2 mul 0 V
  hpt neg vpt -1.62 mul V closepath stroke } def
/PentE { stroke [] 0 setdash gsave
  translate 0 hpt M 4 {72 rotate 0 hpt L} repeat
  closepath stroke grestore } def
/CircE { stroke [] 0 setdash 
  hpt 0 360 arc stroke } def
/Opaque { gsave closepath 1 setgray fill grestore 0 setgray closepath } def
/DiaW { stroke [] 0 setdash vpt add M
  hpt neg vpt neg V hpt vpt neg V
  hpt vpt V hpt neg vpt V Opaque stroke } def
/BoxW { stroke [] 0 setdash exch hpt sub exch vpt add M
  0 vpt2 neg V hpt2 0 V 0 vpt2 V
  hpt2 neg 0 V Opaque stroke } def
/TriUW { stroke [] 0 setdash vpt 1.12 mul add M
  hpt neg vpt -1.62 mul V
  hpt 2 mul 0 V
  hpt neg vpt 1.62 mul V Opaque stroke } def
/TriDW { stroke [] 0 setdash vpt 1.12 mul sub M
  hpt neg vpt 1.62 mul V
  hpt 2 mul 0 V
  hpt neg vpt -1.62 mul V Opaque stroke } def
/PentW { stroke [] 0 setdash gsave
  translate 0 hpt M 4 {72 rotate 0 hpt L} repeat
  Opaque stroke grestore } def
/CircW { stroke [] 0 setdash 
  hpt 0 360 arc Opaque stroke } def
/BoxFill { gsave Rec 1 setgray fill grestore } def
/BoxColFill {
  gsave Rec
  /Fillden exch def
  currentrgbcolor
  /ColB exch def /ColG exch def /ColR exch def
  /ColR ColR Fillden mul Fillden sub 1 add def
  /ColG ColG Fillden mul Fillden sub 1 add def
  /ColB ColB Fillden mul Fillden sub 1 add def
  ColR ColG ColB setrgbcolor
  fill grestore } def
%
%
/PatternFill { gsave /PFa [ 9 2 roll ] def
    PFa 0 get PFa 2 get 2 div add PFa 1 get PFa 3 get 2 div add translate
    PFa 2 get -2 div PFa 3 get -2 div PFa 2 get PFa 3 get Rec
    gsave 1 setgray fill grestore clip
    currentlinewidth 0.5 mul setlinewidth
    /PFs PFa 2 get dup mul PFa 3 get dup mul add sqrt def
    0 0 M PFa 5 get rotate PFs -2 div dup translate
	0 1 PFs PFa 4 get div 1 add floor cvi
	{ PFa 4 get mul 0 M 0 PFs V } for
    0 PFa 6 get ne {
	0 1 PFs PFa 4 get div 1 add floor cvi
	{ PFa 4 get mul 0 2 1 roll M PFs 0 V } for
    } if
    stroke grestore } def
/Symbol-Oblique /Symbol findfont [1 0 .167 1 0 0] makefont
dup length dict begin {1 index /FID eq {pop pop} {def} ifelse} forall
currentdict end definefont pop
end
gnudict begin
gsave
0 0 translate
0.100 0.100 scale
0 setgray
newpath
1.000 UL
LTb
450 300 M
63 0 V
2937 0 R
-63 0 V
1.000 UL
LTb
450 551 M
63 0 V
2937 0 R
-63 0 V
1.000 UL
LTb
450 803 M
63 0 V
2937 0 R
-63 0 V
1.000 UL
LTb
450 1054 M
63 0 V
2937 0 R
-63 0 V
1.000 UL
LTb
450 1306 M
63 0 V
2937 0 R
-63 0 V
1.000 UL
LTb
450 1557 M
63 0 V
2937 0 R
-63 0 V
1.000 UL
LTb
450 1809 M
63 0 V
2937 0 R
-63 0 V
1.000 UL
LTb
450 2060 M
63 0 V
2937 0 R
-63 0 V
1.000 UL
LTb
450 300 M
0 63 V
0 1697 R
0 -63 V
1.000 UL
LTb
1200 300 M
0 63 V
0 1697 R
0 -63 V
1.000 UL
LTb
1950 300 M
0 63 V
0 1697 R
0 -63 V
1.000 UL
LTb
2700 300 M
0 63 V
0 1697 R
0 -63 V
1.000 UL
LTb
3450 300 M
0 63 V
0 1697 R
0 -63 V
1.000 UL
LTb
1.000 UL
LTb
450 300 M
3000 0 V
0 1760 V
-3000 0 V
450 300 L
LTb
LTb
1.000 UP
1.000 UP
1.000 UL
LT0
450 2060 M
600 1431 L
750 519 L
900 306 L
150 191 V
150 184 V
150 85 V
150 28 V
150 7 V
150 2 V
150 0 V
150 0 V
150 0 V
150 0 V
150 0 V
150 0 V
150 0 V
150 0 V
150 0 V
150 0 V
150 0 V
450 2060 Pls
600 1431 Pls
750 519 Pls
900 306 Pls
1050 497 Pls
1200 681 Pls
1350 766 Pls
1500 794 Pls
1650 801 Pls
1800 803 Pls
1950 803 Pls
2100 803 Pls
2250 803 Pls
2400 803 Pls
2550 803 Pls
2700 803 Pls
2850 803 Pls
3000 803 Pls
3150 803 Pls
3300 803 Pls
3450 803 Pls
1.000 UL
LTb
450 300 M
3000 0 V
0 1760 V
-3000 0 V
450 300 L
1.000 UP
stroke
grestore
end
showpage
}}%
\put(1950,50){\makebox(0,0){$k$}}%
\put(100,1180){%
\special{ps: gsave currentpoint currentpoint translate
270 rotate neg exch neg exch translate}%
\makebox(0,0)[b]{\shortstack{$F_k$}}%
\special{ps: currentpoint grestore moveto}%
}%
\put(3450,200){\makebox(0,0){ 20}}%
\put(2700,200){\makebox(0,0){ 15}}%
\put(1950,200){\makebox(0,0){ 10}}%
\put(1200,200){\makebox(0,0){ 5}}%
\put(450,200){\makebox(0,0){ 0}}%
\put(400,2060){\makebox(0,0)[r]{ 1}}%
\put(400,1809){\makebox(0,0)[r]{ 0.8}}%
\put(400,1557){\makebox(0,0)[r]{ 0.6}}%
\put(400,1306){\makebox(0,0)[r]{ 0.4}}%
\put(400,1054){\makebox(0,0)[r]{ 0.2}}%
\put(400,803){\makebox(0,0)[r]{ 0}}%
\put(400,551){\makebox(0,0)[r]{-0.2}}%
\put(400,300){\makebox(0,0)[r]{-0.4}}%
\end{picture}%
\endgroup
 

%% file: figures/su2plaq_0.7densities.tex
\begingroup%
  \makeatletter%
  \newcommand{\GNUPLOTspecial}{%
    \@sanitize\catcode`\%=14\relax\special}%
  \setlength{\unitlength}{0.1bp}%
\begin{picture}(3600,2160)(0,0)%
{\GNUPLOTspecial{"
/gnudict 256 dict def
gnudict begin
/Color false def
/Solid false def
/gnulinewidth 5.000 def
/userlinewidth gnulinewidth def
/vshift -33 def
/dl {10.0 mul} def
/hpt_ 31.5 def
/vpt_ 31.5 def
/hpt hpt_ def
/vpt vpt_ def
/Rounded false def
/M {moveto} bind def
/L {lineto} bind def
/R {rmoveto} bind def
/V {rlineto} bind def
/N {newpath moveto} bind def
/C {setrgbcolor} bind def
/f {rlineto fill} bind def
/vpt2 vpt 2 mul def
/hpt2 hpt 2 mul def
/Lshow { currentpoint stroke M
  0 vshift R show } def
/Rshow { currentpoint stroke M
  dup stringwidth pop neg vshift R show } def
/Cshow { currentpoint stroke M
  dup stringwidth pop -2 div vshift R show } def
/UP { dup vpt_ mul /vpt exch def hpt_ mul /hpt exch def
  /hpt2 hpt 2 mul def /vpt2 vpt 2 mul def } def
/DL { Color {setrgbcolor Solid {pop []} if 0 setdash }
 {pop pop pop 0 setgray Solid {pop []} if 0 setdash} ifelse } def
/BL { stroke userlinewidth 2 mul setlinewidth
      Rounded { 1 setlinejoin 1 setlinecap } if } def
/AL { stroke userlinewidth 2 div setlinewidth
      Rounded { 1 setlinejoin 1 setlinecap } if } def
/UL { dup gnulinewidth mul /userlinewidth exch def
      dup 1 lt {pop 1} if 10 mul /udl exch def } def
/PL { stroke userlinewidth setlinewidth
      Rounded { 1 setlinejoin 1 setlinecap } if } def
/LTw { PL [] 1 setgray } def
/LTb { BL [] 0 0 0 DL } def
/LTa { AL [1 udl mul 2 udl mul] 0 setdash 0 0 0 setrgbcolor } def
/LT0 { PL [] 1 0 0 DL } def
/LT1 { PL [4 dl 2 dl] 0 1 0 DL } def
/LT2 { PL [2 dl 3 dl] 0 0 1 DL } def
/LT3 { PL [1 dl 1.5 dl] 1 0 1 DL } def
/LT4 { PL [5 dl 2 dl 1 dl 2 dl] 0 1 1 DL } def
/LT5 { PL [4 dl 3 dl 1 dl 3 dl] 1 1 0 DL } def
/LT6 { PL [2 dl 2 dl 2 dl 4 dl] 0 0 0 DL } def
/LT7 { PL [2 dl 2 dl 2 dl 2 dl 2 dl 4 dl] 1 0.3 0 DL } def
/LT8 { PL [2 dl 2 dl 2 dl 2 dl 2 dl 2 dl 2 dl 4 dl] 0.5 0.5 0.5 DL } def
/Pnt { stroke [] 0 setdash
   gsave 1 setlinecap M 0 0 V stroke grestore } def
/Dia { stroke [] 0 setdash 2 copy vpt add M
  hpt neg vpt neg V hpt vpt neg V
  hpt vpt V hpt neg vpt V closepath stroke
  Pnt } def
/Pls { stroke [] 0 setdash vpt sub M 0 vpt2 V
  currentpoint stroke M
  hpt neg vpt neg R hpt2 0 V stroke
  } def
/Box { stroke [] 0 setdash 2 copy exch hpt sub exch vpt add M
  0 vpt2 neg V hpt2 0 V 0 vpt2 V
  hpt2 neg 0 V closepath stroke
  Pnt } def
/Crs { stroke [] 0 setdash exch hpt sub exch vpt add M
  hpt2 vpt2 neg V currentpoint stroke M
  hpt2 neg 0 R hpt2 vpt2 V stroke } def
/TriU { stroke [] 0 setdash 2 copy vpt 1.12 mul add M
  hpt neg vpt -1.62 mul V
  hpt 2 mul 0 V
  hpt neg vpt 1.62 mul V closepath stroke
  Pnt  } def
/Star { 2 copy Pls Crs } def
/BoxF { stroke [] 0 setdash exch hpt sub exch vpt add M
  0 vpt2 neg V  hpt2 0 V  0 vpt2 V
  hpt2 neg 0 V  closepath fill } def
/TriUF { stroke [] 0 setdash vpt 1.12 mul add M
  hpt neg vpt -1.62 mul V
  hpt 2 mul 0 V
  hpt neg vpt 1.62 mul V closepath fill } def
/TriD { stroke [] 0 setdash 2 copy vpt 1.12 mul sub M
  hpt neg vpt 1.62 mul V
  hpt 2 mul 0 V
  hpt neg vpt -1.62 mul V closepath stroke
  Pnt  } def
/TriDF { stroke [] 0 setdash vpt 1.12 mul sub M
  hpt neg vpt 1.62 mul V
  hpt 2 mul 0 V
  hpt neg vpt -1.62 mul V closepath fill} def
/DiaF { stroke [] 0 setdash vpt add M
  hpt neg vpt neg V hpt vpt neg V
  hpt vpt V hpt neg vpt V closepath fill } def
/Pent { stroke [] 0 setdash 2 copy gsave
  translate 0 hpt M 4 {72 rotate 0 hpt L} repeat
  closepath stroke grestore Pnt } def
/PentF { stroke [] 0 setdash gsave
  translate 0 hpt M 4 {72 rotate 0 hpt L} repeat
  closepath fill grestore } def
/Circle { stroke [] 0 setdash 2 copy
  hpt 0 360 arc stroke Pnt } def
/CircleF { stroke [] 0 setdash hpt 0 360 arc fill } def
/C0 { BL [] 0 setdash 2 copy moveto vpt 90 450  arc } bind def
/C1 { BL [] 0 setdash 2 copy        moveto
       2 copy  vpt 0 90 arc closepath fill
               vpt 0 360 arc closepath } bind def
/C2 { BL [] 0 setdash 2 copy moveto
       2 copy  vpt 90 180 arc closepath fill
               vpt 0 360 arc closepath } bind def
/C3 { BL [] 0 setdash 2 copy moveto
       2 copy  vpt 0 180 arc closepath fill
               vpt 0 360 arc closepath } bind def
/C4 { BL [] 0 setdash 2 copy moveto
       2 copy  vpt 180 270 arc closepath fill
               vpt 0 360 arc closepath } bind def
/C5 { BL [] 0 setdash 2 copy moveto
       2 copy  vpt 0 90 arc
       2 copy moveto
       2 copy  vpt 180 270 arc closepath fill
               vpt 0 360 arc } bind def
/C6 { BL [] 0 setdash 2 copy moveto
      2 copy  vpt 90 270 arc closepath fill
              vpt 0 360 arc closepath } bind def
/C7 { BL [] 0 setdash 2 copy moveto
      2 copy  vpt 0 270 arc closepath fill
              vpt 0 360 arc closepath } bind def
/C8 { BL [] 0 setdash 2 copy moveto
      2 copy vpt 270 360 arc closepath fill
              vpt 0 360 arc closepath } bind def
/C9 { BL [] 0 setdash 2 copy moveto
      2 copy  vpt 270 450 arc closepath fill
              vpt 0 360 arc closepath } bind def
/C10 { BL [] 0 setdash 2 copy 2 copy moveto vpt 270 360 arc closepath fill
       2 copy moveto
       2 copy vpt 90 180 arc closepath fill
               vpt 0 360 arc closepath } bind def
/C11 { BL [] 0 setdash 2 copy moveto
       2 copy  vpt 0 180 arc closepath fill
       2 copy moveto
       2 copy  vpt 270 360 arc closepath fill
               vpt 0 360 arc closepath } bind def
/C12 { BL [] 0 setdash 2 copy moveto
       2 copy  vpt 180 360 arc closepath fill
               vpt 0 360 arc closepath } bind def
/C13 { BL [] 0 setdash  2 copy moveto
       2 copy  vpt 0 90 arc closepath fill
       2 copy moveto
       2 copy  vpt 180 360 arc closepath fill
               vpt 0 360 arc closepath } bind def
/C14 { BL [] 0 setdash 2 copy moveto
       2 copy  vpt 90 360 arc closepath fill
               vpt 0 360 arc } bind def
/C15 { BL [] 0 setdash 2 copy vpt 0 360 arc closepath fill
               vpt 0 360 arc closepath } bind def
/Rec   { newpath 4 2 roll moveto 1 index 0 rlineto 0 exch rlineto
       neg 0 rlineto closepath } bind def
/Square { dup Rec } bind def
/Bsquare { vpt sub exch vpt sub exch vpt2 Square } bind def
/S0 { BL [] 0 setdash 2 copy moveto 0 vpt rlineto BL Bsquare } bind def
/S1 { BL [] 0 setdash 2 copy vpt Square fill Bsquare } bind def
/S2 { BL [] 0 setdash 2 copy exch vpt sub exch vpt Square fill Bsquare } bind def
/S3 { BL [] 0 setdash 2 copy exch vpt sub exch vpt2 vpt Rec fill Bsquare } bind def
/S4 { BL [] 0 setdash 2 copy exch vpt sub exch vpt sub vpt Square fill Bsquare } bind def
/S5 { BL [] 0 setdash 2 copy 2 copy vpt Square fill
       exch vpt sub exch vpt sub vpt Square fill Bsquare } bind def
/S6 { BL [] 0 setdash 2 copy exch vpt sub exch vpt sub vpt vpt2 Rec fill Bsquare } bind def
/S7 { BL [] 0 setdash 2 copy exch vpt sub exch vpt sub vpt vpt2 Rec fill
       2 copy vpt Square fill
       Bsquare } bind def
/S8 { BL [] 0 setdash 2 copy vpt sub vpt Square fill Bsquare } bind def
/S9 { BL [] 0 setdash 2 copy vpt sub vpt vpt2 Rec fill Bsquare } bind def
/S10 { BL [] 0 setdash 2 copy vpt sub vpt Square fill 2 copy exch vpt sub exch vpt Square fill
       Bsquare } bind def
/S11 { BL [] 0 setdash 2 copy vpt sub vpt Square fill 2 copy exch vpt sub exch vpt2 vpt Rec fill
       Bsquare } bind def
/S12 { BL [] 0 setdash 2 copy exch vpt sub exch vpt sub vpt2 vpt Rec fill Bsquare } bind def
/S13 { BL [] 0 setdash 2 copy exch vpt sub exch vpt sub vpt2 vpt Rec fill
       2 copy vpt Square fill Bsquare } bind def
/S14 { BL [] 0 setdash 2 copy exch vpt sub exch vpt sub vpt2 vpt Rec fill
       2 copy exch vpt sub exch vpt Square fill Bsquare } bind def
/S15 { BL [] 0 setdash 2 copy Bsquare fill Bsquare } bind def
/D0 { gsave translate 45 rotate 0 0 S0 stroke grestore } bind def
/D1 { gsave translate 45 rotate 0 0 S1 stroke grestore } bind def
/D2 { gsave translate 45 rotate 0 0 S2 stroke grestore } bind def
/D3 { gsave translate 45 rotate 0 0 S3 stroke grestore } bind def
/D4 { gsave translate 45 rotate 0 0 S4 stroke grestore } bind def
/D5 { gsave translate 45 rotate 0 0 S5 stroke grestore } bind def
/D6 { gsave translate 45 rotate 0 0 S6 stroke grestore } bind def
/D7 { gsave translate 45 rotate 0 0 S7 stroke grestore } bind def
/D8 { gsave translate 45 rotate 0 0 S8 stroke grestore } bind def
/D9 { gsave translate 45 rotate 0 0 S9 stroke grestore } bind def
/D10 { gsave translate 45 rotate 0 0 S10 stroke grestore } bind def
/D11 { gsave translate 45 rotate 0 0 S11 stroke grestore } bind def
/D12 { gsave translate 45 rotate 0 0 S12 stroke grestore } bind def
/D13 { gsave translate 45 rotate 0 0 S13 stroke grestore } bind def
/D14 { gsave translate 45 rotate 0 0 S14 stroke grestore } bind def
/D15 { gsave translate 45 rotate 0 0 S15 stroke grestore } bind def
/DiaE { stroke [] 0 setdash vpt add M
  hpt neg vpt neg V hpt vpt neg V
  hpt vpt V hpt neg vpt V closepath stroke } def
/BoxE { stroke [] 0 setdash exch hpt sub exch vpt add M
  0 vpt2 neg V hpt2 0 V 0 vpt2 V
  hpt2 neg 0 V closepath stroke } def
/TriUE { stroke [] 0 setdash vpt 1.12 mul add M
  hpt neg vpt -1.62 mul V
  hpt 2 mul 0 V
  hpt neg vpt 1.62 mul V closepath stroke } def
/TriDE { stroke [] 0 setdash vpt 1.12 mul sub M
  hpt neg vpt 1.62 mul V
  hpt 2 mul 0 V
  hpt neg vpt -1.62 mul V closepath stroke } def
/PentE { stroke [] 0 setdash gsave
  translate 0 hpt M 4 {72 rotate 0 hpt L} repeat
  closepath stroke grestore } def
/CircE { stroke [] 0 setdash 
  hpt 0 360 arc stroke } def
/Opaque { gsave closepath 1 setgray fill grestore 0 setgray closepath } def
/DiaW { stroke [] 0 setdash vpt add M
  hpt neg vpt neg V hpt vpt neg V
  hpt vpt V hpt neg vpt V Opaque stroke } def
/BoxW { stroke [] 0 setdash exch hpt sub exch vpt add M
  0 vpt2 neg V hpt2 0 V 0 vpt2 V
  hpt2 neg 0 V Opaque stroke } def
/TriUW { stroke [] 0 setdash vpt 1.12 mul add M
  hpt neg vpt -1.62 mul V
  hpt 2 mul 0 V
  hpt neg vpt 1.62 mul V Opaque stroke } def
/TriDW { stroke [] 0 setdash vpt 1.12 mul sub M
  hpt neg vpt 1.62 mul V
  hpt 2 mul 0 V
  hpt neg vpt -1.62 mul V Opaque stroke } def
/PentW { stroke [] 0 setdash gsave
  translate 0 hpt M 4 {72 rotate 0 hpt L} repeat
  Opaque stroke grestore } def
/CircW { stroke [] 0 setdash 
  hpt 0 360 arc Opaque stroke } def
/BoxFill { gsave Rec 1 setgray fill grestore } def
/BoxColFill {
  gsave Rec
  /Fillden exch def
  currentrgbcolor
  /ColB exch def /ColG exch def /ColR exch def
  /ColR ColR Fillden mul Fillden sub 1 add def
  /ColG ColG Fillden mul Fillden sub 1 add def
  /ColB ColB Fillden mul Fillden sub 1 add def
  ColR ColG ColB setrgbcolor
  fill grestore } def
%
%
/PatternFill { gsave /PFa [ 9 2 roll ] def
    PFa 0 get PFa 2 get 2 div add PFa 1 get PFa 3 get 2 div add translate
    PFa 2 get -2 div PFa 3 get -2 div PFa 2 get PFa 3 get Rec
    gsave 1 setgray fill grestore clip
    currentlinewidth 0.5 mul setlinewidth
    /PFs PFa 2 get dup mul PFa 3 get dup mul add sqrt def
    0 0 M PFa 5 get rotate PFs -2 div dup translate
	0 1 PFs PFa 4 get div 1 add floor cvi
	{ PFa 4 get mul 0 M 0 PFs V } for
    0 PFa 6 get ne {
	0 1 PFs PFa 4 get div 1 add floor cvi
	{ PFa 4 get mul 0 2 1 roll M PFs 0 V } for
    } if
    stroke grestore } def
/Symbol-Oblique /Symbol findfont [1 0 .167 1 0 0] makefont
dup length dict begin {1 index /FID eq {pop pop} {def} ifelse} forall
currentdict end definefont pop
end
gnudict begin
gsave
0 0 translate
0.100 0.100 scale
0 setgray
newpath
1.000 UL
LTb
450 300 M
63 0 V
2937 0 R
-63 0 V
1.000 UL
LTb
450 551 M
63 0 V
2937 0 R
-63 0 V
1.000 UL
LTb
450 803 M
63 0 V
2937 0 R
-63 0 V
1.000 UL
LTb
450 1054 M
63 0 V
2937 0 R
-63 0 V
1.000 UL
LTb
450 1306 M
63 0 V
2937 0 R
-63 0 V
1.000 UL
LTb
450 1557 M
63 0 V
2937 0 R
-63 0 V
1.000 UL
LTb
450 1809 M
63 0 V
2937 0 R
-63 0 V
1.000 UL
LTb
450 2060 M
63 0 V
2937 0 R
-63 0 V
1.000 UL
LTb
518 300 M
0 63 V
0 1697 R
0 -63 V
1.000 UL
LTb
995 300 M
0 63 V
0 1697 R
0 -63 V
1.000 UL
LTb
1473 300 M
0 63 V
0 1697 R
0 -63 V
1.000 UL
LTb
1950 300 M
0 63 V
0 1697 R
0 -63 V
1.000 UL
LTb
2427 300 M
0 63 V
0 1697 R
0 -63 V
1.000 UL
LTb
2905 300 M
0 63 V
0 1697 R
0 -63 V
1.000 UL
LTb
3382 300 M
0 63 V
0 1697 R
0 -63 V
1.000 UL
LTb
1.000 UL
LTb
450 300 M
3000 0 V
0 1760 V
-3000 0 V
450 300 L
LTb
LTb
1.000 UP
1.000 UP
0.500 UL
LT0
451 300 M
3 0 V
3 0 V
3 0 V
3 0 V
3 0 V
3 0 V
3 0 V
3 0 V
3 0 V
3 0 V
3 0 V
3 0 V
3 0 V
2 0 V
3 0 V
3 0 V
3 0 V
3 0 V
3 0 V
3 0 V
3 0 V
3 0 V
3 0 V
3 0 V
3 0 V
3 0 V
3 0 V
2 0 V
3 0 V
3 0 V
3 0 V
3 0 V
3 0 V
3 0 V
3 0 V
3 0 V
3 0 V
3 0 V
3 0 V
3 0 V
3 0 V
3 0 V
2 0 V
3 0 V
3 0 V
3 0 V
3 0 V
3 0 V
3 0 V
3 0 V
3 0 V
3 0 V
3 0 V
3 0 V
3 0 V
3 0 V
2 0 V
3 0 V
3 0 V
3 0 V
3 0 V
3 0 V
3 0 V
3 0 V
3 0 V
3 0 V
3 0 V
3 0 V
3 0 V
3 0 V
2 0 V
3 0 V
3 0 V
3 0 V
3 0 V
3 0 V
3 0 V
3 0 V
3 0 V
3 0 V
3 0 V
3 0 V
3 0 V
3 0 V
2 0 V
3 0 V
3 1 V
3 0 V
3 0 V
3 0 V
3 0 V
3 0 V
3 0 V
3 0 V
3 0 V
3 0 V
3 0 V
3 0 V
3 0 V
2 0 V
3 0 V
3 0 V
3 0 V
3 0 V
stroke
756 301 M
3 0 V
3 0 V
3 0 V
3 0 V
3 0 V
3 0 V
3 0 V
3 0 V
3 0 V
2 0 V
3 0 V
3 0 V
3 1 V
3 0 V
3 -1 V
3 0 V
3 1 V
3 0 V
3 0 V
3 0 V
3 0 V
3 0 V
3 0 V
2 0 V
3 0 V
3 0 V
3 0 V
3 0 V
3 0 V
3 1 V
3 -1 V
3 1 V
3 0 V
3 0 V
3 0 V
3 0 V
3 0 V
2 0 V
3 0 V
3 0 V
3 1 V
3 0 V
3 0 V
3 0 V
3 0 V
3 0 V
3 0 V
3 1 V
3 0 V
3 0 V
3 0 V
2 0 V
3 0 V
3 0 V
3 0 V
3 1 V
3 -1 V
3 1 V
3 0 V
3 0 V
3 1 V
3 0 V
3 0 V
3 0 V
3 1 V
3 0 V
2 0 V
3 0 V
3 0 V
3 1 V
3 0 V
3 1 V
3 0 V
3 0 V
3 0 V
3 1 V
3 0 V
3 0 V
3 1 V
3 1 V
2 -1 V
3 1 V
3 0 V
3 1 V
3 0 V
3 1 V
3 1 V
3 -1 V
3 0 V
3 1 V
3 1 V
3 0 V
3 2 V
3 0 V
2 0 V
3 2 V
3 -1 V
3 1 V
3 1 V
3 1 V
3 1 V
3 0 V
3 1 V
3 0 V
stroke
1061 325 M
3 1 V
3 2 V
3 1 V
3 -1 V
2 1 V
3 2 V
3 1 V
3 0 V
3 2 V
3 0 V
3 2 V
3 1 V
3 2 V
3 1 V
3 0 V
3 0 V
3 2 V
3 3 V
3 0 V
2 2 V
3 2 V
3 1 V
3 2 V
3 2 V
3 2 V
3 2 V
3 1 V
3 1 V
3 2 V
3 0 V
3 3 V
3 2 V
3 3 V
2 2 V
3 2 V
3 1 V
3 4 V
3 1 V
3 3 V
3 3 V
3 3 V
3 2 V
3 3 V
3 1 V
3 3 V
3 5 V
3 3 V
2 -2 V
3 6 V
3 3 V
3 3 V
3 3 V
3 4 V
3 5 V
3 5 V
3 0 V
3 6 V
3 3 V
3 5 V
3 2 V
3 5 V
2 4 V
3 6 V
3 2 V
3 4 V
3 10 V
3 2 V
3 5 V
3 4 V
3 4 V
3 6 V
3 8 V
3 2 V
3 11 V
3 0 V
2 9 V
3 3 V
3 10 V
3 5 V
3 7 V
3 6 V
3 7 V
3 6 V
3 8 V
3 9 V
3 5 V
3 7 V
3 7 V
3 9 V
3 8 V
2 9 V
3 6 V
3 8 V
3 8 V
3 10 V
3 9 V
3 10 V
3 10 V
3 7 V
3 9 V
3 10 V
3 8 V
3 12 V
3 10 V
stroke
1366 754 M
2 6 V
3 12 V
3 13 V
3 10 V
3 13 V
3 8 V
3 14 V
3 8 V
3 17 V
3 9 V
3 12 V
3 11 V
3 13 V
3 9 V
2 18 V
3 10 V
3 9 V
3 16 V
3 13 V
3 13 V
3 9 V
3 20 V
3 12 V
3 13 V
3 11 V
3 19 V
3 12 V
3 16 V
2 13 V
3 10 V
3 17 V
3 19 V
3 10 V
3 13 V
3 14 V
3 19 V
3 22 V
3 8 V
3 11 V
3 16 V
3 14 V
3 14 V
3 15 V
2 22 V
3 9 V
3 20 V
3 15 V
3 10 V
3 11 V
3 23 V
3 12 V
3 15 V
3 17 V
3 14 V
3 20 V
3 6 V
3 13 V
2 13 V
3 24 V
3 10 V
3 9 V
3 16 V
3 14 V
3 16 V
3 12 V
3 14 V
3 13 V
3 13 V
3 7 V
3 11 V
3 13 V
2 14 V
3 4 V
3 10 V
3 17 V
3 6 V
3 13 V
3 6 V
3 10 V
3 6 V
3 21 V
3 -1 V
3 7 V
3 4 V
3 11 V
2 -3 V
3 13 V
3 -2 V
3 5 V
3 9 V
3 -5 V
3 4 V
3 2 V
3 7 V
3 -8 V
3 11 V
3 1 V
3 -4 V
3 0 V
2 -1 V
3 -6 V
3 -5 V
3 8 V
3 -14 V
stroke
1670 1847 M
3 1 V
3 -9 V
3 -1 V
3 -10 V
3 -8 V
3 -11 V
3 -1 V
3 -13 V
3 -1 V
3 -15 V
2 -6 V
3 -15 V
3 -18 V
3 -12 V
3 -8 V
3 -20 V
3 -9 V
3 -16 V
3 -15 V
3 -17 V
3 -18 V
3 -2 V
3 -26 V
3 -9 V
2 -23 V
3 -19 V
3 -17 V
3 -28 V
3 -19 V
3 -22 V
3 -18 V
3 -19 V
3 -19 V
3 -31 V
3 -17 V
3 -29 V
3 -21 V
3 -20 V
2 -18 V
3 -30 V
3 -26 V
3 -16 V
3 -28 V
3 -22 V
3 -29 V
3 -17 V
3 -32 V
3 -21 V
3 -25 V
3 -26 V
3 -25 V
3 -25 V
2 -21 V
3 -24 V
3 -27 V
3 -17 V
3 -26 V
3 -22 V
3 -25 V
3 -20 V
3 -26 V
3 -22 V
3 -21 V
3 -22 V
3 -25 V
3 -17 V
3 -22 V
2 -17 V
3 -21 V
3 -18 V
3 -18 V
3 -19 V
3 -17 V
3 -17 V
3 -17 V
3 -14 V
3 -12 V
3 -16 V
3 -14 V
3 -10 V
3 -13 V
2 -13 V
3 -10 V
3 -9 V
3 -9 V
3 -8 V
3 -7 V
3 -7 V
3 -5 V
3 -6 V
3 -4 V
3 -3 V
3 -3 V
3 -1 V
3 -1 V
2 0 V
3 1 V
3 1 V
3 3 V
3 3 V
3 4 V
3 6 V
3 5 V
3 7 V
stroke
1975 330 M
3 7 V
3 8 V
3 9 V
3 9 V
3 10 V
2 13 V
3 13 V
3 10 V
3 14 V
3 16 V
3 12 V
3 14 V
3 17 V
3 17 V
3 17 V
3 19 V
3 18 V
3 18 V
3 21 V
2 17 V
3 22 V
3 17 V
3 25 V
3 22 V
3 21 V
3 22 V
3 26 V
3 20 V
3 25 V
3 22 V
3 26 V
3 17 V
3 27 V
3 24 V
2 21 V
3 25 V
3 25 V
3 26 V
3 25 V
3 21 V
3 32 V
3 17 V
3 29 V
3 22 V
3 28 V
3 16 V
3 26 V
3 30 V
2 18 V
3 20 V
3 21 V
3 29 V
3 17 V
3 31 V
3 19 V
3 19 V
3 18 V
3 22 V
3 19 V
3 28 V
3 17 V
3 19 V
2 23 V
3 9 V
3 26 V
3 2 V
3 18 V
3 17 V
3 15 V
3 16 V
3 9 V
3 20 V
3 8 V
3 12 V
3 18 V
3 15 V
2 6 V
3 15 V
3 1 V
3 13 V
3 1 V
3 11 V
3 8 V
3 10 V
3 1 V
3 9 V
3 -1 V
3 14 V
3 -8 V
3 5 V
3 6 V
2 1 V
3 0 V
3 4 V
3 -1 V
3 -11 V
3 8 V
3 -7 V
3 -2 V
3 -4 V
3 5 V
3 -9 V
3 -5 V
3 2 V
stroke
2280 1845 M
3 -13 V
2 3 V
3 -11 V
3 -4 V
3 -7 V
3 1 V
3 -21 V
3 -6 V
3 -10 V
3 -6 V
3 -13 V
3 -6 V
3 -17 V
3 -10 V
3 -4 V
2 -14 V
3 -13 V
3 -11 V
3 -7 V
3 -13 V
3 -13 V
3 -14 V
3 -12 V
3 -16 V
3 -14 V
3 -16 V
3 -9 V
3 -10 V
3 -24 V
2 -13 V
3 -13 V
3 -6 V
3 -20 V
3 -14 V
3 -17 V
3 -15 V
3 -12 V
3 -23 V
3 -11 V
3 -10 V
3 -15 V
3 -20 V
3 -9 V
2 -22 V
3 -15 V
3 -14 V
3 -14 V
3 -16 V
3 -11 V
3 -8 V
3 -22 V
3 -19 V
3 -14 V
3 -13 V
3 -10 V
3 -19 V
3 -17 V
3 -10 V
2 -13 V
3 -16 V
3 -12 V
3 -19 V
3 -11 V
3 -13 V
3 -12 V
3 -20 V
3 -9 V
3 -13 V
3 -13 V
3 -16 V
3 -9 V
3 -10 V
2 -18 V
3 -9 V
3 -13 V
3 -11 V
3 -12 V
3 -9 V
3 -17 V
3 -8 V
3 -14 V
3 -8 V
3 -13 V
3 -10 V
3 -13 V
3 -12 V
2 -6 V
3 -10 V
3 -12 V
3 -8 V
3 -10 V
3 -9 V
3 -7 V
3 -10 V
3 -10 V
3 -9 V
3 -10 V
3 -8 V
3 -8 V
3 -6 V
2 -9 V
3 -8 V
3 -9 V
3 -7 V
stroke
2584 604 M
3 -7 V
3 -5 V
3 -9 V
3 -8 V
3 -6 V
3 -7 V
3 -6 V
3 -7 V
3 -5 V
3 -10 V
3 -3 V
2 -9 V
3 0 V
3 -11 V
3 -2 V
3 -8 V
3 -6 V
3 -4 V
3 -4 V
3 -5 V
3 -2 V
3 -10 V
3 -4 V
3 -2 V
3 -6 V
2 -4 V
3 -5 V
3 -2 V
3 -5 V
3 -3 V
3 -6 V
3 0 V
3 -5 V
3 -5 V
3 -4 V
3 -3 V
3 -3 V
3 -3 V
3 -6 V
2 2 V
3 -3 V
3 -5 V
3 -3 V
3 -1 V
3 -3 V
3 -2 V
3 -3 V
3 -3 V
3 -3 V
3 -1 V
3 -4 V
3 -1 V
3 -2 V
2 -2 V
3 -3 V
3 -2 V
3 -3 V
3 0 V
3 -2 V
3 -1 V
3 -1 V
3 -2 V
3 -2 V
3 -2 V
3 -2 V
3 -1 V
3 -2 V
2 -2 V
3 0 V
3 -3 V
3 -2 V
3 0 V
3 0 V
3 -1 V
3 -2 V
3 -1 V
3 -2 V
3 0 V
3 -2 V
3 0 V
3 -1 V
3 -2 V
2 -1 V
3 1 V
3 -1 V
3 -2 V
3 -1 V
3 0 V
3 -1 V
3 0 V
3 -1 V
3 -1 V
3 -1 V
3 -1 V
3 1 V
3 -2 V
2 0 V
3 0 V
3 -2 V
3 0 V
3 -1 V
3 -1 V
3 0 V
3 1 V
stroke
2889 316 M
3 -1 V
3 -1 V
3 0 V
3 -1 V
3 0 V
3 -1 V
2 1 V
3 -1 V
3 -1 V
3 0 V
3 0 V
3 -1 V
3 0 V
3 0 V
3 0 V
3 -1 V
3 0 V
3 -1 V
3 0 V
3 0 V
2 0 V
3 0 V
3 -1 V
3 0 V
3 0 V
3 0 V
3 -1 V
3 0 V
3 0 V
3 -1 V
3 1 V
3 -1 V
3 0 V
3 0 V
3 0 V
2 0 V
3 0 V
3 0 V
3 0 V
3 -1 V
3 0 V
3 0 V
3 0 V
3 0 V
3 0 V
3 0 V
3 -1 V
3 0 V
3 0 V
2 0 V
3 0 V
3 0 V
3 0 V
3 0 V
3 0 V
3 -1 V
3 1 V
3 -1 V
3 0 V
3 0 V
3 0 V
3 0 V
3 0 V
2 0 V
3 0 V
3 0 V
3 0 V
3 0 V
3 0 V
3 0 V
3 -1 V
3 0 V
3 1 V
3 0 V
3 -1 V
3 0 V
3 0 V
2 0 V
3 0 V
3 0 V
3 0 V
3 0 V
3 0 V
3 0 V
3 0 V
3 0 V
3 0 V
3 0 V
3 0 V
3 0 V
3 0 V
2 0 V
3 0 V
3 0 V
3 0 V
3 0 V
3 0 V
3 0 V
3 0 V
3 0 V
3 0 V
3 0 V
3 0 V
3 0 V
stroke
3194 301 M
3 -1 V
3 0 V
2 0 V
3 0 V
3 0 V
3 0 V
3 0 V
3 0 V
3 0 V
3 0 V
3 0 V
3 0 V
3 0 V
3 0 V
3 0 V
3 0 V
2 0 V
3 0 V
3 0 V
3 0 V
3 0 V
3 0 V
3 0 V
3 0 V
3 0 V
3 0 V
3 0 V
3 0 V
3 0 V
3 0 V
2 0 V
3 0 V
3 0 V
3 0 V
3 0 V
3 0 V
3 0 V
3 0 V
3 0 V
3 0 V
3 0 V
3 0 V
3 0 V
3 0 V
2 0 V
3 0 V
3 0 V
3 0 V
3 0 V
3 0 V
3 0 V
3 0 V
3 0 V
3 0 V
3 0 V
3 0 V
3 0 V
3 0 V
3 0 V
2 0 V
3 0 V
3 0 V
3 0 V
3 0 V
3 0 V
3 0 V
3 0 V
3 0 V
3 0 V
3 0 V
3 0 V
3 0 V
3 0 V
2 0 V
3 0 V
3 0 V
3 0 V
3 0 V
3 0 V
3 0 V
3 0 V
3 0 V
3 0 V
3 0 V
3 0 V
3 0 V
3 0 V
1.000 UP
0.500 UL
LT1
451 300 M
3 0 V
3 0 V
3 0 V
3 0 V
3 0 V
3 0 V
3 0 V
3 0 V
3 0 V
3 0 V
3 0 V
3 0 V
3 0 V
2 0 V
3 0 V
3 0 V
3 0 V
3 0 V
3 0 V
3 0 V
3 0 V
3 0 V
3 0 V
3 0 V
3 0 V
3 0 V
3 0 V
2 0 V
3 0 V
3 0 V
3 0 V
3 0 V
3 0 V
3 0 V
3 0 V
3 0 V
3 0 V
3 0 V
3 0 V
3 0 V
3 0 V
3 0 V
2 0 V
3 0 V
3 0 V
3 0 V
3 0 V
3 0 V
3 0 V
3 0 V
3 0 V
3 0 V
3 0 V
3 0 V
3 0 V
3 0 V
2 0 V
3 0 V
3 0 V
3 0 V
3 0 V
3 0 V
3 0 V
3 0 V
3 0 V
3 0 V
3 0 V
3 0 V
3 0 V
3 0 V
2 0 V
3 0 V
3 0 V
3 0 V
3 0 V
3 0 V
3 0 V
3 0 V
3 0 V
3 0 V
3 0 V
3 0 V
3 0 V
3 0 V
2 0 V
3 0 V
3 0 V
3 0 V
3 0 V
3 0 V
3 0 V
3 0 V
3 0 V
3 0 V
3 1 V
3 0 V
3 0 V
3 -1 V
3 1 V
2 0 V
3 0 V
3 0 V
3 0 V
3 0 V
stroke
756 301 M
3 0 V
3 0 V
3 0 V
3 0 V
3 0 V
3 0 V
3 0 V
3 0 V
3 0 V
2 0 V
3 0 V
3 0 V
3 0 V
3 0 V
3 0 V
3 0 V
3 0 V
3 0 V
3 0 V
3 0 V
3 1 V
3 -1 V
3 1 V
2 0 V
3 0 V
3 0 V
3 0 V
3 0 V
3 0 V
3 0 V
3 0 V
3 0 V
3 0 V
3 0 V
3 0 V
3 1 V
3 0 V
2 0 V
3 0 V
3 0 V
3 0 V
3 0 V
3 0 V
3 0 V
3 0 V
3 0 V
3 0 V
3 1 V
3 0 V
3 0 V
3 0 V
2 0 V
3 0 V
3 0 V
3 1 V
3 0 V
3 0 V
3 0 V
3 0 V
3 1 V
3 0 V
3 0 V
3 0 V
3 0 V
3 1 V
3 0 V
2 0 V
3 0 V
3 0 V
3 1 V
3 0 V
3 0 V
3 1 V
3 0 V
3 0 V
3 1 V
3 0 V
3 0 V
3 0 V
3 1 V
2 0 V
3 0 V
3 1 V
3 0 V
3 1 V
3 0 V
3 0 V
3 1 V
3 0 V
3 1 V
3 1 V
3 0 V
3 0 V
3 1 V
2 1 V
3 0 V
3 1 V
3 0 V
3 1 V
3 1 V
3 0 V
3 1 V
3 1 V
3 1 V
stroke
1061 324 M
3 0 V
3 1 V
3 1 V
3 0 V
2 2 V
3 1 V
3 1 V
3 1 V
3 0 V
3 2 V
3 0 V
3 2 V
3 1 V
3 1 V
3 2 V
3 0 V
3 2 V
3 1 V
3 1 V
2 2 V
3 1 V
3 2 V
3 1 V
3 2 V
3 2 V
3 1 V
3 1 V
3 3 V
3 2 V
3 1 V
3 2 V
3 3 V
3 1 V
2 2 V
3 2 V
3 2 V
3 3 V
3 2 V
3 3 V
3 1 V
3 4 V
3 2 V
3 3 V
3 3 V
3 3 V
3 2 V
3 3 V
2 3 V
3 3 V
3 4 V
3 3 V
3 4 V
3 3 V
3 3 V
3 4 V
3 4 V
3 4 V
3 3 V
3 3 V
3 5 V
3 5 V
2 5 V
3 3 V
3 5 V
3 5 V
3 5 V
3 4 V
3 6 V
3 5 V
3 4 V
3 7 V
3 5 V
3 6 V
3 5 V
3 6 V
2 6 V
3 6 V
3 6 V
3 7 V
3 7 V
3 6 V
3 6 V
3 6 V
3 8 V
3 8 V
3 7 V
3 9 V
3 6 V
3 9 V
3 9 V
2 7 V
3 7 V
3 9 V
3 8 V
3 12 V
3 8 V
3 7 V
3 10 V
3 9 V
3 11 V
3 10 V
3 10 V
3 10 V
3 9 V
stroke
1366 757 M
2 10 V
3 11 V
3 11 V
3 11 V
3 11 V
3 12 V
3 11 V
3 10 V
3 15 V
3 9 V
3 14 V
3 13 V
3 9 V
3 13 V
2 12 V
3 13 V
3 16 V
3 12 V
3 10 V
3 15 V
3 15 V
3 13 V
3 13 V
3 16 V
3 14 V
3 11 V
3 14 V
3 17 V
2 12 V
3 15 V
3 13 V
3 15 V
3 18 V
3 15 V
3 14 V
3 15 V
3 14 V
3 12 V
3 18 V
3 13 V
3 17 V
3 15 V
3 18 V
2 12 V
3 16 V
3 17 V
3 12 V
3 18 V
3 13 V
3 15 V
3 14 V
3 16 V
3 14 V
3 15 V
3 15 V
3 13 V
3 13 V
2 18 V
3 15 V
3 7 V
3 21 V
3 10 V
3 10 V
3 20 V
3 8 V
3 13 V
3 12 V
3 12 V
3 12 V
3 11 V
3 10 V
2 14 V
3 8 V
3 11 V
3 12 V
3 6 V
3 10 V
3 10 V
3 8 V
3 11 V
3 1 V
3 12 V
3 6 V
3 4 V
3 7 V
2 5 V
3 4 V
3 7 V
3 3 V
3 4 V
3 2 V
3 3 V
3 1 V
3 4 V
3 -1 V
3 2 V
3 0 V
3 -3 V
3 -2 V
2 1 V
3 -5 V
3 -1 V
3 -8 V
3 -1 V
stroke
1670 1849 M
3 -5 V
3 -6 V
3 -9 V
3 -4 V
3 -6 V
3 -11 V
3 -4 V
3 -12 V
3 -8 V
3 -10 V
2 -16 V
3 -6 V
3 -15 V
3 -9 V
3 -17 V
3 -12 V
3 -15 V
3 -14 V
3 -16 V
3 -15 V
3 -18 V
3 -15 V
3 -22 V
3 -15 V
2 -19 V
3 -18 V
3 -18 V
3 -22 V
3 -16 V
3 -21 V
3 -25 V
3 -25 V
3 -12 V
3 -27 V
3 -22 V
3 -21 V
3 -22 V
3 -24 V
2 -24 V
3 -24 V
3 -23 V
3 -23 V
3 -22 V
3 -28 V
3 -20 V
3 -28 V
3 -23 V
3 -23 V
3 -25 V
3 -24 V
3 -24 V
3 -24 V
2 -25 V
3 -23 V
3 -25 V
3 -22 V
3 -24 V
3 -24 V
3 -22 V
3 -22 V
3 -23 V
3 -21 V
3 -22 V
3 -25 V
3 -17 V
3 -21 V
3 -20 V
2 -20 V
3 -19 V
3 -18 V
3 -18 V
3 -17 V
3 -17 V
3 -17 V
3 -16 V
3 -16 V
3 -13 V
3 -14 V
3 -12 V
3 -13 V
3 -11 V
2 -12 V
3 -10 V
3 -10 V
3 -8 V
3 -8 V
3 -8 V
3 -6 V
3 -6 V
3 -4 V
3 -5 V
3 -3 V
3 -2 V
3 -2 V
3 -1 V
2 0 V
3 1 V
3 2 V
3 2 V
3 3 V
3 5 V
3 4 V
3 6 V
3 6 V
stroke
1975 329 M
3 8 V
3 8 V
3 8 V
3 10 V
3 10 V
2 12 V
3 11 V
3 13 V
3 12 V
3 14 V
3 13 V
3 16 V
3 16 V
3 17 V
3 17 V
3 17 V
3 18 V
3 18 V
3 19 V
2 20 V
3 20 V
3 21 V
3 17 V
3 25 V
3 22 V
3 21 V
3 23 V
3 22 V
3 22 V
3 24 V
3 24 V
3 22 V
3 25 V
3 23 V
2 25 V
3 24 V
3 24 V
3 24 V
3 25 V
3 23 V
3 23 V
3 28 V
3 20 V
3 28 V
3 22 V
3 23 V
3 23 V
3 24 V
2 24 V
3 24 V
3 22 V
3 21 V
3 22 V
3 27 V
3 12 V
3 25 V
3 25 V
3 21 V
3 16 V
3 22 V
3 18 V
3 18 V
2 19 V
3 15 V
3 22 V
3 15 V
3 18 V
3 15 V
3 16 V
3 14 V
3 15 V
3 12 V
3 17 V
3 9 V
3 15 V
3 6 V
2 16 V
3 10 V
3 8 V
3 12 V
3 4 V
3 11 V
3 6 V
3 4 V
3 9 V
3 6 V
3 5 V
3 1 V
3 8 V
3 1 V
3 5 V
2 -1 V
3 2 V
3 3 V
3 0 V
3 -2 V
3 1 V
3 -4 V
3 -1 V
3 -3 V
3 -2 V
3 -4 V
3 -3 V
3 -7 V
stroke
2280 1843 M
3 -4 V
2 -5 V
3 -7 V
3 -4 V
3 -6 V
3 -12 V
3 -1 V
3 -11 V
3 -8 V
3 -10 V
3 -10 V
3 -6 V
3 -12 V
3 -11 V
3 -8 V
2 -14 V
3 -10 V
3 -11 V
3 -12 V
3 -12 V
3 -12 V
3 -13 V
3 -8 V
3 -20 V
3 -10 V
3 -10 V
3 -21 V
3 -7 V
3 -15 V
2 -18 V
3 -13 V
3 -13 V
3 -15 V
3 -15 V
3 -14 V
3 -16 V
3 -14 V
3 -15 V
3 -13 V
3 -18 V
3 -12 V
3 -17 V
3 -16 V
2 -12 V
3 -18 V
3 -15 V
3 -17 V
3 -13 V
3 -18 V
3 -12 V
3 -14 V
3 -15 V
3 -14 V
3 -15 V
3 -18 V
3 -15 V
3 -13 V
3 -15 V
2 -12 V
3 -17 V
3 -14 V
3 -11 V
3 -14 V
3 -16 V
3 -13 V
3 -13 V
3 -15 V
3 -15 V
3 -10 V
3 -12 V
3 -16 V
3 -13 V
2 -12 V
3 -13 V
3 -9 V
3 -13 V
3 -14 V
3 -9 V
3 -15 V
3 -10 V
3 -11 V
3 -12 V
3 -11 V
3 -11 V
3 -11 V
3 -11 V
2 -10 V
3 -9 V
3 -10 V
3 -10 V
3 -10 V
3 -11 V
3 -9 V
3 -10 V
3 -7 V
3 -8 V
3 -12 V
3 -8 V
3 -9 V
3 -7 V
2 -7 V
3 -9 V
3 -9 V
3 -6 V
stroke
2584 606 M
3 -9 V
3 -7 V
3 -8 V
3 -8 V
3 -6 V
3 -6 V
3 -6 V
3 -7 V
3 -7 V
3 -6 V
3 -6 V
2 -6 V
3 -6 V
3 -5 V
3 -6 V
3 -5 V
3 -7 V
3 -4 V
3 -5 V
3 -6 V
3 -4 V
3 -5 V
3 -5 V
3 -5 V
3 -3 V
2 -5 V
3 -5 V
3 -5 V
3 -3 V
3 -3 V
3 -4 V
3 -4 V
3 -4 V
3 -3 V
3 -3 V
3 -4 V
3 -3 V
3 -4 V
3 -3 V
2 -3 V
3 -3 V
3 -2 V
3 -3 V
3 -3 V
3 -3 V
3 -2 V
3 -4 V
3 -1 V
3 -3 V
3 -2 V
3 -3 V
3 -2 V
3 -2 V
2 -2 V
3 -1 V
3 -3 V
3 -2 V
3 -1 V
3 -2 V
3 -3 V
3 -1 V
3 -1 V
3 -2 V
3 -2 V
3 -1 V
3 -2 V
3 -1 V
2 -2 V
3 -1 V
3 -1 V
3 -2 V
3 0 V
3 -2 V
3 -1 V
3 -1 V
3 -2 V
3 0 V
3 -2 V
3 0 V
3 -1 V
3 -1 V
3 -1 V
2 -2 V
3 0 V
3 -1 V
3 -1 V
3 0 V
3 -1 V
3 -1 V
3 -1 V
3 0 V
3 -1 V
3 -1 V
3 0 V
3 -1 V
3 0 V
2 -1 V
3 -1 V
3 0 V
3 0 V
3 -1 V
3 -1 V
3 0 V
3 -1 V
stroke
2889 313 M
3 0 V
3 0 V
3 -1 V
3 0 V
3 -1 V
3 0 V
2 0 V
3 -1 V
3 0 V
3 0 V
3 0 V
3 -1 V
3 0 V
3 0 V
3 -1 V
3 0 V
3 0 V
3 -1 V
3 0 V
3 0 V
2 0 V
3 0 V
3 -1 V
3 0 V
3 0 V
3 0 V
3 0 V
3 -1 V
3 0 V
3 0 V
3 0 V
3 0 V
3 -1 V
3 0 V
3 0 V
2 0 V
3 0 V
3 0 V
3 0 V
3 -1 V
3 0 V
3 0 V
3 0 V
3 0 V
3 0 V
3 0 V
3 0 V
3 0 V
3 0 V
2 0 V
3 0 V
3 -1 V
3 0 V
3 0 V
3 0 V
3 0 V
3 0 V
3 0 V
3 0 V
3 0 V
3 0 V
3 0 V
3 0 V
2 0 V
3 -1 V
3 1 V
3 -1 V
3 0 V
3 0 V
3 0 V
3 0 V
3 0 V
3 0 V
3 0 V
3 0 V
3 0 V
3 0 V
2 0 V
3 0 V
3 0 V
3 0 V
3 0 V
3 0 V
3 0 V
3 0 V
3 0 V
3 0 V
3 0 V
3 0 V
3 0 V
3 0 V
2 0 V
3 -1 V
3 1 V
3 0 V
3 0 V
3 -1 V
3 0 V
3 0 V
3 0 V
3 0 V
3 0 V
3 0 V
3 0 V
stroke
3194 300 M
3 0 V
3 0 V
2 0 V
3 0 V
3 0 V
3 0 V
3 0 V
3 0 V
3 0 V
3 0 V
3 0 V
3 0 V
3 0 V
3 0 V
3 0 V
3 0 V
2 0 V
3 0 V
3 0 V
3 0 V
3 0 V
3 0 V
3 0 V
3 0 V
3 0 V
3 0 V
3 0 V
3 0 V
3 0 V
3 0 V
2 0 V
3 0 V
3 0 V
3 0 V
3 0 V
3 0 V
3 0 V
3 0 V
3 0 V
3 0 V
3 0 V
3 0 V
3 0 V
3 0 V
2 0 V
3 0 V
3 0 V
3 0 V
3 0 V
3 0 V
3 0 V
3 0 V
3 0 V
3 0 V
3 0 V
3 0 V
3 0 V
3 0 V
3 0 V
2 0 V
3 0 V
3 0 V
3 0 V
3 0 V
3 0 V
3 0 V
3 0 V
3 0 V
3 0 V
3 0 V
3 0 V
3 0 V
3 0 V
2 0 V
3 0 V
3 0 V
3 0 V
3 0 V
3 0 V
3 0 V
3 0 V
3 0 V
3 0 V
3 0 V
3 0 V
3 0 V
3 0 V
1.000 UP
0.500 UL
LT2
451 300 M
3 0 V
3 0 V
3 0 V
3 0 V
3 0 V
3 0 V
3 0 V
3 0 V
3 0 V
3 0 V
3 0 V
3 0 V
3 0 V
2 0 V
3 0 V
3 0 V
3 0 V
3 0 V
3 0 V
3 0 V
3 0 V
3 0 V
3 0 V
3 0 V
3 0 V
3 0 V
3 0 V
2 0 V
3 0 V
3 0 V
3 0 V
3 0 V
3 0 V
3 0 V
3 0 V
3 0 V
3 0 V
3 0 V
3 0 V
3 0 V
3 0 V
3 0 V
2 0 V
3 0 V
3 0 V
3 0 V
3 0 V
3 0 V
3 0 V
3 0 V
3 0 V
3 0 V
3 0 V
3 0 V
3 0 V
3 0 V
2 0 V
3 0 V
3 0 V
3 0 V
3 0 V
3 0 V
3 0 V
3 0 V
3 0 V
3 0 V
3 0 V
3 0 V
3 0 V
3 0 V
2 0 V
3 0 V
3 0 V
3 0 V
3 0 V
3 0 V
3 0 V
3 0 V
3 0 V
3 0 V
3 0 V
3 0 V
3 0 V
3 0 V
2 0 V
3 0 V
3 0 V
3 0 V
3 0 V
3 0 V
3 0 V
3 0 V
3 0 V
3 0 V
3 0 V
3 0 V
3 0 V
3 0 V
3 0 V
2 0 V
3 0 V
3 0 V
3 0 V
3 0 V
stroke
756 300 M
3 0 V
3 0 V
3 0 V
3 0 V
3 0 V
3 0 V
3 0 V
3 1 V
3 -1 V
2 1 V
3 0 V
3 0 V
3 0 V
3 0 V
3 0 V
3 0 V
3 0 V
3 0 V
3 0 V
3 0 V
3 0 V
3 0 V
3 0 V
2 0 V
3 0 V
3 0 V
3 0 V
3 0 V
3 0 V
3 0 V
3 0 V
3 0 V
3 0 V
3 0 V
3 0 V
3 1 V
3 0 V
2 0 V
3 0 V
3 0 V
3 0 V
3 0 V
3 0 V
3 0 V
3 0 V
3 0 V
3 0 V
3 1 V
3 0 V
3 0 V
3 0 V
2 0 V
3 0 V
3 0 V
3 0 V
3 0 V
3 1 V
3 0 V
3 0 V
3 0 V
3 0 V
3 0 V
3 0 V
3 1 V
3 0 V
3 0 V
2 0 V
3 0 V
3 1 V
3 0 V
3 0 V
3 0 V
3 1 V
3 0 V
3 0 V
3 0 V
3 1 V
3 0 V
3 0 V
3 0 V
2 1 V
3 0 V
3 0 V
3 1 V
3 0 V
3 1 V
3 0 V
3 1 V
3 0 V
3 1 V
3 0 V
3 0 V
3 1 V
3 0 V
2 1 V
3 0 V
3 1 V
3 0 V
3 1 V
3 0 V
3 1 V
3 1 V
3 1 V
3 0 V
stroke
1061 320 M
3 1 V
3 1 V
3 0 V
3 1 V
2 1 V
3 1 V
3 1 V
3 1 V
3 1 V
3 1 V
3 0 V
3 2 V
3 1 V
3 2 V
3 1 V
3 1 V
3 1 V
3 1 V
3 2 V
2 0 V
3 2 V
3 2 V
3 1 V
3 2 V
3 2 V
3 1 V
3 2 V
3 2 V
3 1 V
3 3 V
3 1 V
3 2 V
3 2 V
2 2 V
3 3 V
3 2 V
3 2 V
3 2 V
3 2 V
3 3 V
3 2 V
3 3 V
3 3 V
3 3 V
3 2 V
3 3 V
3 3 V
2 3 V
3 4 V
3 3 V
3 4 V
3 2 V
3 4 V
3 4 V
3 3 V
3 3 V
3 5 V
3 5 V
3 3 V
3 5 V
3 3 V
2 5 V
3 4 V
3 6 V
3 6 V
3 4 V
3 5 V
3 4 V
3 6 V
3 5 V
3 6 V
3 5 V
3 8 V
3 4 V
3 6 V
2 6 V
3 7 V
3 6 V
3 5 V
3 8 V
3 8 V
3 7 V
3 7 V
3 7 V
3 6 V
3 9 V
3 7 V
3 10 V
3 6 V
3 9 V
2 7 V
3 10 V
3 8 V
3 10 V
3 10 V
3 9 V
3 8 V
3 10 V
3 12 V
3 8 V
3 10 V
3 10 V
3 10 V
3 11 V
stroke
1366 764 M
2 13 V
3 8 V
3 12 V
3 11 V
3 13 V
3 11 V
3 10 V
3 12 V
3 14 V
3 11 V
3 13 V
3 12 V
3 14 V
3 13 V
2 11 V
3 14 V
3 14 V
3 13 V
3 12 V
3 12 V
3 16 V
3 14 V
3 15 V
3 14 V
3 11 V
3 16 V
3 15 V
3 17 V
2 11 V
3 16 V
3 15 V
3 13 V
3 15 V
3 17 V
3 14 V
3 15 V
3 17 V
3 15 V
3 15 V
3 16 V
3 14 V
3 14 V
3 18 V
2 14 V
3 13 V
3 16 V
3 16 V
3 14 V
3 17 V
3 15 V
3 10 V
3 20 V
3 14 V
3 11 V
3 18 V
3 10 V
3 14 V
2 19 V
3 11 V
3 14 V
3 12 V
3 15 V
3 11 V
3 12 V
3 15 V
3 10 V
3 15 V
3 12 V
3 7 V
3 14 V
3 9 V
2 13 V
3 9 V
3 11 V
3 10 V
3 9 V
3 10 V
3 5 V
3 8 V
3 6 V
3 12 V
3 5 V
3 5 V
3 5 V
3 6 V
2 5 V
3 6 V
3 2 V
3 5 V
3 0 V
3 6 V
3 3 V
3 -4 V
3 10 V
3 -6 V
3 1 V
3 -3 V
3 1 V
3 -6 V
2 -1 V
3 -3 V
3 -3 V
3 -5 V
3 -7 V
stroke
1670 1843 M
3 -5 V
3 -4 V
3 -11 V
3 -3 V
3 -8 V
3 -12 V
3 -9 V
3 -9 V
3 -8 V
3 -14 V
2 -8 V
3 -10 V
3 -18 V
3 -10 V
3 -14 V
3 -13 V
3 -12 V
3 -20 V
3 -13 V
3 -21 V
3 -17 V
3 -15 V
3 -17 V
3 -18 V
2 -21 V
3 -17 V
3 -19 V
3 -21 V
3 -24 V
3 -17 V
3 -20 V
3 -20 V
3 -20 V
3 -24 V
3 -25 V
3 -19 V
3 -25 V
3 -20 V
2 -22 V
3 -26 V
3 -25 V
3 -24 V
3 -22 V
3 -24 V
3 -22 V
3 -26 V
3 -23 V
3 -25 V
3 -20 V
3 -27 V
3 -22 V
3 -25 V
2 -25 V
3 -22 V
3 -26 V
3 -20 V
3 -26 V
3 -21 V
3 -22 V
3 -24 V
3 -21 V
3 -21 V
3 -21 V
3 -22 V
3 -19 V
3 -21 V
3 -20 V
2 -18 V
3 -20 V
3 -18 V
3 -16 V
3 -18 V
3 -17 V
3 -15 V
3 -16 V
3 -15 V
3 -13 V
3 -14 V
3 -13 V
3 -11 V
3 -12 V
2 -11 V
3 -10 V
3 -9 V
3 -8 V
3 -8 V
3 -7 V
3 -7 V
3 -5 V
3 -5 V
3 -4 V
3 -3 V
3 -3 V
3 -1 V
3 -1 V
2 0 V
3 1 V
3 1 V
3 3 V
3 3 V
3 4 V
3 5 V
3 5 V
3 7 V
stroke
1975 329 M
3 7 V
3 8 V
3 8 V
3 9 V
3 10 V
2 11 V
3 12 V
3 11 V
3 13 V
3 14 V
3 13 V
3 15 V
3 16 V
3 15 V
3 17 V
3 18 V
3 16 V
3 18 V
3 20 V
2 18 V
3 20 V
3 21 V
3 19 V
3 22 V
3 21 V
3 21 V
3 21 V
3 24 V
3 22 V
3 21 V
3 26 V
3 20 V
3 26 V
3 22 V
2 25 V
3 25 V
3 22 V
3 27 V
3 20 V
3 25 V
3 23 V
3 26 V
3 22 V
3 24 V
3 22 V
3 24 V
3 25 V
3 26 V
2 22 V
3 20 V
3 25 V
3 19 V
3 25 V
3 24 V
3 20 V
3 20 V
3 20 V
3 17 V
3 24 V
3 21 V
3 19 V
3 17 V
2 21 V
3 18 V
3 17 V
3 15 V
3 17 V
3 21 V
3 13 V
3 20 V
3 12 V
3 13 V
3 14 V
3 10 V
3 18 V
3 10 V
2 8 V
3 14 V
3 8 V
3 9 V
3 9 V
3 12 V
3 8 V
3 3 V
3 11 V
3 4 V
3 5 V
3 7 V
3 5 V
3 3 V
3 3 V
2 1 V
3 6 V
3 -1 V
3 3 V
3 -1 V
3 6 V
3 -10 V
3 4 V
3 -3 V
3 -6 V
3 0 V
3 -5 V
3 -2 V
stroke
2280 1853 M
3 -6 V
2 -5 V
3 -6 V
3 -5 V
3 -5 V
3 -5 V
3 -12 V
3 -6 V
3 -8 V
3 -5 V
3 -10 V
3 -9 V
3 -10 V
3 -11 V
3 -9 V
2 -13 V
3 -9 V
3 -14 V
3 -7 V
3 -12 V
3 -15 V
3 -10 V
3 -15 V
3 -12 V
3 -11 V
3 -15 V
3 -12 V
3 -14 V
3 -11 V
2 -19 V
3 -14 V
3 -10 V
3 -18 V
3 -11 V
3 -14 V
3 -20 V
3 -10 V
3 -15 V
3 -17 V
3 -14 V
3 -16 V
3 -16 V
3 -13 V
2 -14 V
3 -18 V
3 -14 V
3 -14 V
3 -16 V
3 -15 V
3 -15 V
3 -17 V
3 -15 V
3 -14 V
3 -17 V
3 -15 V
3 -13 V
3 -15 V
3 -16 V
2 -11 V
3 -17 V
3 -15 V
3 -16 V
3 -11 V
3 -14 V
3 -15 V
3 -14 V
3 -16 V
3 -12 V
3 -12 V
3 -13 V
3 -14 V
3 -14 V
2 -11 V
3 -13 V
3 -14 V
3 -12 V
3 -13 V
3 -11 V
3 -14 V
3 -12 V
3 -10 V
3 -11 V
3 -13 V
3 -11 V
3 -12 V
3 -8 V
2 -13 V
3 -11 V
3 -10 V
3 -10 V
3 -10 V
3 -8 V
3 -12 V
3 -10 V
3 -8 V
3 -9 V
3 -10 V
3 -10 V
3 -8 V
3 -10 V
2 -7 V
3 -9 V
3 -6 V
3 -10 V
stroke
2584 606 M
3 -7 V
3 -9 V
3 -6 V
3 -7 V
3 -7 V
3 -7 V
3 -8 V
3 -8 V
3 -5 V
3 -6 V
3 -7 V
2 -6 V
3 -6 V
3 -4 V
3 -8 V
3 -5 V
3 -6 V
3 -5 V
3 -6 V
3 -4 V
3 -5 V
3 -4 V
3 -6 V
3 -6 V
3 -4 V
2 -5 V
3 -3 V
3 -5 V
3 -3 V
3 -5 V
3 -5 V
3 -3 V
3 -3 V
3 -4 V
3 -4 V
3 -2 V
3 -4 V
3 -3 V
3 -4 V
2 -3 V
3 -3 V
3 -3 V
3 -2 V
3 -3 V
3 -3 V
3 -3 V
3 -2 V
3 -3 V
3 -2 V
3 -2 V
3 -2 V
3 -2 V
3 -3 V
2 -2 V
3 -2 V
3 -2 V
3 -1 V
3 -3 V
3 -1 V
3 -2 V
3 -2 V
3 -1 V
3 -2 V
3 -2 V
3 -1 V
3 -2 V
3 -2 V
2 0 V
3 -2 V
3 -1 V
3 -1 V
3 -1 V
3 -1 V
3 -2 V
3 -1 V
3 -2 V
3 0 V
3 -1 V
3 -1 V
3 -1 V
3 -1 V
3 -1 V
2 -1 V
3 -1 V
3 0 V
3 -1 V
3 -1 V
3 0 V
3 -1 V
3 -1 V
3 -1 V
3 0 V
3 -1 V
3 0 V
3 -1 V
3 0 V
2 -1 V
3 0 V
3 -1 V
3 0 V
3 0 V
3 -1 V
3 0 V
3 -1 V
stroke
2889 311 M
3 0 V
3 -1 V
3 0 V
3 -1 V
3 0 V
3 0 V
2 -1 V
3 0 V
3 0 V
3 0 V
3 -1 V
3 0 V
3 0 V
3 0 V
3 -1 V
3 0 V
3 0 V
3 0 V
3 -1 V
3 0 V
2 0 V
3 0 V
3 0 V
3 -1 V
3 0 V
3 0 V
3 0 V
3 0 V
3 0 V
3 0 V
3 -1 V
3 0 V
3 0 V
3 0 V
3 0 V
2 0 V
3 0 V
3 0 V
3 0 V
3 -1 V
3 0 V
3 0 V
3 0 V
3 0 V
3 0 V
3 0 V
3 0 V
3 0 V
3 0 V
2 0 V
3 0 V
3 -1 V
3 0 V
3 0 V
3 0 V
3 0 V
3 0 V
3 0 V
3 0 V
3 0 V
3 0 V
3 0 V
3 0 V
2 0 V
3 0 V
3 0 V
3 0 V
3 0 V
3 0 V
3 0 V
3 0 V
3 0 V
3 0 V
3 0 V
3 0 V
3 0 V
3 0 V
2 -1 V
3 1 V
3 -1 V
3 0 V
3 0 V
3 0 V
3 0 V
3 0 V
3 0 V
3 0 V
3 0 V
3 0 V
3 0 V
3 0 V
2 0 V
3 0 V
3 0 V
3 0 V
3 0 V
3 0 V
3 0 V
3 0 V
3 0 V
3 0 V
3 0 V
3 0 V
3 0 V
stroke
3194 300 M
3 0 V
3 0 V
2 0 V
3 0 V
3 0 V
3 0 V
3 0 V
3 0 V
3 0 V
3 0 V
3 0 V
3 0 V
3 0 V
3 0 V
3 0 V
3 0 V
2 0 V
3 0 V
3 0 V
3 0 V
3 0 V
3 0 V
3 0 V
3 0 V
3 0 V
3 0 V
3 0 V
3 0 V
3 0 V
3 0 V
2 0 V
3 0 V
3 0 V
3 0 V
3 0 V
3 0 V
3 0 V
3 0 V
3 0 V
3 0 V
3 0 V
3 0 V
3 0 V
3 0 V
2 0 V
3 0 V
3 0 V
3 0 V
3 0 V
3 0 V
3 0 V
3 0 V
3 0 V
3 0 V
3 0 V
3 0 V
3 0 V
3 0 V
3 0 V
2 0 V
3 0 V
3 0 V
3 0 V
3 0 V
3 0 V
3 0 V
3 0 V
3 0 V
3 0 V
3 0 V
3 0 V
3 0 V
3 0 V
2 0 V
3 0 V
3 0 V
3 0 V
3 0 V
3 0 V
3 0 V
3 0 V
3 0 V
3 0 V
3 0 V
3 0 V
3 0 V
3 0 V
1.000 UL
LTb
450 300 M
3000 0 V
0 1760 V
-3000 0 V
450 300 L
1.000 UP
stroke
grestore
end
showpage
}}%
\put(1950,50){\makebox(0,0){$\alpha$}}%
\put(100,1180){%
\special{ps: gsave currentpoint currentpoint translate
270 rotate neg exch neg exch translate}%
\makebox(0,0)[b]{\shortstack{$\rho(\alpha)$}}%
\special{ps: currentpoint grestore moveto}%
}%
\put(3382,200){\makebox(0,0){ 3}}%
\put(2905,200){\makebox(0,0){ 2}}%
\put(2427,200){\makebox(0,0){ 1}}%
\put(1950,200){\makebox(0,0){ 0}}%
\put(1473,200){\makebox(0,0){-1}}%
\put(995,200){\makebox(0,0){-2}}%
\put(518,200){\makebox(0,0){-3}}%
\put(400,2060){\makebox(0,0)[r]{ 0.7}}%
\put(400,1809){\makebox(0,0)[r]{ 0.6}}%
\put(400,1557){\makebox(0,0)[r]{ 0.5}}%
\put(400,1306){\makebox(0,0)[r]{ 0.4}}%
\put(400,1054){\makebox(0,0)[r]{ 0.3}}%
\put(400,803){\makebox(0,0)[r]{ 0.2}}%
\put(400,551){\makebox(0,0)[r]{ 0.1}}%
\put(400,300){\makebox(0,0)[r]{ 0}}%
\end{picture}%
\endgroup
 

%% file: figures/su2plaq_0.7fouriers.tex
\begingroup%
  \makeatletter%
  \newcommand{\GNUPLOTspecial}{%
    \@sanitize\catcode`\%=14\relax\special}%
  \setlength{\unitlength}{0.1bp}%
\begin{picture}(3600,2160)(0,0)%
{\GNUPLOTspecial{"
/gnudict 256 dict def
gnudict begin
/Color false def
/Solid false def
/gnulinewidth 5.000 def
/userlinewidth gnulinewidth def
/vshift -33 def
/dl {10.0 mul} def
/hpt_ 31.5 def
/vpt_ 31.5 def
/hpt hpt_ def
/vpt vpt_ def
/Rounded false def
/M {moveto} bind def
/L {lineto} bind def
/R {rmoveto} bind def
/V {rlineto} bind def
/N {newpath moveto} bind def
/C {setrgbcolor} bind def
/f {rlineto fill} bind def
/vpt2 vpt 2 mul def
/hpt2 hpt 2 mul def
/Lshow { currentpoint stroke M
  0 vshift R show } def
/Rshow { currentpoint stroke M
  dup stringwidth pop neg vshift R show } def
/Cshow { currentpoint stroke M
  dup stringwidth pop -2 div vshift R show } def
/UP { dup vpt_ mul /vpt exch def hpt_ mul /hpt exch def
  /hpt2 hpt 2 mul def /vpt2 vpt 2 mul def } def
/DL { Color {setrgbcolor Solid {pop []} if 0 setdash }
 {pop pop pop 0 setgray Solid {pop []} if 0 setdash} ifelse } def
/BL { stroke userlinewidth 2 mul setlinewidth
      Rounded { 1 setlinejoin 1 setlinecap } if } def
/AL { stroke userlinewidth 2 div setlinewidth
      Rounded { 1 setlinejoin 1 setlinecap } if } def
/UL { dup gnulinewidth mul /userlinewidth exch def
      dup 1 lt {pop 1} if 10 mul /udl exch def } def
/PL { stroke userlinewidth setlinewidth
      Rounded { 1 setlinejoin 1 setlinecap } if } def
/LTw { PL [] 1 setgray } def
/LTb { BL [] 0 0 0 DL } def
/LTa { AL [1 udl mul 2 udl mul] 0 setdash 0 0 0 setrgbcolor } def
/LT0 { PL [] 1 0 0 DL } def
/LT1 { PL [4 dl 2 dl] 0 1 0 DL } def
/LT2 { PL [2 dl 3 dl] 0 0 1 DL } def
/LT3 { PL [1 dl 1.5 dl] 1 0 1 DL } def
/LT4 { PL [5 dl 2 dl 1 dl 2 dl] 0 1 1 DL } def
/LT5 { PL [4 dl 3 dl 1 dl 3 dl] 1 1 0 DL } def
/LT6 { PL [2 dl 2 dl 2 dl 4 dl] 0 0 0 DL } def
/LT7 { PL [2 dl 2 dl 2 dl 2 dl 2 dl 4 dl] 1 0.3 0 DL } def
/LT8 { PL [2 dl 2 dl 2 dl 2 dl 2 dl 2 dl 2 dl 4 dl] 0.5 0.5 0.5 DL } def
/Pnt { stroke [] 0 setdash
   gsave 1 setlinecap M 0 0 V stroke grestore } def
/Dia { stroke [] 0 setdash 2 copy vpt add M
  hpt neg vpt neg V hpt vpt neg V
  hpt vpt V hpt neg vpt V closepath stroke
  Pnt } def
/Pls { stroke [] 0 setdash vpt sub M 0 vpt2 V
  currentpoint stroke M
  hpt neg vpt neg R hpt2 0 V stroke
  } def
/Box { stroke [] 0 setdash 2 copy exch hpt sub exch vpt add M
  0 vpt2 neg V hpt2 0 V 0 vpt2 V
  hpt2 neg 0 V closepath stroke
  Pnt } def
/Crs { stroke [] 0 setdash exch hpt sub exch vpt add M
  hpt2 vpt2 neg V currentpoint stroke M
  hpt2 neg 0 R hpt2 vpt2 V stroke } def
/TriU { stroke [] 0 setdash 2 copy vpt 1.12 mul add M
  hpt neg vpt -1.62 mul V
  hpt 2 mul 0 V
  hpt neg vpt 1.62 mul V closepath stroke
  Pnt  } def
/Star { 2 copy Pls Crs } def
/BoxF { stroke [] 0 setdash exch hpt sub exch vpt add M
  0 vpt2 neg V  hpt2 0 V  0 vpt2 V
  hpt2 neg 0 V  closepath fill } def
/TriUF { stroke [] 0 setdash vpt 1.12 mul add M
  hpt neg vpt -1.62 mul V
  hpt 2 mul 0 V
  hpt neg vpt 1.62 mul V closepath fill } def
/TriD { stroke [] 0 setdash 2 copy vpt 1.12 mul sub M
  hpt neg vpt 1.62 mul V
  hpt 2 mul 0 V
  hpt neg vpt -1.62 mul V closepath stroke
  Pnt  } def
/TriDF { stroke [] 0 setdash vpt 1.12 mul sub M
  hpt neg vpt 1.62 mul V
  hpt 2 mul 0 V
  hpt neg vpt -1.62 mul V closepath fill} def
/DiaF { stroke [] 0 setdash vpt add M
  hpt neg vpt neg V hpt vpt neg V
  hpt vpt V hpt neg vpt V closepath fill } def
/Pent { stroke [] 0 setdash 2 copy gsave
  translate 0 hpt M 4 {72 rotate 0 hpt L} repeat
  closepath stroke grestore Pnt } def
/PentF { stroke [] 0 setdash gsave
  translate 0 hpt M 4 {72 rotate 0 hpt L} repeat
  closepath fill grestore } def
/Circle { stroke [] 0 setdash 2 copy
  hpt 0 360 arc stroke Pnt } def
/CircleF { stroke [] 0 setdash hpt 0 360 arc fill } def
/C0 { BL [] 0 setdash 2 copy moveto vpt 90 450  arc } bind def
/C1 { BL [] 0 setdash 2 copy        moveto
       2 copy  vpt 0 90 arc closepath fill
               vpt 0 360 arc closepath } bind def
/C2 { BL [] 0 setdash 2 copy moveto
       2 copy  vpt 90 180 arc closepath fill
               vpt 0 360 arc closepath } bind def
/C3 { BL [] 0 setdash 2 copy moveto
       2 copy  vpt 0 180 arc closepath fill
               vpt 0 360 arc closepath } bind def
/C4 { BL [] 0 setdash 2 copy moveto
       2 copy  vpt 180 270 arc closepath fill
               vpt 0 360 arc closepath } bind def
/C5 { BL [] 0 setdash 2 copy moveto
       2 copy  vpt 0 90 arc
       2 copy moveto
       2 copy  vpt 180 270 arc closepath fill
               vpt 0 360 arc } bind def
/C6 { BL [] 0 setdash 2 copy moveto
      2 copy  vpt 90 270 arc closepath fill
              vpt 0 360 arc closepath } bind def
/C7 { BL [] 0 setdash 2 copy moveto
      2 copy  vpt 0 270 arc closepath fill
              vpt 0 360 arc closepath } bind def
/C8 { BL [] 0 setdash 2 copy moveto
      2 copy vpt 270 360 arc closepath fill
              vpt 0 360 arc closepath } bind def
/C9 { BL [] 0 setdash 2 copy moveto
      2 copy  vpt 270 450 arc closepath fill
              vpt 0 360 arc closepath } bind def
/C10 { BL [] 0 setdash 2 copy 2 copy moveto vpt 270 360 arc closepath fill
       2 copy moveto
       2 copy vpt 90 180 arc closepath fill
               vpt 0 360 arc closepath } bind def
/C11 { BL [] 0 setdash 2 copy moveto
       2 copy  vpt 0 180 arc closepath fill
       2 copy moveto
       2 copy  vpt 270 360 arc closepath fill
               vpt 0 360 arc closepath } bind def
/C12 { BL [] 0 setdash 2 copy moveto
       2 copy  vpt 180 360 arc closepath fill
               vpt 0 360 arc closepath } bind def
/C13 { BL [] 0 setdash  2 copy moveto
       2 copy  vpt 0 90 arc closepath fill
       2 copy moveto
       2 copy  vpt 180 360 arc closepath fill
               vpt 0 360 arc closepath } bind def
/C14 { BL [] 0 setdash 2 copy moveto
       2 copy  vpt 90 360 arc closepath fill
               vpt 0 360 arc } bind def
/C15 { BL [] 0 setdash 2 copy vpt 0 360 arc closepath fill
               vpt 0 360 arc closepath } bind def
/Rec   { newpath 4 2 roll moveto 1 index 0 rlineto 0 exch rlineto
       neg 0 rlineto closepath } bind def
/Square { dup Rec } bind def
/Bsquare { vpt sub exch vpt sub exch vpt2 Square } bind def
/S0 { BL [] 0 setdash 2 copy moveto 0 vpt rlineto BL Bsquare } bind def
/S1 { BL [] 0 setdash 2 copy vpt Square fill Bsquare } bind def
/S2 { BL [] 0 setdash 2 copy exch vpt sub exch vpt Square fill Bsquare } bind def
/S3 { BL [] 0 setdash 2 copy exch vpt sub exch vpt2 vpt Rec fill Bsquare } bind def
/S4 { BL [] 0 setdash 2 copy exch vpt sub exch vpt sub vpt Square fill Bsquare } bind def
/S5 { BL [] 0 setdash 2 copy 2 copy vpt Square fill
       exch vpt sub exch vpt sub vpt Square fill Bsquare } bind def
/S6 { BL [] 0 setdash 2 copy exch vpt sub exch vpt sub vpt vpt2 Rec fill Bsquare } bind def
/S7 { BL [] 0 setdash 2 copy exch vpt sub exch vpt sub vpt vpt2 Rec fill
       2 copy vpt Square fill
       Bsquare } bind def
/S8 { BL [] 0 setdash 2 copy vpt sub vpt Square fill Bsquare } bind def
/S9 { BL [] 0 setdash 2 copy vpt sub vpt vpt2 Rec fill Bsquare } bind def
/S10 { BL [] 0 setdash 2 copy vpt sub vpt Square fill 2 copy exch vpt sub exch vpt Square fill
       Bsquare } bind def
/S11 { BL [] 0 setdash 2 copy vpt sub vpt Square fill 2 copy exch vpt sub exch vpt2 vpt Rec fill
       Bsquare } bind def
/S12 { BL [] 0 setdash 2 copy exch vpt sub exch vpt sub vpt2 vpt Rec fill Bsquare } bind def
/S13 { BL [] 0 setdash 2 copy exch vpt sub exch vpt sub vpt2 vpt Rec fill
       2 copy vpt Square fill Bsquare } bind def
/S14 { BL [] 0 setdash 2 copy exch vpt sub exch vpt sub vpt2 vpt Rec fill
       2 copy exch vpt sub exch vpt Square fill Bsquare } bind def
/S15 { BL [] 0 setdash 2 copy Bsquare fill Bsquare } bind def
/D0 { gsave translate 45 rotate 0 0 S0 stroke grestore } bind def
/D1 { gsave translate 45 rotate 0 0 S1 stroke grestore } bind def
/D2 { gsave translate 45 rotate 0 0 S2 stroke grestore } bind def
/D3 { gsave translate 45 rotate 0 0 S3 stroke grestore } bind def
/D4 { gsave translate 45 rotate 0 0 S4 stroke grestore } bind def
/D5 { gsave translate 45 rotate 0 0 S5 stroke grestore } bind def
/D6 { gsave translate 45 rotate 0 0 S6 stroke grestore } bind def
/D7 { gsave translate 45 rotate 0 0 S7 stroke grestore } bind def
/D8 { gsave translate 45 rotate 0 0 S8 stroke grestore } bind def
/D9 { gsave translate 45 rotate 0 0 S9 stroke grestore } bind def
/D10 { gsave translate 45 rotate 0 0 S10 stroke grestore } bind def
/D11 { gsave translate 45 rotate 0 0 S11 stroke grestore } bind def
/D12 { gsave translate 45 rotate 0 0 S12 stroke grestore } bind def
/D13 { gsave translate 45 rotate 0 0 S13 stroke grestore } bind def
/D14 { gsave translate 45 rotate 0 0 S14 stroke grestore } bind def
/D15 { gsave translate 45 rotate 0 0 S15 stroke grestore } bind def
/DiaE { stroke [] 0 setdash vpt add M
  hpt neg vpt neg V hpt vpt neg V
  hpt vpt V hpt neg vpt V closepath stroke } def
/BoxE { stroke [] 0 setdash exch hpt sub exch vpt add M
  0 vpt2 neg V hpt2 0 V 0 vpt2 V
  hpt2 neg 0 V closepath stroke } def
/TriUE { stroke [] 0 setdash vpt 1.12 mul add M
  hpt neg vpt -1.62 mul V
  hpt 2 mul 0 V
  hpt neg vpt 1.62 mul V closepath stroke } def
/TriDE { stroke [] 0 setdash vpt 1.12 mul sub M
  hpt neg vpt 1.62 mul V
  hpt 2 mul 0 V
  hpt neg vpt -1.62 mul V closepath stroke } def
/PentE { stroke [] 0 setdash gsave
  translate 0 hpt M 4 {72 rotate 0 hpt L} repeat
  closepath stroke grestore } def
/CircE { stroke [] 0 setdash 
  hpt 0 360 arc stroke } def
/Opaque { gsave closepath 1 setgray fill grestore 0 setgray closepath } def
/DiaW { stroke [] 0 setdash vpt add M
  hpt neg vpt neg V hpt vpt neg V
  hpt vpt V hpt neg vpt V Opaque stroke } def
/BoxW { stroke [] 0 setdash exch hpt sub exch vpt add M
  0 vpt2 neg V hpt2 0 V 0 vpt2 V
  hpt2 neg 0 V Opaque stroke } def
/TriUW { stroke [] 0 setdash vpt 1.12 mul add M
  hpt neg vpt -1.62 mul V
  hpt 2 mul 0 V
  hpt neg vpt 1.62 mul V Opaque stroke } def
/TriDW { stroke [] 0 setdash vpt 1.12 mul sub M
  hpt neg vpt 1.62 mul V
  hpt 2 mul 0 V
  hpt neg vpt -1.62 mul V Opaque stroke } def
/PentW { stroke [] 0 setdash gsave
  translate 0 hpt M 4 {72 rotate 0 hpt L} repeat
  Opaque stroke grestore } def
/CircW { stroke [] 0 setdash 
  hpt 0 360 arc Opaque stroke } def
/BoxFill { gsave Rec 1 setgray fill grestore } def
/BoxColFill {
  gsave Rec
  /Fillden exch def
  currentrgbcolor
  /ColB exch def /ColG exch def /ColR exch def
  /ColR ColR Fillden mul Fillden sub 1 add def
  /ColG ColG Fillden mul Fillden sub 1 add def
  /ColB ColB Fillden mul Fillden sub 1 add def
  ColR ColG ColB setrgbcolor
  fill grestore } def
%
%
/PatternFill { gsave /PFa [ 9 2 roll ] def
    PFa 0 get PFa 2 get 2 div add PFa 1 get PFa 3 get 2 div add translate
    PFa 2 get -2 div PFa 3 get -2 div PFa 2 get PFa 3 get Rec
    gsave 1 setgray fill grestore clip
    currentlinewidth 0.5 mul setlinewidth
    /PFs PFa 2 get dup mul PFa 3 get dup mul add sqrt def
    0 0 M PFa 5 get rotate PFs -2 div dup translate
	0 1 PFs PFa 4 get div 1 add floor cvi
	{ PFa 4 get mul 0 M 0 PFs V } for
    0 PFa 6 get ne {
	0 1 PFs PFa 4 get div 1 add floor cvi
	{ PFa 4 get mul 0 2 1 roll M PFs 0 V } for
    } if
    stroke grestore } def
/Symbol-Oblique /Symbol findfont [1 0 .167 1 0 0] makefont
dup length dict begin {1 index /FID eq {pop pop} {def} ifelse} forall
currentdict end definefont pop
end
gnudict begin
gsave
0 0 translate
0.100 0.100 scale
0 setgray
newpath
1.000 UL
LTb
450 300 M
63 0 V
2937 0 R
-63 0 V
1.000 UL
LTb
450 520 M
63 0 V
2937 0 R
-63 0 V
1.000 UL
LTb
450 740 M
63 0 V
2937 0 R
-63 0 V
1.000 UL
LTb
450 960 M
63 0 V
2937 0 R
-63 0 V
1.000 UL
LTb
450 1180 M
63 0 V
2937 0 R
-63 0 V
1.000 UL
LTb
450 1400 M
63 0 V
2937 0 R
-63 0 V
1.000 UL
LTb
450 1620 M
63 0 V
2937 0 R
-63 0 V
1.000 UL
LTb
450 1840 M
63 0 V
2937 0 R
-63 0 V
1.000 UL
LTb
450 2060 M
63 0 V
2937 0 R
-63 0 V
1.000 UL
LTb
450 300 M
0 63 V
0 1697 R
0 -63 V
1.000 UL
LTb
1200 300 M
0 63 V
0 1697 R
0 -63 V
1.000 UL
LTb
1950 300 M
0 63 V
0 1697 R
0 -63 V
1.000 UL
LTb
2700 300 M
0 63 V
0 1697 R
0 -63 V
1.000 UL
LTb
3450 300 M
0 63 V
0 1697 R
0 -63 V
1.000 UL
LTb
1.000 UL
LTb
450 300 M
3000 0 V
0 1760 V
-3000 0 V
450 300 L
LTb
LTb
1.000 UP
1.000 UP
0.500 UL
LT0
450 2060 M
600 1730 L
750 1065 L
900 601 L
150 -90 V
150 123 V
150 151 V
150 99 V
150 48 V
150 19 V
150 7 V
150 1 V
150 1 V
150 0 V
150 0 V
150 0 V
150 0 V
150 0 V
150 0 V
150 0 V
150 0 V
450 2060 Pls
600 1730 Pls
750 1065 Pls
900 601 Pls
1050 511 Pls
1200 634 Pls
1350 785 Pls
1500 884 Pls
1650 932 Pls
1800 951 Pls
1950 958 Pls
2100 959 Pls
2250 960 Pls
2400 960 Pls
2550 960 Pls
2700 960 Pls
2850 960 Pls
3000 960 Pls
3150 960 Pls
3300 960 Pls
3450 960 Pls
1.000 UP
0.500 UL
LT1
450 2060 M
600 1730 L
750 1062 L
900 596 L
150 -89 V
150 128 V
150 153 V
150 100 V
150 47 V
150 17 V
150 6 V
150 1 V
150 1 V
150 0 V
150 0 V
150 0 V
150 0 V
150 0 V
150 0 V
150 0 V
150 0 V
450 2060 Crs
600 1730 Crs
750 1062 Crs
900 596 Crs
1050 507 Crs
1200 635 Crs
1350 788 Crs
1500 888 Crs
1650 935 Crs
1800 952 Crs
1950 958 Crs
2100 959 Crs
2250 960 Crs
2400 960 Crs
2550 960 Crs
2700 960 Crs
2850 960 Crs
3000 960 Crs
3150 960 Crs
3300 960 Crs
3450 960 Crs
1.000 UP
0.500 UL
LT2
450 2060 M
600 1730 L
750 1058 L
900 586 L
150 -85 V
150 136 V
150 158 V
150 98 V
150 45 V
150 16 V
150 4 V
150 2 V
150 0 V
150 0 V
150 0 V
150 0 V
150 0 V
150 0 V
150 0 V
150 0 V
150 0 V
600 1730 Star
750 1058 Star
900 586 Star
1050 501 Star
1200 637 Star
1350 795 Star
1500 893 Star
1650 938 Star
1800 954 Star
1950 958 Star
2100 960 Star
2250 960 Star
2400 960 Star
2550 960 Star
2700 960 Star
2850 960 Star
3000 960 Star
3150 960 Star
3300 960 Star
3450 960 Star
1.000 UL
LTb
450 300 M
3000 0 V
0 1760 V
-3000 0 V
450 300 L
1.000 UP
stroke
grestore
end
showpage
}}%
\put(1950,50){\makebox(0,0){$k$}}%
\put(100,1180){%
\special{ps: gsave currentpoint currentpoint translate
270 rotate neg exch neg exch translate}%
\makebox(0,0)[b]{\shortstack{$F_k$}}%
\special{ps: currentpoint grestore moveto}%
}%
\put(3450,200){\makebox(0,0){ 20}}%
\put(2700,200){\makebox(0,0){ 15}}%
\put(1950,200){\makebox(0,0){ 10}}%
\put(1200,200){\makebox(0,0){ 5}}%
\put(450,200){\makebox(0,0){ 0}}%
\put(400,2060){\makebox(0,0)[r]{ 1}}%
\put(400,1840){\makebox(0,0)[r]{ 0.8}}%
\put(400,1620){\makebox(0,0)[r]{ 0.6}}%
\put(400,1400){\makebox(0,0)[r]{ 0.4}}%
\put(400,1180){\makebox(0,0)[r]{ 0.2}}%
\put(400,960){\makebox(0,0)[r]{ 0}}%
\put(400,740){\makebox(0,0)[r]{-0.2}}%
\put(400,520){\makebox(0,0)[r]{-0.4}}%
\put(400,300){\makebox(0,0)[r]{-0.6}}%
\end{picture}%
\endgroup
 

%% file: figures/su2plaqdiffs.tex
\begingroup%
  \makeatletter%
  \newcommand{\GNUPLOTspecial}{%
    \@sanitize\catcode`\%=14\relax\special}%
  \setlength{\unitlength}{0.1bp}%
\begin{picture}(3600,2160)(0,0)%
{\GNUPLOTspecial{"
/gnudict 256 dict def
gnudict begin
/Color false def
/Solid false def
/gnulinewidth 5.000 def
/userlinewidth gnulinewidth def
/vshift -33 def
/dl {10.0 mul} def
/hpt_ 31.5 def
/vpt_ 31.5 def
/hpt hpt_ def
/vpt vpt_ def
/Rounded false def
/M {moveto} bind def
/L {lineto} bind def
/R {rmoveto} bind def
/V {rlineto} bind def
/N {newpath moveto} bind def
/C {setrgbcolor} bind def
/f {rlineto fill} bind def
/vpt2 vpt 2 mul def
/hpt2 hpt 2 mul def
/Lshow { currentpoint stroke M
  0 vshift R show } def
/Rshow { currentpoint stroke M
  dup stringwidth pop neg vshift R show } def
/Cshow { currentpoint stroke M
  dup stringwidth pop -2 div vshift R show } def
/UP { dup vpt_ mul /vpt exch def hpt_ mul /hpt exch def
  /hpt2 hpt 2 mul def /vpt2 vpt 2 mul def } def
/DL { Color {setrgbcolor Solid {pop []} if 0 setdash }
 {pop pop pop 0 setgray Solid {pop []} if 0 setdash} ifelse } def
/BL { stroke userlinewidth 2 mul setlinewidth
      Rounded { 1 setlinejoin 1 setlinecap } if } def
/AL { stroke userlinewidth 2 div setlinewidth
      Rounded { 1 setlinejoin 1 setlinecap } if } def
/UL { dup gnulinewidth mul /userlinewidth exch def
      dup 1 lt {pop 1} if 10 mul /udl exch def } def
/PL { stroke userlinewidth setlinewidth
      Rounded { 1 setlinejoin 1 setlinecap } if } def
/LTw { PL [] 1 setgray } def
/LTb { BL [] 0 0 0 DL } def
/LTa { AL [1 udl mul 2 udl mul] 0 setdash 0 0 0 setrgbcolor } def
/LT0 { PL [] 1 0 0 DL } def
/LT1 { PL [4 dl 2 dl] 0 1 0 DL } def
/LT2 { PL [2 dl 3 dl] 0 0 1 DL } def
/LT3 { PL [1 dl 1.5 dl] 1 0 1 DL } def
/LT4 { PL [5 dl 2 dl 1 dl 2 dl] 0 1 1 DL } def
/LT5 { PL [4 dl 3 dl 1 dl 3 dl] 1 1 0 DL } def
/LT6 { PL [2 dl 2 dl 2 dl 4 dl] 0 0 0 DL } def
/LT7 { PL [2 dl 2 dl 2 dl 2 dl 2 dl 4 dl] 1 0.3 0 DL } def
/LT8 { PL [2 dl 2 dl 2 dl 2 dl 2 dl 2 dl 2 dl 4 dl] 0.5 0.5 0.5 DL } def
/Pnt { stroke [] 0 setdash
   gsave 1 setlinecap M 0 0 V stroke grestore } def
/Dia { stroke [] 0 setdash 2 copy vpt add M
  hpt neg vpt neg V hpt vpt neg V
  hpt vpt V hpt neg vpt V closepath stroke
  Pnt } def
/Pls { stroke [] 0 setdash vpt sub M 0 vpt2 V
  currentpoint stroke M
  hpt neg vpt neg R hpt2 0 V stroke
  } def
/Box { stroke [] 0 setdash 2 copy exch hpt sub exch vpt add M
  0 vpt2 neg V hpt2 0 V 0 vpt2 V
  hpt2 neg 0 V closepath stroke
  Pnt } def
/Crs { stroke [] 0 setdash exch hpt sub exch vpt add M
  hpt2 vpt2 neg V currentpoint stroke M
  hpt2 neg 0 R hpt2 vpt2 V stroke } def
/TriU { stroke [] 0 setdash 2 copy vpt 1.12 mul add M
  hpt neg vpt -1.62 mul V
  hpt 2 mul 0 V
  hpt neg vpt 1.62 mul V closepath stroke
  Pnt  } def
/Star { 2 copy Pls Crs } def
/BoxF { stroke [] 0 setdash exch hpt sub exch vpt add M
  0 vpt2 neg V  hpt2 0 V  0 vpt2 V
  hpt2 neg 0 V  closepath fill } def
/TriUF { stroke [] 0 setdash vpt 1.12 mul add M
  hpt neg vpt -1.62 mul V
  hpt 2 mul 0 V
  hpt neg vpt 1.62 mul V closepath fill } def
/TriD { stroke [] 0 setdash 2 copy vpt 1.12 mul sub M
  hpt neg vpt 1.62 mul V
  hpt 2 mul 0 V
  hpt neg vpt -1.62 mul V closepath stroke
  Pnt  } def
/TriDF { stroke [] 0 setdash vpt 1.12 mul sub M
  hpt neg vpt 1.62 mul V
  hpt 2 mul 0 V
  hpt neg vpt -1.62 mul V closepath fill} def
/DiaF { stroke [] 0 setdash vpt add M
  hpt neg vpt neg V hpt vpt neg V
  hpt vpt V hpt neg vpt V closepath fill } def
/Pent { stroke [] 0 setdash 2 copy gsave
  translate 0 hpt M 4 {72 rotate 0 hpt L} repeat
  closepath stroke grestore Pnt } def
/PentF { stroke [] 0 setdash gsave
  translate 0 hpt M 4 {72 rotate 0 hpt L} repeat
  closepath fill grestore } def
/Circle { stroke [] 0 setdash 2 copy
  hpt 0 360 arc stroke Pnt } def
/CircleF { stroke [] 0 setdash hpt 0 360 arc fill } def
/C0 { BL [] 0 setdash 2 copy moveto vpt 90 450  arc } bind def
/C1 { BL [] 0 setdash 2 copy        moveto
       2 copy  vpt 0 90 arc closepath fill
               vpt 0 360 arc closepath } bind def
/C2 { BL [] 0 setdash 2 copy moveto
       2 copy  vpt 90 180 arc closepath fill
               vpt 0 360 arc closepath } bind def
/C3 { BL [] 0 setdash 2 copy moveto
       2 copy  vpt 0 180 arc closepath fill
               vpt 0 360 arc closepath } bind def
/C4 { BL [] 0 setdash 2 copy moveto
       2 copy  vpt 180 270 arc closepath fill
               vpt 0 360 arc closepath } bind def
/C5 { BL [] 0 setdash 2 copy moveto
       2 copy  vpt 0 90 arc
       2 copy moveto
       2 copy  vpt 180 270 arc closepath fill
               vpt 0 360 arc } bind def
/C6 { BL [] 0 setdash 2 copy moveto
      2 copy  vpt 90 270 arc closepath fill
              vpt 0 360 arc closepath } bind def
/C7 { BL [] 0 setdash 2 copy moveto
      2 copy  vpt 0 270 arc closepath fill
              vpt 0 360 arc closepath } bind def
/C8 { BL [] 0 setdash 2 copy moveto
      2 copy vpt 270 360 arc closepath fill
              vpt 0 360 arc closepath } bind def
/C9 { BL [] 0 setdash 2 copy moveto
      2 copy  vpt 270 450 arc closepath fill
              vpt 0 360 arc closepath } bind def
/C10 { BL [] 0 setdash 2 copy 2 copy moveto vpt 270 360 arc closepath fill
       2 copy moveto
       2 copy vpt 90 180 arc closepath fill
               vpt 0 360 arc closepath } bind def
/C11 { BL [] 0 setdash 2 copy moveto
       2 copy  vpt 0 180 arc closepath fill
       2 copy moveto
       2 copy  vpt 270 360 arc closepath fill
               vpt 0 360 arc closepath } bind def
/C12 { BL [] 0 setdash 2 copy moveto
       2 copy  vpt 180 360 arc closepath fill
               vpt 0 360 arc closepath } bind def
/C13 { BL [] 0 setdash  2 copy moveto
       2 copy  vpt 0 90 arc closepath fill
       2 copy moveto
       2 copy  vpt 180 360 arc closepath fill
               vpt 0 360 arc closepath } bind def
/C14 { BL [] 0 setdash 2 copy moveto
       2 copy  vpt 90 360 arc closepath fill
               vpt 0 360 arc } bind def
/C15 { BL [] 0 setdash 2 copy vpt 0 360 arc closepath fill
               vpt 0 360 arc closepath } bind def
/Rec   { newpath 4 2 roll moveto 1 index 0 rlineto 0 exch rlineto
       neg 0 rlineto closepath } bind def
/Square { dup Rec } bind def
/Bsquare { vpt sub exch vpt sub exch vpt2 Square } bind def
/S0 { BL [] 0 setdash 2 copy moveto 0 vpt rlineto BL Bsquare } bind def
/S1 { BL [] 0 setdash 2 copy vpt Square fill Bsquare } bind def
/S2 { BL [] 0 setdash 2 copy exch vpt sub exch vpt Square fill Bsquare } bind def
/S3 { BL [] 0 setdash 2 copy exch vpt sub exch vpt2 vpt Rec fill Bsquare } bind def
/S4 { BL [] 0 setdash 2 copy exch vpt sub exch vpt sub vpt Square fill Bsquare } bind def
/S5 { BL [] 0 setdash 2 copy 2 copy vpt Square fill
       exch vpt sub exch vpt sub vpt Square fill Bsquare } bind def
/S6 { BL [] 0 setdash 2 copy exch vpt sub exch vpt sub vpt vpt2 Rec fill Bsquare } bind def
/S7 { BL [] 0 setdash 2 copy exch vpt sub exch vpt sub vpt vpt2 Rec fill
       2 copy vpt Square fill
       Bsquare } bind def
/S8 { BL [] 0 setdash 2 copy vpt sub vpt Square fill Bsquare } bind def
/S9 { BL [] 0 setdash 2 copy vpt sub vpt vpt2 Rec fill Bsquare } bind def
/S10 { BL [] 0 setdash 2 copy vpt sub vpt Square fill 2 copy exch vpt sub exch vpt Square fill
       Bsquare } bind def
/S11 { BL [] 0 setdash 2 copy vpt sub vpt Square fill 2 copy exch vpt sub exch vpt2 vpt Rec fill
       Bsquare } bind def
/S12 { BL [] 0 setdash 2 copy exch vpt sub exch vpt sub vpt2 vpt Rec fill Bsquare } bind def
/S13 { BL [] 0 setdash 2 copy exch vpt sub exch vpt sub vpt2 vpt Rec fill
       2 copy vpt Square fill Bsquare } bind def
/S14 { BL [] 0 setdash 2 copy exch vpt sub exch vpt sub vpt2 vpt Rec fill
       2 copy exch vpt sub exch vpt Square fill Bsquare } bind def
/S15 { BL [] 0 setdash 2 copy Bsquare fill Bsquare } bind def
/D0 { gsave translate 45 rotate 0 0 S0 stroke grestore } bind def
/D1 { gsave translate 45 rotate 0 0 S1 stroke grestore } bind def
/D2 { gsave translate 45 rotate 0 0 S2 stroke grestore } bind def
/D3 { gsave translate 45 rotate 0 0 S3 stroke grestore } bind def
/D4 { gsave translate 45 rotate 0 0 S4 stroke grestore } bind def
/D5 { gsave translate 45 rotate 0 0 S5 stroke grestore } bind def
/D6 { gsave translate 45 rotate 0 0 S6 stroke grestore } bind def
/D7 { gsave translate 45 rotate 0 0 S7 stroke grestore } bind def
/D8 { gsave translate 45 rotate 0 0 S8 stroke grestore } bind def
/D9 { gsave translate 45 rotate 0 0 S9 stroke grestore } bind def
/D10 { gsave translate 45 rotate 0 0 S10 stroke grestore } bind def
/D11 { gsave translate 45 rotate 0 0 S11 stroke grestore } bind def
/D12 { gsave translate 45 rotate 0 0 S12 stroke grestore } bind def
/D13 { gsave translate 45 rotate 0 0 S13 stroke grestore } bind def
/D14 { gsave translate 45 rotate 0 0 S14 stroke grestore } bind def
/D15 { gsave translate 45 rotate 0 0 S15 stroke grestore } bind def
/DiaE { stroke [] 0 setdash vpt add M
  hpt neg vpt neg V hpt vpt neg V
  hpt vpt V hpt neg vpt V closepath stroke } def
/BoxE { stroke [] 0 setdash exch hpt sub exch vpt add M
  0 vpt2 neg V hpt2 0 V 0 vpt2 V
  hpt2 neg 0 V closepath stroke } def
/TriUE { stroke [] 0 setdash vpt 1.12 mul add M
  hpt neg vpt -1.62 mul V
  hpt 2 mul 0 V
  hpt neg vpt 1.62 mul V closepath stroke } def
/TriDE { stroke [] 0 setdash vpt 1.12 mul sub M
  hpt neg vpt 1.62 mul V
  hpt 2 mul 0 V
  hpt neg vpt -1.62 mul V closepath stroke } def
/PentE { stroke [] 0 setdash gsave
  translate 0 hpt M 4 {72 rotate 0 hpt L} repeat
  closepath stroke grestore } def
/CircE { stroke [] 0 setdash 
  hpt 0 360 arc stroke } def
/Opaque { gsave closepath 1 setgray fill grestore 0 setgray closepath } def
/DiaW { stroke [] 0 setdash vpt add M
  hpt neg vpt neg V hpt vpt neg V
  hpt vpt V hpt neg vpt V Opaque stroke } def
/BoxW { stroke [] 0 setdash exch hpt sub exch vpt add M
  0 vpt2 neg V hpt2 0 V 0 vpt2 V
  hpt2 neg 0 V Opaque stroke } def
/TriUW { stroke [] 0 setdash vpt 1.12 mul add M
  hpt neg vpt -1.62 mul V
  hpt 2 mul 0 V
  hpt neg vpt 1.62 mul V Opaque stroke } def
/TriDW { stroke [] 0 setdash vpt 1.12 mul sub M
  hpt neg vpt 1.62 mul V
  hpt 2 mul 0 V
  hpt neg vpt -1.62 mul V Opaque stroke } def
/PentW { stroke [] 0 setdash gsave
  translate 0 hpt M 4 {72 rotate 0 hpt L} repeat
  Opaque stroke grestore } def
/CircW { stroke [] 0 setdash 
  hpt 0 360 arc Opaque stroke } def
/BoxFill { gsave Rec 1 setgray fill grestore } def
/BoxColFill {
  gsave Rec
  /Fillden exch def
  currentrgbcolor
  /ColB exch def /ColG exch def /ColR exch def
  /ColR ColR Fillden mul Fillden sub 1 add def
  /ColG ColG Fillden mul Fillden sub 1 add def
  /ColB ColB Fillden mul Fillden sub 1 add def
  ColR ColG ColB setrgbcolor
  fill grestore } def
%
%
/PatternFill { gsave /PFa [ 9 2 roll ] def
    PFa 0 get PFa 2 get 2 div add PFa 1 get PFa 3 get 2 div add translate
    PFa 2 get -2 div PFa 3 get -2 div PFa 2 get PFa 3 get Rec
    gsave 1 setgray fill grestore clip
    currentlinewidth 0.5 mul setlinewidth
    /PFs PFa 2 get dup mul PFa 3 get dup mul add sqrt def
    0 0 M PFa 5 get rotate PFs -2 div dup translate
	0 1 PFs PFa 4 get div 1 add floor cvi
	{ PFa 4 get mul 0 M 0 PFs V } for
    0 PFa 6 get ne {
	0 1 PFs PFa 4 get div 1 add floor cvi
	{ PFa 4 get mul 0 2 1 roll M PFs 0 V } for
    } if
    stroke grestore } def
/Symbol-Oblique /Symbol findfont [1 0 .167 1 0 0] makefont
dup length dict begin {1 index /FID eq {pop pop} {def} ifelse} forall
currentdict end definefont pop
end
gnudict begin
gsave
0 0 translate
0.100 0.100 scale
0 setgray
newpath
1.000 UL
LTb
550 300 M
63 0 V
2837 0 R
-63 0 V
1.000 UL
LTb
550 460 M
63 0 V
2837 0 R
-63 0 V
1.000 UL
LTb
550 620 M
63 0 V
2837 0 R
-63 0 V
1.000 UL
LTb
550 780 M
63 0 V
2837 0 R
-63 0 V
1.000 UL
LTb
550 940 M
63 0 V
2837 0 R
-63 0 V
1.000 UL
LTb
550 1100 M
63 0 V
2837 0 R
-63 0 V
1.000 UL
LTb
550 1260 M
63 0 V
2837 0 R
-63 0 V
1.000 UL
LTb
550 1420 M
63 0 V
2837 0 R
-63 0 V
1.000 UL
LTb
550 1580 M
63 0 V
2837 0 R
-63 0 V
1.000 UL
LTb
550 1740 M
63 0 V
2837 0 R
-63 0 V
1.000 UL
LTb
550 1900 M
63 0 V
2837 0 R
-63 0 V
1.000 UL
LTb
550 2060 M
63 0 V
2837 0 R
-63 0 V
1.000 UL
LTb
550 300 M
0 63 V
0 1697 R
0 -63 V
1.000 UL
LTb
1130 300 M
0 63 V
0 1697 R
0 -63 V
1.000 UL
LTb
1710 300 M
0 63 V
0 1697 R
0 -63 V
1.000 UL
LTb
2290 300 M
0 63 V
0 1697 R
0 -63 V
1.000 UL
LTb
2870 300 M
0 63 V
0 1697 R
0 -63 V
1.000 UL
LTb
3450 300 M
0 63 V
0 1697 R
0 -63 V
1.000 UL
LTb
1.000 UL
LTb
550 300 M
2900 0 V
0 1760 V
-2900 0 V
550 300 L
LTb
LTb
1.000 UP
1.000 UP
1.000 UL
LT0
753 460 M
0 7 V
-31 -7 R
62 0 V
-62 7 R
62 0 V
636 -5 R
0 16 V
-31 -16 R
62 0 V
-62 16 R
62 0 V
259 -13 R
0 13 V
-31 -13 R
62 0 V
-62 13 R
62 0 V
259 80 R
0 57 V
-31 -57 R
62 0 V
-62 57 R
62 0 V
259 211 R
0 74 V
-31 -74 R
62 0 V
-62 74 R
62 0 V
259 64 R
0 48 V
-31 -48 R
62 0 V
-62 48 R
62 0 V
259 45 R
0 121 V
-31 -121 R
62 0 V
-62 121 R
62 0 V
3160 866 M
0 122 V
3129 866 M
62 0 V
-62 122 R
62 0 V
3363 615 M
0 48 V
-31 -48 R
62 0 V
-62 48 R
62 0 V
27 -155 R
0 27 V
-31 -27 R
62 0 V
-62 27 R
62 0 V
753 463 Pls
1420 470 Pls
1710 471 Pls
2000 586 Pls
2290 863 Pls
2580 988 Pls
2870 1118 Pls
3160 927 Pls
3363 639 Pls
3421 521 Pls
1.000 UP
1.000 UL
LT0
753 460 M
0 7 V
-31 -7 R
62 0 V
-62 7 R
62 0 V
636 -5 R
0 18 V
-31 -18 R
62 0 V
-62 18 R
62 0 V
259 6 R
0 14 V
-31 -14 R
62 0 V
-62 14 R
62 0 V
259 264 R
0 93 V
-31 -93 R
62 0 V
-62 93 R
62 0 V
259 976 R
0 70 V
-31 -70 R
62 0 V
-62 70 R
62 0 V
259 -61 R
0 55 V
-31 -55 R
62 0 V
-62 55 R
62 0 V
259 -301 R
0 128 V
-31 -128 R
62 0 V
-62 128 R
62 0 V
259 -616 R
0 118 V
-31 -118 R
62 0 V
-62 118 R
62 0 V
3363 700 M
0 48 V
-31 -48 R
62 0 V
-62 48 R
62 0 V
27 -214 R
0 29 V
-31 -29 R
62 0 V
-62 29 R
62 0 V
753 463 Crs
1420 471 Crs
1710 493 Crs
2000 810 Crs
2290 1868 Crs
2580 1870 Crs
2870 1660 Crs
3160 1167 Crs
3363 724 Crs
3421 549 Crs
1.000 UL
LT2
550 460 M
29 0 V
30 0 V
29 0 V
29 0 V
29 0 V
30 0 V
29 0 V
29 0 V
30 0 V
29 0 V
29 0 V
30 0 V
29 0 V
29 0 V
29 0 V
30 0 V
29 0 V
29 0 V
30 0 V
29 0 V
29 0 V
29 0 V
30 0 V
29 0 V
29 0 V
30 0 V
29 0 V
29 0 V
29 0 V
30 0 V
29 0 V
29 0 V
30 0 V
29 0 V
29 0 V
30 0 V
29 0 V
29 0 V
29 0 V
30 0 V
29 0 V
29 0 V
30 0 V
29 0 V
29 0 V
29 0 V
30 0 V
29 0 V
29 0 V
30 0 V
29 0 V
29 0 V
30 0 V
29 0 V
29 0 V
29 0 V
30 0 V
29 0 V
29 0 V
30 0 V
29 0 V
29 0 V
29 0 V
30 0 V
29 0 V
29 0 V
30 0 V
29 0 V
29 0 V
30 0 V
29 0 V
29 0 V
29 0 V
30 0 V
29 0 V
29 0 V
30 0 V
29 0 V
29 0 V
29 0 V
30 0 V
29 0 V
29 0 V
30 0 V
29 0 V
29 0 V
29 0 V
30 0 V
29 0 V
29 0 V
30 0 V
29 0 V
29 0 V
30 0 V
29 0 V
29 0 V
29 0 V
30 0 V
29 0 V
1.000 UL
LTb
550 300 M
2900 0 V
0 1760 V
-2900 0 V
550 300 L
1.000 UP
stroke
grestore
end
showpage
}}%
\put(2000,50){\makebox(0,0){$\langle u_p \rangle$}}%
\put(100,1180){%
\special{ps: gsave currentpoint currentpoint translate
270 rotate neg exch neg exch translate}%
\makebox(0,0)[b]{\shortstack{$F_{k^\prime}^{D=1+1}-F_{k^\prime}^{D^\prime}$}}%
\special{ps: currentpoint grestore moveto}%
}%
\put(3450,200){\makebox(0,0){ 1}}%
\put(2870,200){\makebox(0,0){ 0.8}}%
\put(2290,200){\makebox(0,0){ 0.6}}%
\put(1710,200){\makebox(0,0){ 0.4}}%
\put(1130,200){\makebox(0,0){ 0.2}}%
\put(550,200){\makebox(0,0){ 0}}%
\put(500,2060){\makebox(0,0)[r]{ 0.01}}%
\put(500,1900){\makebox(0,0)[r]{ 0.009}}%
\put(500,1740){\makebox(0,0)[r]{ 0.008}}%
\put(500,1580){\makebox(0,0)[r]{ 0.007}}%
\put(500,1420){\makebox(0,0)[r]{ 0.006}}%
\put(500,1260){\makebox(0,0)[r]{ 0.005}}%
\put(500,1100){\makebox(0,0)[r]{ 0.004}}%
\put(500,940){\makebox(0,0)[r]{ 0.003}}%
\put(500,780){\makebox(0,0)[r]{ 0.002}}%
\put(500,620){\makebox(0,0)[r]{ 0.001}}%
\put(500,460){\makebox(0,0)[r]{ 0}}%
\put(500,300){\makebox(0,0)[r]{-0.001}}%
\end{picture}%
\endgroup
 

%% file: figures/su2_2x2diffs.tex
\begingroup%
  \makeatletter%
  \newcommand{\GNUPLOTspecial}{%
    \@sanitize\catcode`\%=14\relax\special}%
  \setlength{\unitlength}{0.1bp}%
\begin{picture}(3600,2160)(0,0)%
{\GNUPLOTspecial{"
/gnudict 256 dict def
gnudict begin
/Color false def
/Solid false def
/gnulinewidth 5.000 def
/userlinewidth gnulinewidth def
/vshift -33 def
/dl {10.0 mul} def
/hpt_ 31.5 def
/vpt_ 31.5 def
/hpt hpt_ def
/vpt vpt_ def
/Rounded false def
/M {moveto} bind def
/L {lineto} bind def
/R {rmoveto} bind def
/V {rlineto} bind def
/N {newpath moveto} bind def
/C {setrgbcolor} bind def
/f {rlineto fill} bind def
/vpt2 vpt 2 mul def
/hpt2 hpt 2 mul def
/Lshow { currentpoint stroke M
  0 vshift R show } def
/Rshow { currentpoint stroke M
  dup stringwidth pop neg vshift R show } def
/Cshow { currentpoint stroke M
  dup stringwidth pop -2 div vshift R show } def
/UP { dup vpt_ mul /vpt exch def hpt_ mul /hpt exch def
  /hpt2 hpt 2 mul def /vpt2 vpt 2 mul def } def
/DL { Color {setrgbcolor Solid {pop []} if 0 setdash }
 {pop pop pop 0 setgray Solid {pop []} if 0 setdash} ifelse } def
/BL { stroke userlinewidth 2 mul setlinewidth
      Rounded { 1 setlinejoin 1 setlinecap } if } def
/AL { stroke userlinewidth 2 div setlinewidth
      Rounded { 1 setlinejoin 1 setlinecap } if } def
/UL { dup gnulinewidth mul /userlinewidth exch def
      dup 1 lt {pop 1} if 10 mul /udl exch def } def
/PL { stroke userlinewidth setlinewidth
      Rounded { 1 setlinejoin 1 setlinecap } if } def
/LTw { PL [] 1 setgray } def
/LTb { BL [] 0 0 0 DL } def
/LTa { AL [1 udl mul 2 udl mul] 0 setdash 0 0 0 setrgbcolor } def
/LT0 { PL [] 1 0 0 DL } def
/LT1 { PL [4 dl 2 dl] 0 1 0 DL } def
/LT2 { PL [2 dl 3 dl] 0 0 1 DL } def
/LT3 { PL [1 dl 1.5 dl] 1 0 1 DL } def
/LT4 { PL [5 dl 2 dl 1 dl 2 dl] 0 1 1 DL } def
/LT5 { PL [4 dl 3 dl 1 dl 3 dl] 1 1 0 DL } def
/LT6 { PL [2 dl 2 dl 2 dl 4 dl] 0 0 0 DL } def
/LT7 { PL [2 dl 2 dl 2 dl 2 dl 2 dl 4 dl] 1 0.3 0 DL } def
/LT8 { PL [2 dl 2 dl 2 dl 2 dl 2 dl 2 dl 2 dl 4 dl] 0.5 0.5 0.5 DL } def
/Pnt { stroke [] 0 setdash
   gsave 1 setlinecap M 0 0 V stroke grestore } def
/Dia { stroke [] 0 setdash 2 copy vpt add M
  hpt neg vpt neg V hpt vpt neg V
  hpt vpt V hpt neg vpt V closepath stroke
  Pnt } def
/Pls { stroke [] 0 setdash vpt sub M 0 vpt2 V
  currentpoint stroke M
  hpt neg vpt neg R hpt2 0 V stroke
  } def
/Box { stroke [] 0 setdash 2 copy exch hpt sub exch vpt add M
  0 vpt2 neg V hpt2 0 V 0 vpt2 V
  hpt2 neg 0 V closepath stroke
  Pnt } def
/Crs { stroke [] 0 setdash exch hpt sub exch vpt add M
  hpt2 vpt2 neg V currentpoint stroke M
  hpt2 neg 0 R hpt2 vpt2 V stroke } def
/TriU { stroke [] 0 setdash 2 copy vpt 1.12 mul add M
  hpt neg vpt -1.62 mul V
  hpt 2 mul 0 V
  hpt neg vpt 1.62 mul V closepath stroke
  Pnt  } def
/Star { 2 copy Pls Crs } def
/BoxF { stroke [] 0 setdash exch hpt sub exch vpt add M
  0 vpt2 neg V  hpt2 0 V  0 vpt2 V
  hpt2 neg 0 V  closepath fill } def
/TriUF { stroke [] 0 setdash vpt 1.12 mul add M
  hpt neg vpt -1.62 mul V
  hpt 2 mul 0 V
  hpt neg vpt 1.62 mul V closepath fill } def
/TriD { stroke [] 0 setdash 2 copy vpt 1.12 mul sub M
  hpt neg vpt 1.62 mul V
  hpt 2 mul 0 V
  hpt neg vpt -1.62 mul V closepath stroke
  Pnt  } def
/TriDF { stroke [] 0 setdash vpt 1.12 mul sub M
  hpt neg vpt 1.62 mul V
  hpt 2 mul 0 V
  hpt neg vpt -1.62 mul V closepath fill} def
/DiaF { stroke [] 0 setdash vpt add M
  hpt neg vpt neg V hpt vpt neg V
  hpt vpt V hpt neg vpt V closepath fill } def
/Pent { stroke [] 0 setdash 2 copy gsave
  translate 0 hpt M 4 {72 rotate 0 hpt L} repeat
  closepath stroke grestore Pnt } def
/PentF { stroke [] 0 setdash gsave
  translate 0 hpt M 4 {72 rotate 0 hpt L} repeat
  closepath fill grestore } def
/Circle { stroke [] 0 setdash 2 copy
  hpt 0 360 arc stroke Pnt } def
/CircleF { stroke [] 0 setdash hpt 0 360 arc fill } def
/C0 { BL [] 0 setdash 2 copy moveto vpt 90 450  arc } bind def
/C1 { BL [] 0 setdash 2 copy        moveto
       2 copy  vpt 0 90 arc closepath fill
               vpt 0 360 arc closepath } bind def
/C2 { BL [] 0 setdash 2 copy moveto
       2 copy  vpt 90 180 arc closepath fill
               vpt 0 360 arc closepath } bind def
/C3 { BL [] 0 setdash 2 copy moveto
       2 copy  vpt 0 180 arc closepath fill
               vpt 0 360 arc closepath } bind def
/C4 { BL [] 0 setdash 2 copy moveto
       2 copy  vpt 180 270 arc closepath fill
               vpt 0 360 arc closepath } bind def
/C5 { BL [] 0 setdash 2 copy moveto
       2 copy  vpt 0 90 arc
       2 copy moveto
       2 copy  vpt 180 270 arc closepath fill
               vpt 0 360 arc } bind def
/C6 { BL [] 0 setdash 2 copy moveto
      2 copy  vpt 90 270 arc closepath fill
              vpt 0 360 arc closepath } bind def
/C7 { BL [] 0 setdash 2 copy moveto
      2 copy  vpt 0 270 arc closepath fill
              vpt 0 360 arc closepath } bind def
/C8 { BL [] 0 setdash 2 copy moveto
      2 copy vpt 270 360 arc closepath fill
              vpt 0 360 arc closepath } bind def
/C9 { BL [] 0 setdash 2 copy moveto
      2 copy  vpt 270 450 arc closepath fill
              vpt 0 360 arc closepath } bind def
/C10 { BL [] 0 setdash 2 copy 2 copy moveto vpt 270 360 arc closepath fill
       2 copy moveto
       2 copy vpt 90 180 arc closepath fill
               vpt 0 360 arc closepath } bind def
/C11 { BL [] 0 setdash 2 copy moveto
       2 copy  vpt 0 180 arc closepath fill
       2 copy moveto
       2 copy  vpt 270 360 arc closepath fill
               vpt 0 360 arc closepath } bind def
/C12 { BL [] 0 setdash 2 copy moveto
       2 copy  vpt 180 360 arc closepath fill
               vpt 0 360 arc closepath } bind def
/C13 { BL [] 0 setdash  2 copy moveto
       2 copy  vpt 0 90 arc closepath fill
       2 copy moveto
       2 copy  vpt 180 360 arc closepath fill
               vpt 0 360 arc closepath } bind def
/C14 { BL [] 0 setdash 2 copy moveto
       2 copy  vpt 90 360 arc closepath fill
               vpt 0 360 arc } bind def
/C15 { BL [] 0 setdash 2 copy vpt 0 360 arc closepath fill
               vpt 0 360 arc closepath } bind def
/Rec   { newpath 4 2 roll moveto 1 index 0 rlineto 0 exch rlineto
       neg 0 rlineto closepath } bind def
/Square { dup Rec } bind def
/Bsquare { vpt sub exch vpt sub exch vpt2 Square } bind def
/S0 { BL [] 0 setdash 2 copy moveto 0 vpt rlineto BL Bsquare } bind def
/S1 { BL [] 0 setdash 2 copy vpt Square fill Bsquare } bind def
/S2 { BL [] 0 setdash 2 copy exch vpt sub exch vpt Square fill Bsquare } bind def
/S3 { BL [] 0 setdash 2 copy exch vpt sub exch vpt2 vpt Rec fill Bsquare } bind def
/S4 { BL [] 0 setdash 2 copy exch vpt sub exch vpt sub vpt Square fill Bsquare } bind def
/S5 { BL [] 0 setdash 2 copy 2 copy vpt Square fill
       exch vpt sub exch vpt sub vpt Square fill Bsquare } bind def
/S6 { BL [] 0 setdash 2 copy exch vpt sub exch vpt sub vpt vpt2 Rec fill Bsquare } bind def
/S7 { BL [] 0 setdash 2 copy exch vpt sub exch vpt sub vpt vpt2 Rec fill
       2 copy vpt Square fill
       Bsquare } bind def
/S8 { BL [] 0 setdash 2 copy vpt sub vpt Square fill Bsquare } bind def
/S9 { BL [] 0 setdash 2 copy vpt sub vpt vpt2 Rec fill Bsquare } bind def
/S10 { BL [] 0 setdash 2 copy vpt sub vpt Square fill 2 copy exch vpt sub exch vpt Square fill
       Bsquare } bind def
/S11 { BL [] 0 setdash 2 copy vpt sub vpt Square fill 2 copy exch vpt sub exch vpt2 vpt Rec fill
       Bsquare } bind def
/S12 { BL [] 0 setdash 2 copy exch vpt sub exch vpt sub vpt2 vpt Rec fill Bsquare } bind def
/S13 { BL [] 0 setdash 2 copy exch vpt sub exch vpt sub vpt2 vpt Rec fill
       2 copy vpt Square fill Bsquare } bind def
/S14 { BL [] 0 setdash 2 copy exch vpt sub exch vpt sub vpt2 vpt Rec fill
       2 copy exch vpt sub exch vpt Square fill Bsquare } bind def
/S15 { BL [] 0 setdash 2 copy Bsquare fill Bsquare } bind def
/D0 { gsave translate 45 rotate 0 0 S0 stroke grestore } bind def
/D1 { gsave translate 45 rotate 0 0 S1 stroke grestore } bind def
/D2 { gsave translate 45 rotate 0 0 S2 stroke grestore } bind def
/D3 { gsave translate 45 rotate 0 0 S3 stroke grestore } bind def
/D4 { gsave translate 45 rotate 0 0 S4 stroke grestore } bind def
/D5 { gsave translate 45 rotate 0 0 S5 stroke grestore } bind def
/D6 { gsave translate 45 rotate 0 0 S6 stroke grestore } bind def
/D7 { gsave translate 45 rotate 0 0 S7 stroke grestore } bind def
/D8 { gsave translate 45 rotate 0 0 S8 stroke grestore } bind def
/D9 { gsave translate 45 rotate 0 0 S9 stroke grestore } bind def
/D10 { gsave translate 45 rotate 0 0 S10 stroke grestore } bind def
/D11 { gsave translate 45 rotate 0 0 S11 stroke grestore } bind def
/D12 { gsave translate 45 rotate 0 0 S12 stroke grestore } bind def
/D13 { gsave translate 45 rotate 0 0 S13 stroke grestore } bind def
/D14 { gsave translate 45 rotate 0 0 S14 stroke grestore } bind def
/D15 { gsave translate 45 rotate 0 0 S15 stroke grestore } bind def
/DiaE { stroke [] 0 setdash vpt add M
  hpt neg vpt neg V hpt vpt neg V
  hpt vpt V hpt neg vpt V closepath stroke } def
/BoxE { stroke [] 0 setdash exch hpt sub exch vpt add M
  0 vpt2 neg V hpt2 0 V 0 vpt2 V
  hpt2 neg 0 V closepath stroke } def
/TriUE { stroke [] 0 setdash vpt 1.12 mul add M
  hpt neg vpt -1.62 mul V
  hpt 2 mul 0 V
  hpt neg vpt 1.62 mul V closepath stroke } def
/TriDE { stroke [] 0 setdash vpt 1.12 mul sub M
  hpt neg vpt 1.62 mul V
  hpt 2 mul 0 V
  hpt neg vpt -1.62 mul V closepath stroke } def
/PentE { stroke [] 0 setdash gsave
  translate 0 hpt M 4 {72 rotate 0 hpt L} repeat
  closepath stroke grestore } def
/CircE { stroke [] 0 setdash 
  hpt 0 360 arc stroke } def
/Opaque { gsave closepath 1 setgray fill grestore 0 setgray closepath } def
/DiaW { stroke [] 0 setdash vpt add M
  hpt neg vpt neg V hpt vpt neg V
  hpt vpt V hpt neg vpt V Opaque stroke } def
/BoxW { stroke [] 0 setdash exch hpt sub exch vpt add M
  0 vpt2 neg V hpt2 0 V 0 vpt2 V
  hpt2 neg 0 V Opaque stroke } def
/TriUW { stroke [] 0 setdash vpt 1.12 mul add M
  hpt neg vpt -1.62 mul V
  hpt 2 mul 0 V
  hpt neg vpt 1.62 mul V Opaque stroke } def
/TriDW { stroke [] 0 setdash vpt 1.12 mul sub M
  hpt neg vpt 1.62 mul V
  hpt 2 mul 0 V
  hpt neg vpt -1.62 mul V Opaque stroke } def
/PentW { stroke [] 0 setdash gsave
  translate 0 hpt M 4 {72 rotate 0 hpt L} repeat
  Opaque stroke grestore } def
/CircW { stroke [] 0 setdash 
  hpt 0 360 arc Opaque stroke } def
/BoxFill { gsave Rec 1 setgray fill grestore } def
/BoxColFill {
  gsave Rec
  /Fillden exch def
  currentrgbcolor
  /ColB exch def /ColG exch def /ColR exch def
  /ColR ColR Fillden mul Fillden sub 1 add def
  /ColG ColG Fillden mul Fillden sub 1 add def
  /ColB ColB Fillden mul Fillden sub 1 add def
  ColR ColG ColB setrgbcolor
  fill grestore } def
%
%
/PatternFill { gsave /PFa [ 9 2 roll ] def
    PFa 0 get PFa 2 get 2 div add PFa 1 get PFa 3 get 2 div add translate
    PFa 2 get -2 div PFa 3 get -2 div PFa 2 get PFa 3 get Rec
    gsave 1 setgray fill grestore clip
    currentlinewidth 0.5 mul setlinewidth
    /PFs PFa 2 get dup mul PFa 3 get dup mul add sqrt def
    0 0 M PFa 5 get rotate PFs -2 div dup translate
	0 1 PFs PFa 4 get div 1 add floor cvi
	{ PFa 4 get mul 0 M 0 PFs V } for
    0 PFa 6 get ne {
	0 1 PFs PFa 4 get div 1 add floor cvi
	{ PFa 4 get mul 0 2 1 roll M PFs 0 V } for
    } if
    stroke grestore } def
/Symbol-Oblique /Symbol findfont [1 0 .167 1 0 0] makefont
dup length dict begin {1 index /FID eq {pop pop} {def} ifelse} forall
currentdict end definefont pop
end
gnudict begin
gsave
0 0 translate
0.100 0.100 scale
0 setgray
newpath
1.000 UL
LTb
600 300 M
63 0 V
2787 0 R
-63 0 V
1.000 UL
LTb
600 652 M
63 0 V
2787 0 R
-63 0 V
1.000 UL
LTb
600 1004 M
63 0 V
2787 0 R
-63 0 V
1.000 UL
LTb
600 1356 M
63 0 V
2787 0 R
-63 0 V
1.000 UL
LTb
600 1708 M
63 0 V
2787 0 R
-63 0 V
1.000 UL
LTb
600 2060 M
63 0 V
2787 0 R
-63 0 V
1.000 UL
LTb
600 300 M
0 63 V
0 1697 R
0 -63 V
1.000 UL
LTb
1170 300 M
0 63 V
0 1697 R
0 -63 V
1.000 UL
LTb
1740 300 M
0 63 V
0 1697 R
0 -63 V
1.000 UL
LTb
2310 300 M
0 63 V
0 1697 R
0 -63 V
1.000 UL
LTb
2880 300 M
0 63 V
0 1697 R
0 -63 V
1.000 UL
LTb
3450 300 M
0 63 V
0 1697 R
0 -63 V
1.000 UL
LTb
1.000 UL
LTb
600 300 M
2850 0 V
0 1760 V
-2850 0 V
600 300 L
LTb
LTb
1.000 UP
1.000 UP
1.000 UL
LT0
714 632 M
0 23 V
683 632 M
62 0 V
-62 23 R
62 0 V
55 -93 R
0 31 V
769 562 M
62 0 V
-62 31 R
62 0 V
1027 425 M
0 54 V
996 425 M
62 0 V
-62 54 R
62 0 V
397 12 R
0 45 V
-31 -45 R
62 0 V
-62 45 R
62 0 V
254 134 R
0 104 V
1709 670 M
62 0 V
-62 104 R
62 0 V
254 223 R
0 155 V
1994 997 M
62 0 V
-62 155 R
62 0 V
254 -42 R
0 211 V
-31 -211 R
62 0 V
-62 211 R
62 0 V
2595 962 M
0 352 V
2564 962 M
62 0 V
-62 352 R
62 0 V
2880 948 M
0 95 V
-31 -95 R
62 0 V
-62 95 R
62 0 V
3165 789 M
0 83 V
-31 -83 R
62 0 V
-62 83 R
62 0 V
3364 690 M
0 41 V
-31 -41 R
62 0 V
-62 41 R
62 0 V
714 644 Pls
800 577 Pls
1027 452 Pls
1455 513 Pls
1740 722 Pls
2025 1074 Pls
2310 1215 Pls
2595 1138 Pls
2880 996 Pls
3165 831 Pls
3364 710 Pls
1.000 UP
1.000 UL
LT0
714 624 M
0 20 V
683 624 M
62 0 V
-62 20 R
62 0 V
800 499 M
0 39 V
769 499 M
62 0 V
-62 39 R
62 0 V
1027 395 M
0 63 V
996 395 M
62 0 V
-62 63 R
62 0 V
397 -45 R
0 112 V
1424 413 M
62 0 V
-62 112 R
62 0 V
254 818 R
0 142 V
-31 -142 R
62 0 V
-62 142 R
62 0 V
254 251 R
0 282 V
-31 -282 R
62 0 V
-62 282 R
62 0 V
254 -416 R
0 198 V
-31 -198 R
62 0 V
-62 198 R
62 0 V
254 -627 R
0 324 V
-31 -324 R
62 0 V
-62 324 R
62 0 V
254 -411 R
0 86 V
-31 -86 R
62 0 V
-62 86 R
62 0 V
3165 839 M
0 89 V
-31 -89 R
62 0 V
-62 89 R
62 0 V
714 634 Crs
800 518 Crs
1027 427 Crs
1455 469 Crs
1740 1414 Crs
2025 1877 Crs
2310 1701 Crs
2595 1335 Crs
2880 1129 Crs
3165 884 Crs
1.000 UL
LT2
600 652 M
29 0 V
29 0 V
28 0 V
29 0 V
29 0 V
29 0 V
29 0 V
28 0 V
29 0 V
29 0 V
29 0 V
28 0 V
29 0 V
29 0 V
29 0 V
29 0 V
28 0 V
29 0 V
29 0 V
29 0 V
29 0 V
28 0 V
29 0 V
29 0 V
29 0 V
28 0 V
29 0 V
29 0 V
29 0 V
29 0 V
28 0 V
29 0 V
29 0 V
29 0 V
29 0 V
28 0 V
29 0 V
29 0 V
29 0 V
29 0 V
28 0 V
29 0 V
29 0 V
29 0 V
28 0 V
29 0 V
29 0 V
29 0 V
29 0 V
28 0 V
29 0 V
29 0 V
29 0 V
29 0 V
28 0 V
29 0 V
29 0 V
29 0 V
28 0 V
29 0 V
29 0 V
29 0 V
29 0 V
28 0 V
29 0 V
29 0 V
29 0 V
29 0 V
28 0 V
29 0 V
29 0 V
29 0 V
29 0 V
28 0 V
29 0 V
29 0 V
29 0 V
28 0 V
29 0 V
29 0 V
29 0 V
29 0 V
28 0 V
29 0 V
29 0 V
29 0 V
29 0 V
28 0 V
29 0 V
29 0 V
29 0 V
28 0 V
29 0 V
29 0 V
29 0 V
29 0 V
28 0 V
29 0 V
29 0 V
1.000 UL
LTb
600 300 M
2850 0 V
0 1760 V
-2850 0 V
600 300 L
1.000 UP
stroke
grestore
end
showpage
}}%
\put(2025,50){\makebox(0,0){$\langle u_w \rangle$}}%
\put(100,1180){%
\special{ps: gsave currentpoint currentpoint translate
270 rotate neg exch neg exch translate}%
\makebox(0,0)[b]{\shortstack{$F_{k^\prime}^{D=1+1}-F_{k^\prime}^{D^\prime}$}}%
\special{ps: currentpoint grestore moveto}%
}%
\put(3450,200){\makebox(0,0){ 1}}%
\put(2880,200){\makebox(0,0){ 0.8}}%
\put(2310,200){\makebox(0,0){ 0.6}}%
\put(1740,200){\makebox(0,0){ 0.4}}%
\put(1170,200){\makebox(0,0){ 0.2}}%
\put(600,200){\makebox(0,0){ 0}}%
\put(550,2060){\makebox(0,0)[r]{ 0.002}}%
\put(550,1708){\makebox(0,0)[r]{ 0.0015}}%
\put(550,1356){\makebox(0,0)[r]{ 0.001}}%
\put(550,1004){\makebox(0,0)[r]{ 0.0005}}%
\put(550,652){\makebox(0,0)[r]{ 0}}%
\put(550,300){\makebox(0,0)[r]{-0.0005}}%
\end{picture}%
\endgroup
 

%% file: figures/su2_4x4diffs.tex
\begingroup%
  \makeatletter%
  \newcommand{\GNUPLOTspecial}{%
    \@sanitize\catcode`\%=14\relax\special}%
  \setlength{\unitlength}{0.1bp}%
\begin{picture}(3600,2160)(0,0)%
{\GNUPLOTspecial{"
/gnudict 256 dict def
gnudict begin
/Color false def
/Solid false def
/gnulinewidth 5.000 def
/userlinewidth gnulinewidth def
/vshift -33 def
/dl {10.0 mul} def
/hpt_ 31.5 def
/vpt_ 31.5 def
/hpt hpt_ def
/vpt vpt_ def
/Rounded false def
/M {moveto} bind def
/L {lineto} bind def
/R {rmoveto} bind def
/V {rlineto} bind def
/N {newpath moveto} bind def
/C {setrgbcolor} bind def
/f {rlineto fill} bind def
/vpt2 vpt 2 mul def
/hpt2 hpt 2 mul def
/Lshow { currentpoint stroke M
  0 vshift R show } def
/Rshow { currentpoint stroke M
  dup stringwidth pop neg vshift R show } def
/Cshow { currentpoint stroke M
  dup stringwidth pop -2 div vshift R show } def
/UP { dup vpt_ mul /vpt exch def hpt_ mul /hpt exch def
  /hpt2 hpt 2 mul def /vpt2 vpt 2 mul def } def
/DL { Color {setrgbcolor Solid {pop []} if 0 setdash }
 {pop pop pop 0 setgray Solid {pop []} if 0 setdash} ifelse } def
/BL { stroke userlinewidth 2 mul setlinewidth
      Rounded { 1 setlinejoin 1 setlinecap } if } def
/AL { stroke userlinewidth 2 div setlinewidth
      Rounded { 1 setlinejoin 1 setlinecap } if } def
/UL { dup gnulinewidth mul /userlinewidth exch def
      dup 1 lt {pop 1} if 10 mul /udl exch def } def
/PL { stroke userlinewidth setlinewidth
      Rounded { 1 setlinejoin 1 setlinecap } if } def
/LTw { PL [] 1 setgray } def
/LTb { BL [] 0 0 0 DL } def
/LTa { AL [1 udl mul 2 udl mul] 0 setdash 0 0 0 setrgbcolor } def
/LT0 { PL [] 1 0 0 DL } def
/LT1 { PL [4 dl 2 dl] 0 1 0 DL } def
/LT2 { PL [2 dl 3 dl] 0 0 1 DL } def
/LT3 { PL [1 dl 1.5 dl] 1 0 1 DL } def
/LT4 { PL [5 dl 2 dl 1 dl 2 dl] 0 1 1 DL } def
/LT5 { PL [4 dl 3 dl 1 dl 3 dl] 1 1 0 DL } def
/LT6 { PL [2 dl 2 dl 2 dl 4 dl] 0 0 0 DL } def
/LT7 { PL [2 dl 2 dl 2 dl 2 dl 2 dl 4 dl] 1 0.3 0 DL } def
/LT8 { PL [2 dl 2 dl 2 dl 2 dl 2 dl 2 dl 2 dl 4 dl] 0.5 0.5 0.5 DL } def
/Pnt { stroke [] 0 setdash
   gsave 1 setlinecap M 0 0 V stroke grestore } def
/Dia { stroke [] 0 setdash 2 copy vpt add M
  hpt neg vpt neg V hpt vpt neg V
  hpt vpt V hpt neg vpt V closepath stroke
  Pnt } def
/Pls { stroke [] 0 setdash vpt sub M 0 vpt2 V
  currentpoint stroke M
  hpt neg vpt neg R hpt2 0 V stroke
  } def
/Box { stroke [] 0 setdash 2 copy exch hpt sub exch vpt add M
  0 vpt2 neg V hpt2 0 V 0 vpt2 V
  hpt2 neg 0 V closepath stroke
  Pnt } def
/Crs { stroke [] 0 setdash exch hpt sub exch vpt add M
  hpt2 vpt2 neg V currentpoint stroke M
  hpt2 neg 0 R hpt2 vpt2 V stroke } def
/TriU { stroke [] 0 setdash 2 copy vpt 1.12 mul add M
  hpt neg vpt -1.62 mul V
  hpt 2 mul 0 V
  hpt neg vpt 1.62 mul V closepath stroke
  Pnt  } def
/Star { 2 copy Pls Crs } def
/BoxF { stroke [] 0 setdash exch hpt sub exch vpt add M
  0 vpt2 neg V  hpt2 0 V  0 vpt2 V
  hpt2 neg 0 V  closepath fill } def
/TriUF { stroke [] 0 setdash vpt 1.12 mul add M
  hpt neg vpt -1.62 mul V
  hpt 2 mul 0 V
  hpt neg vpt 1.62 mul V closepath fill } def
/TriD { stroke [] 0 setdash 2 copy vpt 1.12 mul sub M
  hpt neg vpt 1.62 mul V
  hpt 2 mul 0 V
  hpt neg vpt -1.62 mul V closepath stroke
  Pnt  } def
/TriDF { stroke [] 0 setdash vpt 1.12 mul sub M
  hpt neg vpt 1.62 mul V
  hpt 2 mul 0 V
  hpt neg vpt -1.62 mul V closepath fill} def
/DiaF { stroke [] 0 setdash vpt add M
  hpt neg vpt neg V hpt vpt neg V
  hpt vpt V hpt neg vpt V closepath fill } def
/Pent { stroke [] 0 setdash 2 copy gsave
  translate 0 hpt M 4 {72 rotate 0 hpt L} repeat
  closepath stroke grestore Pnt } def
/PentF { stroke [] 0 setdash gsave
  translate 0 hpt M 4 {72 rotate 0 hpt L} repeat
  closepath fill grestore } def
/Circle { stroke [] 0 setdash 2 copy
  hpt 0 360 arc stroke Pnt } def
/CircleF { stroke [] 0 setdash hpt 0 360 arc fill } def
/C0 { BL [] 0 setdash 2 copy moveto vpt 90 450  arc } bind def
/C1 { BL [] 0 setdash 2 copy        moveto
       2 copy  vpt 0 90 arc closepath fill
               vpt 0 360 arc closepath } bind def
/C2 { BL [] 0 setdash 2 copy moveto
       2 copy  vpt 90 180 arc closepath fill
               vpt 0 360 arc closepath } bind def
/C3 { BL [] 0 setdash 2 copy moveto
       2 copy  vpt 0 180 arc closepath fill
               vpt 0 360 arc closepath } bind def
/C4 { BL [] 0 setdash 2 copy moveto
       2 copy  vpt 180 270 arc closepath fill
               vpt 0 360 arc closepath } bind def
/C5 { BL [] 0 setdash 2 copy moveto
       2 copy  vpt 0 90 arc
       2 copy moveto
       2 copy  vpt 180 270 arc closepath fill
               vpt 0 360 arc } bind def
/C6 { BL [] 0 setdash 2 copy moveto
      2 copy  vpt 90 270 arc closepath fill
              vpt 0 360 arc closepath } bind def
/C7 { BL [] 0 setdash 2 copy moveto
      2 copy  vpt 0 270 arc closepath fill
              vpt 0 360 arc closepath } bind def
/C8 { BL [] 0 setdash 2 copy moveto
      2 copy vpt 270 360 arc closepath fill
              vpt 0 360 arc closepath } bind def
/C9 { BL [] 0 setdash 2 copy moveto
      2 copy  vpt 270 450 arc closepath fill
              vpt 0 360 arc closepath } bind def
/C10 { BL [] 0 setdash 2 copy 2 copy moveto vpt 270 360 arc closepath fill
       2 copy moveto
       2 copy vpt 90 180 arc closepath fill
               vpt 0 360 arc closepath } bind def
/C11 { BL [] 0 setdash 2 copy moveto
       2 copy  vpt 0 180 arc closepath fill
       2 copy moveto
       2 copy  vpt 270 360 arc closepath fill
               vpt 0 360 arc closepath } bind def
/C12 { BL [] 0 setdash 2 copy moveto
       2 copy  vpt 180 360 arc closepath fill
               vpt 0 360 arc closepath } bind def
/C13 { BL [] 0 setdash  2 copy moveto
       2 copy  vpt 0 90 arc closepath fill
       2 copy moveto
       2 copy  vpt 180 360 arc closepath fill
               vpt 0 360 arc closepath } bind def
/C14 { BL [] 0 setdash 2 copy moveto
       2 copy  vpt 90 360 arc closepath fill
               vpt 0 360 arc } bind def
/C15 { BL [] 0 setdash 2 copy vpt 0 360 arc closepath fill
               vpt 0 360 arc closepath } bind def
/Rec   { newpath 4 2 roll moveto 1 index 0 rlineto 0 exch rlineto
       neg 0 rlineto closepath } bind def
/Square { dup Rec } bind def
/Bsquare { vpt sub exch vpt sub exch vpt2 Square } bind def
/S0 { BL [] 0 setdash 2 copy moveto 0 vpt rlineto BL Bsquare } bind def
/S1 { BL [] 0 setdash 2 copy vpt Square fill Bsquare } bind def
/S2 { BL [] 0 setdash 2 copy exch vpt sub exch vpt Square fill Bsquare } bind def
/S3 { BL [] 0 setdash 2 copy exch vpt sub exch vpt2 vpt Rec fill Bsquare } bind def
/S4 { BL [] 0 setdash 2 copy exch vpt sub exch vpt sub vpt Square fill Bsquare } bind def
/S5 { BL [] 0 setdash 2 copy 2 copy vpt Square fill
       exch vpt sub exch vpt sub vpt Square fill Bsquare } bind def
/S6 { BL [] 0 setdash 2 copy exch vpt sub exch vpt sub vpt vpt2 Rec fill Bsquare } bind def
/S7 { BL [] 0 setdash 2 copy exch vpt sub exch vpt sub vpt vpt2 Rec fill
       2 copy vpt Square fill
       Bsquare } bind def
/S8 { BL [] 0 setdash 2 copy vpt sub vpt Square fill Bsquare } bind def
/S9 { BL [] 0 setdash 2 copy vpt sub vpt vpt2 Rec fill Bsquare } bind def
/S10 { BL [] 0 setdash 2 copy vpt sub vpt Square fill 2 copy exch vpt sub exch vpt Square fill
       Bsquare } bind def
/S11 { BL [] 0 setdash 2 copy vpt sub vpt Square fill 2 copy exch vpt sub exch vpt2 vpt Rec fill
       Bsquare } bind def
/S12 { BL [] 0 setdash 2 copy exch vpt sub exch vpt sub vpt2 vpt Rec fill Bsquare } bind def
/S13 { BL [] 0 setdash 2 copy exch vpt sub exch vpt sub vpt2 vpt Rec fill
       2 copy vpt Square fill Bsquare } bind def
/S14 { BL [] 0 setdash 2 copy exch vpt sub exch vpt sub vpt2 vpt Rec fill
       2 copy exch vpt sub exch vpt Square fill Bsquare } bind def
/S15 { BL [] 0 setdash 2 copy Bsquare fill Bsquare } bind def
/D0 { gsave translate 45 rotate 0 0 S0 stroke grestore } bind def
/D1 { gsave translate 45 rotate 0 0 S1 stroke grestore } bind def
/D2 { gsave translate 45 rotate 0 0 S2 stroke grestore } bind def
/D3 { gsave translate 45 rotate 0 0 S3 stroke grestore } bind def
/D4 { gsave translate 45 rotate 0 0 S4 stroke grestore } bind def
/D5 { gsave translate 45 rotate 0 0 S5 stroke grestore } bind def
/D6 { gsave translate 45 rotate 0 0 S6 stroke grestore } bind def
/D7 { gsave translate 45 rotate 0 0 S7 stroke grestore } bind def
/D8 { gsave translate 45 rotate 0 0 S8 stroke grestore } bind def
/D9 { gsave translate 45 rotate 0 0 S9 stroke grestore } bind def
/D10 { gsave translate 45 rotate 0 0 S10 stroke grestore } bind def
/D11 { gsave translate 45 rotate 0 0 S11 stroke grestore } bind def
/D12 { gsave translate 45 rotate 0 0 S12 stroke grestore } bind def
/D13 { gsave translate 45 rotate 0 0 S13 stroke grestore } bind def
/D14 { gsave translate 45 rotate 0 0 S14 stroke grestore } bind def
/D15 { gsave translate 45 rotate 0 0 S15 stroke grestore } bind def
/DiaE { stroke [] 0 setdash vpt add M
  hpt neg vpt neg V hpt vpt neg V
  hpt vpt V hpt neg vpt V closepath stroke } def
/BoxE { stroke [] 0 setdash exch hpt sub exch vpt add M
  0 vpt2 neg V hpt2 0 V 0 vpt2 V
  hpt2 neg 0 V closepath stroke } def
/TriUE { stroke [] 0 setdash vpt 1.12 mul add M
  hpt neg vpt -1.62 mul V
  hpt 2 mul 0 V
  hpt neg vpt 1.62 mul V closepath stroke } def
/TriDE { stroke [] 0 setdash vpt 1.12 mul sub M
  hpt neg vpt 1.62 mul V
  hpt 2 mul 0 V
  hpt neg vpt -1.62 mul V closepath stroke } def
/PentE { stroke [] 0 setdash gsave
  translate 0 hpt M 4 {72 rotate 0 hpt L} repeat
  closepath stroke grestore } def
/CircE { stroke [] 0 setdash 
  hpt 0 360 arc stroke } def
/Opaque { gsave closepath 1 setgray fill grestore 0 setgray closepath } def
/DiaW { stroke [] 0 setdash vpt add M
  hpt neg vpt neg V hpt vpt neg V
  hpt vpt V hpt neg vpt V Opaque stroke } def
/BoxW { stroke [] 0 setdash exch hpt sub exch vpt add M
  0 vpt2 neg V hpt2 0 V 0 vpt2 V
  hpt2 neg 0 V Opaque stroke } def
/TriUW { stroke [] 0 setdash vpt 1.12 mul add M
  hpt neg vpt -1.62 mul V
  hpt 2 mul 0 V
  hpt neg vpt 1.62 mul V Opaque stroke } def
/TriDW { stroke [] 0 setdash vpt 1.12 mul sub M
  hpt neg vpt 1.62 mul V
  hpt 2 mul 0 V
  hpt neg vpt -1.62 mul V Opaque stroke } def
/PentW { stroke [] 0 setdash gsave
  translate 0 hpt M 4 {72 rotate 0 hpt L} repeat
  Opaque stroke grestore } def
/CircW { stroke [] 0 setdash 
  hpt 0 360 arc Opaque stroke } def
/BoxFill { gsave Rec 1 setgray fill grestore } def
/BoxColFill {
  gsave Rec
  /Fillden exch def
  currentrgbcolor
  /ColB exch def /ColG exch def /ColR exch def
  /ColR ColR Fillden mul Fillden sub 1 add def
  /ColG ColG Fillden mul Fillden sub 1 add def
  /ColB ColB Fillden mul Fillden sub 1 add def
  ColR ColG ColB setrgbcolor
  fill grestore } def
%
%
/PatternFill { gsave /PFa [ 9 2 roll ] def
    PFa 0 get PFa 2 get 2 div add PFa 1 get PFa 3 get 2 div add translate
    PFa 2 get -2 div PFa 3 get -2 div PFa 2 get PFa 3 get Rec
    gsave 1 setgray fill grestore clip
    currentlinewidth 0.5 mul setlinewidth
    /PFs PFa 2 get dup mul PFa 3 get dup mul add sqrt def
    0 0 M PFa 5 get rotate PFs -2 div dup translate
	0 1 PFs PFa 4 get div 1 add floor cvi
	{ PFa 4 get mul 0 M 0 PFs V } for
    0 PFa 6 get ne {
	0 1 PFs PFa 4 get div 1 add floor cvi
	{ PFa 4 get mul 0 2 1 roll M PFs 0 V } for
    } if
    stroke grestore } def
/Symbol-Oblique /Symbol findfont [1 0 .167 1 0 0] makefont
dup length dict begin {1 index /FID eq {pop pop} {def} ifelse} forall
currentdict end definefont pop
end
gnudict begin
gsave
0 0 translate
0.100 0.100 scale
0 setgray
newpath
1.000 UL
LTb
650 300 M
63 0 V
2737 0 R
-63 0 V
1.000 UL
LTb
650 593 M
63 0 V
2737 0 R
-63 0 V
1.000 UL
LTb
650 887 M
63 0 V
2737 0 R
-63 0 V
1.000 UL
LTb
650 1180 M
63 0 V
2737 0 R
-63 0 V
1.000 UL
LTb
650 1473 M
63 0 V
2737 0 R
-63 0 V
1.000 UL
LTb
650 1767 M
63 0 V
2737 0 R
-63 0 V
1.000 UL
LTb
650 2060 M
63 0 V
2737 0 R
-63 0 V
1.000 UL
LTb
650 300 M
0 63 V
0 1697 R
0 -63 V
1.000 UL
LTb
1210 300 M
0 63 V
0 1697 R
0 -63 V
1.000 UL
LTb
1770 300 M
0 63 V
0 1697 R
0 -63 V
1.000 UL
LTb
2330 300 M
0 63 V
0 1697 R
0 -63 V
1.000 UL
LTb
2890 300 M
0 63 V
0 1697 R
0 -63 V
1.000 UL
LTb
3450 300 M
0 63 V
0 1697 R
0 -63 V
1.000 UL
LTb
1.000 UL
LTb
650 300 M
2800 0 V
0 1760 V
-2800 0 V
650 300 L
LTb
LTb
1.000 UP
1.000 UP
1.000 UL
LT0
762 1045 M
0 188 V
731 1045 M
62 0 V
-62 188 R
62 0 V
846 969 M
0 152 V
815 969 M
62 0 V
-62 152 R
62 0 V
1210 546 M
0 165 V
1179 546 M
62 0 V
-62 165 R
62 0 V
529 563 R
0 270 V
-31 -270 R
62 0 V
-62 270 R
62 0 V
249 -346 R
0 269 V
-31 -269 R
62 0 V
-62 269 R
62 0 V
529 88 R
0 329 V
-31 -329 R
62 0 V
-62 329 R
62 0 V
762 1139 Pls
846 1045 Pls
1210 629 Pls
1770 1409 Pls
2050 1333 Pls
2610 1720 Pls
1.000 UP
1.000 UL
LT0
762 1333 M
0 152 V
731 1333 M
62 0 V
-62 152 R
62 0 V
53 -64 R
0 140 V
815 1421 M
62 0 V
-62 140 R
62 0 V
333 -41 R
0 188 V
-31 -188 R
62 0 V
-62 188 R
62 0 V
529 -153 R
0 294 V
-31 -294 R
62 0 V
-62 294 R
62 0 V
249 -112 R
0 235 V
-31 -235 R
62 0 V
-62 235 R
62 0 V
529 -458 R
0 294 V
-31 -294 R
62 0 V
-62 294 R
62 0 V
762 1409 Crs
846 1491 Crs
1210 1614 Crs
1770 1702 Crs
2050 1855 Crs
2610 1661 Crs
1.000 UL
LT2
650 1473 M
28 0 V
29 0 V
28 0 V
28 0 V
28 0 V
29 0 V
28 0 V
28 0 V
29 0 V
28 0 V
28 0 V
28 0 V
29 0 V
28 0 V
28 0 V
29 0 V
28 0 V
28 0 V
28 0 V
29 0 V
28 0 V
28 0 V
29 0 V
28 0 V
28 0 V
28 0 V
29 0 V
28 0 V
28 0 V
28 0 V
29 0 V
28 0 V
28 0 V
29 0 V
28 0 V
28 0 V
28 0 V
29 0 V
28 0 V
28 0 V
29 0 V
28 0 V
28 0 V
28 0 V
29 0 V
28 0 V
28 0 V
29 0 V
28 0 V
28 0 V
28 0 V
29 0 V
28 0 V
28 0 V
29 0 V
28 0 V
28 0 V
28 0 V
29 0 V
28 0 V
28 0 V
29 0 V
28 0 V
28 0 V
28 0 V
29 0 V
28 0 V
28 0 V
29 0 V
28 0 V
28 0 V
28 0 V
29 0 V
28 0 V
28 0 V
28 0 V
29 0 V
28 0 V
28 0 V
29 0 V
28 0 V
28 0 V
28 0 V
29 0 V
28 0 V
28 0 V
29 0 V
28 0 V
28 0 V
28 0 V
29 0 V
28 0 V
28 0 V
29 0 V
28 0 V
28 0 V
28 0 V
29 0 V
28 0 V
1.000 UL
LTb
650 300 M
2800 0 V
0 1760 V
-2800 0 V
650 300 L
1.000 UP
stroke
grestore
end
showpage
}}%
\put(2050,50){\makebox(0,0){$\langle u_w \rangle$}}%
\put(100,1180){%
\special{ps: gsave currentpoint currentpoint translate
270 rotate neg exch neg exch translate}%
\makebox(0,0)[b]{\shortstack{$F_{k^\prime}^{D=1+1}-F_{k^\prime}^{D^\prime}$}}%
\special{ps: currentpoint grestore moveto}%
}%
\put(3450,200){\makebox(0,0){ 1}}%
\put(2890,200){\makebox(0,0){ 0.8}}%
\put(2330,200){\makebox(0,0){ 0.6}}%
\put(1770,200){\makebox(0,0){ 0.4}}%
\put(1210,200){\makebox(0,0){ 0.2}}%
\put(650,200){\makebox(0,0){ 0}}%
\put(600,2060){\makebox(0,0)[r]{ 1e-04}}%
\put(600,1767){\makebox(0,0)[r]{ 5e-05}}%
\put(600,1473){\makebox(0,0)[r]{ 0}}%
\put(600,1180){\makebox(0,0)[r]{-5e-05}}%
\put(600,887){\makebox(0,0)[r]{-0.0001}}%
\put(600,593){\makebox(0,0)[r]{-0.00015}}%
\put(600,300){\makebox(0,0)[r]{-0.0002}}%
\end{picture}%
\endgroup
 

%% file: figures/su3plaqdiffs.tex
\begingroup%
  \makeatletter%
  \newcommand{\GNUPLOTspecial}{%
    \@sanitize\catcode`\%=14\relax\special}%
  \setlength{\unitlength}{0.1bp}%
\begin{picture}(3600,2160)(0,0)%
{\GNUPLOTspecial{"
/gnudict 256 dict def
gnudict begin
/Color false def
/Solid false def
/gnulinewidth 5.000 def
/userlinewidth gnulinewidth def
/vshift -33 def
/dl {10.0 mul} def
/hpt_ 31.5 def
/vpt_ 31.5 def
/hpt hpt_ def
/vpt vpt_ def
/Rounded false def
/M {moveto} bind def
/L {lineto} bind def
/R {rmoveto} bind def
/V {rlineto} bind def
/N {newpath moveto} bind def
/C {setrgbcolor} bind def
/f {rlineto fill} bind def
/vpt2 vpt 2 mul def
/hpt2 hpt 2 mul def
/Lshow { currentpoint stroke M
  0 vshift R show } def
/Rshow { currentpoint stroke M
  dup stringwidth pop neg vshift R show } def
/Cshow { currentpoint stroke M
  dup stringwidth pop -2 div vshift R show } def
/UP { dup vpt_ mul /vpt exch def hpt_ mul /hpt exch def
  /hpt2 hpt 2 mul def /vpt2 vpt 2 mul def } def
/DL { Color {setrgbcolor Solid {pop []} if 0 setdash }
 {pop pop pop 0 setgray Solid {pop []} if 0 setdash} ifelse } def
/BL { stroke userlinewidth 2 mul setlinewidth
      Rounded { 1 setlinejoin 1 setlinecap } if } def
/AL { stroke userlinewidth 2 div setlinewidth
      Rounded { 1 setlinejoin 1 setlinecap } if } def
/UL { dup gnulinewidth mul /userlinewidth exch def
      dup 1 lt {pop 1} if 10 mul /udl exch def } def
/PL { stroke userlinewidth setlinewidth
      Rounded { 1 setlinejoin 1 setlinecap } if } def
/LTw { PL [] 1 setgray } def
/LTb { BL [] 0 0 0 DL } def
/LTa { AL [1 udl mul 2 udl mul] 0 setdash 0 0 0 setrgbcolor } def
/LT0 { PL [] 1 0 0 DL } def
/LT1 { PL [4 dl 2 dl] 0 1 0 DL } def
/LT2 { PL [2 dl 3 dl] 0 0 1 DL } def
/LT3 { PL [1 dl 1.5 dl] 1 0 1 DL } def
/LT4 { PL [5 dl 2 dl 1 dl 2 dl] 0 1 1 DL } def
/LT5 { PL [4 dl 3 dl 1 dl 3 dl] 1 1 0 DL } def
/LT6 { PL [2 dl 2 dl 2 dl 4 dl] 0 0 0 DL } def
/LT7 { PL [2 dl 2 dl 2 dl 2 dl 2 dl 4 dl] 1 0.3 0 DL } def
/LT8 { PL [2 dl 2 dl 2 dl 2 dl 2 dl 2 dl 2 dl 4 dl] 0.5 0.5 0.5 DL } def
/Pnt { stroke [] 0 setdash
   gsave 1 setlinecap M 0 0 V stroke grestore } def
/Dia { stroke [] 0 setdash 2 copy vpt add M
  hpt neg vpt neg V hpt vpt neg V
  hpt vpt V hpt neg vpt V closepath stroke
  Pnt } def
/Pls { stroke [] 0 setdash vpt sub M 0 vpt2 V
  currentpoint stroke M
  hpt neg vpt neg R hpt2 0 V stroke
  } def
/Box { stroke [] 0 setdash 2 copy exch hpt sub exch vpt add M
  0 vpt2 neg V hpt2 0 V 0 vpt2 V
  hpt2 neg 0 V closepath stroke
  Pnt } def
/Crs { stroke [] 0 setdash exch hpt sub exch vpt add M
  hpt2 vpt2 neg V currentpoint stroke M
  hpt2 neg 0 R hpt2 vpt2 V stroke } def
/TriU { stroke [] 0 setdash 2 copy vpt 1.12 mul add M
  hpt neg vpt -1.62 mul V
  hpt 2 mul 0 V
  hpt neg vpt 1.62 mul V closepath stroke
  Pnt  } def
/Star { 2 copy Pls Crs } def
/BoxF { stroke [] 0 setdash exch hpt sub exch vpt add M
  0 vpt2 neg V  hpt2 0 V  0 vpt2 V
  hpt2 neg 0 V  closepath fill } def
/TriUF { stroke [] 0 setdash vpt 1.12 mul add M
  hpt neg vpt -1.62 mul V
  hpt 2 mul 0 V
  hpt neg vpt 1.62 mul V closepath fill } def
/TriD { stroke [] 0 setdash 2 copy vpt 1.12 mul sub M
  hpt neg vpt 1.62 mul V
  hpt 2 mul 0 V
  hpt neg vpt -1.62 mul V closepath stroke
  Pnt  } def
/TriDF { stroke [] 0 setdash vpt 1.12 mul sub M
  hpt neg vpt 1.62 mul V
  hpt 2 mul 0 V
  hpt neg vpt -1.62 mul V closepath fill} def
/DiaF { stroke [] 0 setdash vpt add M
  hpt neg vpt neg V hpt vpt neg V
  hpt vpt V hpt neg vpt V closepath fill } def
/Pent { stroke [] 0 setdash 2 copy gsave
  translate 0 hpt M 4 {72 rotate 0 hpt L} repeat
  closepath stroke grestore Pnt } def
/PentF { stroke [] 0 setdash gsave
  translate 0 hpt M 4 {72 rotate 0 hpt L} repeat
  closepath fill grestore } def
/Circle { stroke [] 0 setdash 2 copy
  hpt 0 360 arc stroke Pnt } def
/CircleF { stroke [] 0 setdash hpt 0 360 arc fill } def
/C0 { BL [] 0 setdash 2 copy moveto vpt 90 450  arc } bind def
/C1 { BL [] 0 setdash 2 copy        moveto
       2 copy  vpt 0 90 arc closepath fill
               vpt 0 360 arc closepath } bind def
/C2 { BL [] 0 setdash 2 copy moveto
       2 copy  vpt 90 180 arc closepath fill
               vpt 0 360 arc closepath } bind def
/C3 { BL [] 0 setdash 2 copy moveto
       2 copy  vpt 0 180 arc closepath fill
               vpt 0 360 arc closepath } bind def
/C4 { BL [] 0 setdash 2 copy moveto
       2 copy  vpt 180 270 arc closepath fill
               vpt 0 360 arc closepath } bind def
/C5 { BL [] 0 setdash 2 copy moveto
       2 copy  vpt 0 90 arc
       2 copy moveto
       2 copy  vpt 180 270 arc closepath fill
               vpt 0 360 arc } bind def
/C6 { BL [] 0 setdash 2 copy moveto
      2 copy  vpt 90 270 arc closepath fill
              vpt 0 360 arc closepath } bind def
/C7 { BL [] 0 setdash 2 copy moveto
      2 copy  vpt 0 270 arc closepath fill
              vpt 0 360 arc closepath } bind def
/C8 { BL [] 0 setdash 2 copy moveto
      2 copy vpt 270 360 arc closepath fill
              vpt 0 360 arc closepath } bind def
/C9 { BL [] 0 setdash 2 copy moveto
      2 copy  vpt 270 450 arc closepath fill
              vpt 0 360 arc closepath } bind def
/C10 { BL [] 0 setdash 2 copy 2 copy moveto vpt 270 360 arc closepath fill
       2 copy moveto
       2 copy vpt 90 180 arc closepath fill
               vpt 0 360 arc closepath } bind def
/C11 { BL [] 0 setdash 2 copy moveto
       2 copy  vpt 0 180 arc closepath fill
       2 copy moveto
       2 copy  vpt 270 360 arc closepath fill
               vpt 0 360 arc closepath } bind def
/C12 { BL [] 0 setdash 2 copy moveto
       2 copy  vpt 180 360 arc closepath fill
               vpt 0 360 arc closepath } bind def
/C13 { BL [] 0 setdash  2 copy moveto
       2 copy  vpt 0 90 arc closepath fill
       2 copy moveto
       2 copy  vpt 180 360 arc closepath fill
               vpt 0 360 arc closepath } bind def
/C14 { BL [] 0 setdash 2 copy moveto
       2 copy  vpt 90 360 arc closepath fill
               vpt 0 360 arc } bind def
/C15 { BL [] 0 setdash 2 copy vpt 0 360 arc closepath fill
               vpt 0 360 arc closepath } bind def
/Rec   { newpath 4 2 roll moveto 1 index 0 rlineto 0 exch rlineto
       neg 0 rlineto closepath } bind def
/Square { dup Rec } bind def
/Bsquare { vpt sub exch vpt sub exch vpt2 Square } bind def
/S0 { BL [] 0 setdash 2 copy moveto 0 vpt rlineto BL Bsquare } bind def
/S1 { BL [] 0 setdash 2 copy vpt Square fill Bsquare } bind def
/S2 { BL [] 0 setdash 2 copy exch vpt sub exch vpt Square fill Bsquare } bind def
/S3 { BL [] 0 setdash 2 copy exch vpt sub exch vpt2 vpt Rec fill Bsquare } bind def
/S4 { BL [] 0 setdash 2 copy exch vpt sub exch vpt sub vpt Square fill Bsquare } bind def
/S5 { BL [] 0 setdash 2 copy 2 copy vpt Square fill
       exch vpt sub exch vpt sub vpt Square fill Bsquare } bind def
/S6 { BL [] 0 setdash 2 copy exch vpt sub exch vpt sub vpt vpt2 Rec fill Bsquare } bind def
/S7 { BL [] 0 setdash 2 copy exch vpt sub exch vpt sub vpt vpt2 Rec fill
       2 copy vpt Square fill
       Bsquare } bind def
/S8 { BL [] 0 setdash 2 copy vpt sub vpt Square fill Bsquare } bind def
/S9 { BL [] 0 setdash 2 copy vpt sub vpt vpt2 Rec fill Bsquare } bind def
/S10 { BL [] 0 setdash 2 copy vpt sub vpt Square fill 2 copy exch vpt sub exch vpt Square fill
       Bsquare } bind def
/S11 { BL [] 0 setdash 2 copy vpt sub vpt Square fill 2 copy exch vpt sub exch vpt2 vpt Rec fill
       Bsquare } bind def
/S12 { BL [] 0 setdash 2 copy exch vpt sub exch vpt sub vpt2 vpt Rec fill Bsquare } bind def
/S13 { BL [] 0 setdash 2 copy exch vpt sub exch vpt sub vpt2 vpt Rec fill
       2 copy vpt Square fill Bsquare } bind def
/S14 { BL [] 0 setdash 2 copy exch vpt sub exch vpt sub vpt2 vpt Rec fill
       2 copy exch vpt sub exch vpt Square fill Bsquare } bind def
/S15 { BL [] 0 setdash 2 copy Bsquare fill Bsquare } bind def
/D0 { gsave translate 45 rotate 0 0 S0 stroke grestore } bind def
/D1 { gsave translate 45 rotate 0 0 S1 stroke grestore } bind def
/D2 { gsave translate 45 rotate 0 0 S2 stroke grestore } bind def
/D3 { gsave translate 45 rotate 0 0 S3 stroke grestore } bind def
/D4 { gsave translate 45 rotate 0 0 S4 stroke grestore } bind def
/D5 { gsave translate 45 rotate 0 0 S5 stroke grestore } bind def
/D6 { gsave translate 45 rotate 0 0 S6 stroke grestore } bind def
/D7 { gsave translate 45 rotate 0 0 S7 stroke grestore } bind def
/D8 { gsave translate 45 rotate 0 0 S8 stroke grestore } bind def
/D9 { gsave translate 45 rotate 0 0 S9 stroke grestore } bind def
/D10 { gsave translate 45 rotate 0 0 S10 stroke grestore } bind def
/D11 { gsave translate 45 rotate 0 0 S11 stroke grestore } bind def
/D12 { gsave translate 45 rotate 0 0 S12 stroke grestore } bind def
/D13 { gsave translate 45 rotate 0 0 S13 stroke grestore } bind def
/D14 { gsave translate 45 rotate 0 0 S14 stroke grestore } bind def
/D15 { gsave translate 45 rotate 0 0 S15 stroke grestore } bind def
/DiaE { stroke [] 0 setdash vpt add M
  hpt neg vpt neg V hpt vpt neg V
  hpt vpt V hpt neg vpt V closepath stroke } def
/BoxE { stroke [] 0 setdash exch hpt sub exch vpt add M
  0 vpt2 neg V hpt2 0 V 0 vpt2 V
  hpt2 neg 0 V closepath stroke } def
/TriUE { stroke [] 0 setdash vpt 1.12 mul add M
  hpt neg vpt -1.62 mul V
  hpt 2 mul 0 V
  hpt neg vpt 1.62 mul V closepath stroke } def
/TriDE { stroke [] 0 setdash vpt 1.12 mul sub M
  hpt neg vpt 1.62 mul V
  hpt 2 mul 0 V
  hpt neg vpt -1.62 mul V closepath stroke } def
/PentE { stroke [] 0 setdash gsave
  translate 0 hpt M 4 {72 rotate 0 hpt L} repeat
  closepath stroke grestore } def
/CircE { stroke [] 0 setdash 
  hpt 0 360 arc stroke } def
/Opaque { gsave closepath 1 setgray fill grestore 0 setgray closepath } def
/DiaW { stroke [] 0 setdash vpt add M
  hpt neg vpt neg V hpt vpt neg V
  hpt vpt V hpt neg vpt V Opaque stroke } def
/BoxW { stroke [] 0 setdash exch hpt sub exch vpt add M
  0 vpt2 neg V hpt2 0 V 0 vpt2 V
  hpt2 neg 0 V Opaque stroke } def
/TriUW { stroke [] 0 setdash vpt 1.12 mul add M
  hpt neg vpt -1.62 mul V
  hpt 2 mul 0 V
  hpt neg vpt 1.62 mul V Opaque stroke } def
/TriDW { stroke [] 0 setdash vpt 1.12 mul sub M
  hpt neg vpt 1.62 mul V
  hpt 2 mul 0 V
  hpt neg vpt -1.62 mul V Opaque stroke } def
/PentW { stroke [] 0 setdash gsave
  translate 0 hpt M 4 {72 rotate 0 hpt L} repeat
  Opaque stroke grestore } def
/CircW { stroke [] 0 setdash 
  hpt 0 360 arc Opaque stroke } def
/BoxFill { gsave Rec 1 setgray fill grestore } def
/BoxColFill {
  gsave Rec
  /Fillden exch def
  currentrgbcolor
  /ColB exch def /ColG exch def /ColR exch def
  /ColR ColR Fillden mul Fillden sub 1 add def
  /ColG ColG Fillden mul Fillden sub 1 add def
  /ColB ColB Fillden mul Fillden sub 1 add def
  ColR ColG ColB setrgbcolor
  fill grestore } def
%
%
/PatternFill { gsave /PFa [ 9 2 roll ] def
    PFa 0 get PFa 2 get 2 div add PFa 1 get PFa 3 get 2 div add translate
    PFa 2 get -2 div PFa 3 get -2 div PFa 2 get PFa 3 get Rec
    gsave 1 setgray fill grestore clip
    currentlinewidth 0.5 mul setlinewidth
    /PFs PFa 2 get dup mul PFa 3 get dup mul add sqrt def
    0 0 M PFa 5 get rotate PFs -2 div dup translate
	0 1 PFs PFa 4 get div 1 add floor cvi
	{ PFa 4 get mul 0 M 0 PFs V } for
    0 PFa 6 get ne {
	0 1 PFs PFa 4 get div 1 add floor cvi
	{ PFa 4 get mul 0 2 1 roll M PFs 0 V } for
    } if
    stroke grestore } def
/Symbol-Oblique /Symbol findfont [1 0 .167 1 0 0] makefont
dup length dict begin {1 index /FID eq {pop pop} {def} ifelse} forall
currentdict end definefont pop
end
gnudict begin
gsave
0 0 translate
0.100 0.100 scale
0 setgray
newpath
1.000 UL
LTb
550 300 M
63 0 V
2837 0 R
-63 0 V
1.000 UL
LTb
550 551 M
63 0 V
2837 0 R
-63 0 V
1.000 UL
LTb
550 803 M
63 0 V
2837 0 R
-63 0 V
1.000 UL
LTb
550 1054 M
63 0 V
2837 0 R
-63 0 V
1.000 UL
LTb
550 1306 M
63 0 V
2837 0 R
-63 0 V
1.000 UL
LTb
550 1557 M
63 0 V
2837 0 R
-63 0 V
1.000 UL
LTb
550 1809 M
63 0 V
2837 0 R
-63 0 V
1.000 UL
LTb
550 2060 M
63 0 V
2837 0 R
-63 0 V
1.000 UL
LTb
550 300 M
0 63 V
0 1697 R
0 -63 V
1.000 UL
LTb
1130 300 M
0 63 V
0 1697 R
0 -63 V
1.000 UL
LTb
1710 300 M
0 63 V
0 1697 R
0 -63 V
1.000 UL
LTb
2290 300 M
0 63 V
0 1697 R
0 -63 V
1.000 UL
LTb
2870 300 M
0 63 V
0 1697 R
0 -63 V
1.000 UL
LTb
3450 300 M
0 63 V
0 1697 R
0 -63 V
1.000 UL
LTb
1.000 UL
LTb
550 300 M
2900 0 V
0 1760 V
-2900 0 V
550 300 L
LTb
LTb
1.000 UP
1.000 UP
1.000 UL
LT0
753 545 M
0 7 V
-31 -7 R
62 0 V
-62 7 R
62 0 V
636 -1 R
0 5 V
-31 -5 R
62 0 V
-62 5 R
62 0 V
549 124 R
0 27 V
-31 -27 R
62 0 V
-62 27 R
62 0 V
549 470 R
0 26 V
-31 -26 R
62 0 V
-62 26 R
62 0 V
3363 620 M
0 11 V
-31 -11 R
62 0 V
-62 11 R
62 0 V
753 549 Pls
1420 553 Pls
2000 693 Pls
2580 1190 Pls
3363 625 Pls
1.000 UP
1.000 UL
LT0
753 546 M
0 7 V
-31 -7 R
62 0 V
-62 7 R
62 0 V
636 2 R
0 4 V
-31 -4 R
62 0 V
-62 4 R
62 0 V
549 473 R
0 70 V
-31 -70 R
62 0 V
-62 70 R
62 0 V
549 735 R
0 28 V
-31 -28 R
62 0 V
-62 28 R
62 0 V
3363 658 M
0 12 V
-31 -12 R
62 0 V
-62 12 R
62 0 V
753 549 Crs
1420 557 Crs
2000 1067 Crs
2580 1851 Crs
3363 664 Crs
1.000 UL
LT2
550 551 M
29 0 V
30 0 V
29 0 V
29 0 V
29 0 V
30 0 V
29 0 V
29 0 V
30 0 V
29 0 V
29 0 V
30 0 V
29 0 V
29 0 V
29 0 V
30 0 V
29 0 V
29 0 V
30 0 V
29 0 V
29 0 V
29 0 V
30 0 V
29 0 V
29 0 V
30 0 V
29 0 V
29 0 V
29 0 V
30 0 V
29 0 V
29 0 V
30 0 V
29 0 V
29 0 V
30 0 V
29 0 V
29 0 V
29 0 V
30 0 V
29 0 V
29 0 V
30 0 V
29 0 V
29 0 V
29 0 V
30 0 V
29 0 V
29 0 V
30 0 V
29 0 V
29 0 V
30 0 V
29 0 V
29 0 V
29 0 V
30 0 V
29 0 V
29 0 V
30 0 V
29 0 V
29 0 V
29 0 V
30 0 V
29 0 V
29 0 V
30 0 V
29 0 V
29 0 V
30 0 V
29 0 V
29 0 V
29 0 V
30 0 V
29 0 V
29 0 V
30 0 V
29 0 V
29 0 V
29 0 V
30 0 V
29 0 V
29 0 V
30 0 V
29 0 V
29 0 V
29 0 V
30 0 V
29 0 V
29 0 V
30 0 V
29 0 V
29 0 V
30 0 V
29 0 V
29 0 V
29 0 V
30 0 V
29 0 V
1.000 UL
LTb
550 300 M
2900 0 V
0 1760 V
-2900 0 V
550 300 L
1.000 UP
stroke
grestore
end
showpage
}}%
\put(2000,50){\makebox(0,0){$\langle u_p \rangle$}}%
\put(100,1180){%
\special{ps: gsave currentpoint currentpoint translate
270 rotate neg exch neg exch translate}%
\makebox(0,0)[b]{\shortstack{$F_{k^\prime}^{D=1+1}-F_{k^\prime}^{D^\prime}$}}%
\special{ps: currentpoint grestore moveto}%
}%
\put(3450,200){\makebox(0,0){ 1}}%
\put(2870,200){\makebox(0,0){ 0.8}}%
\put(2290,200){\makebox(0,0){ 0.6}}%
\put(1710,200){\makebox(0,0){ 0.4}}%
\put(1130,200){\makebox(0,0){ 0.2}}%
\put(550,200){\makebox(0,0){ 0}}%
\put(500,2060){\makebox(0,0)[r]{ 0.012}}%
\put(500,1809){\makebox(0,0)[r]{ 0.01}}%
\put(500,1557){\makebox(0,0)[r]{ 0.008}}%
\put(500,1306){\makebox(0,0)[r]{ 0.006}}%
\put(500,1054){\makebox(0,0)[r]{ 0.004}}%
\put(500,803){\makebox(0,0)[r]{ 0.002}}%
\put(500,551){\makebox(0,0)[r]{ 0}}%
\put(500,300){\makebox(0,0)[r]{-0.002}}%
\end{picture}%
\endgroup
 

%% file: figures/su6plaqdiffs.tex
\begingroup%
  \makeatletter%
  \newcommand{\GNUPLOTspecial}{%
    \@sanitize\catcode`\%=14\relax\special}%
  \setlength{\unitlength}{0.1bp}%
\begin{picture}(3600,2160)(0,0)%
{\GNUPLOTspecial{"
/gnudict 256 dict def
gnudict begin
/Color false def
/Solid false def
/gnulinewidth 5.000 def
/userlinewidth gnulinewidth def
/vshift -33 def
/dl {10.0 mul} def
/hpt_ 31.5 def
/vpt_ 31.5 def
/hpt hpt_ def
/vpt vpt_ def
/Rounded false def
/M {moveto} bind def
/L {lineto} bind def
/R {rmoveto} bind def
/V {rlineto} bind def
/N {newpath moveto} bind def
/C {setrgbcolor} bind def
/f {rlineto fill} bind def
/vpt2 vpt 2 mul def
/hpt2 hpt 2 mul def
/Lshow { currentpoint stroke M
  0 vshift R show } def
/Rshow { currentpoint stroke M
  dup stringwidth pop neg vshift R show } def
/Cshow { currentpoint stroke M
  dup stringwidth pop -2 div vshift R show } def
/UP { dup vpt_ mul /vpt exch def hpt_ mul /hpt exch def
  /hpt2 hpt 2 mul def /vpt2 vpt 2 mul def } def
/DL { Color {setrgbcolor Solid {pop []} if 0 setdash }
 {pop pop pop 0 setgray Solid {pop []} if 0 setdash} ifelse } def
/BL { stroke userlinewidth 2 mul setlinewidth
      Rounded { 1 setlinejoin 1 setlinecap } if } def
/AL { stroke userlinewidth 2 div setlinewidth
      Rounded { 1 setlinejoin 1 setlinecap } if } def
/UL { dup gnulinewidth mul /userlinewidth exch def
      dup 1 lt {pop 1} if 10 mul /udl exch def } def
/PL { stroke userlinewidth setlinewidth
      Rounded { 1 setlinejoin 1 setlinecap } if } def
/LTw { PL [] 1 setgray } def
/LTb { BL [] 0 0 0 DL } def
/LTa { AL [1 udl mul 2 udl mul] 0 setdash 0 0 0 setrgbcolor } def
/LT0 { PL [] 1 0 0 DL } def
/LT1 { PL [4 dl 2 dl] 0 1 0 DL } def
/LT2 { PL [2 dl 3 dl] 0 0 1 DL } def
/LT3 { PL [1 dl 1.5 dl] 1 0 1 DL } def
/LT4 { PL [5 dl 2 dl 1 dl 2 dl] 0 1 1 DL } def
/LT5 { PL [4 dl 3 dl 1 dl 3 dl] 1 1 0 DL } def
/LT6 { PL [2 dl 2 dl 2 dl 4 dl] 0 0 0 DL } def
/LT7 { PL [2 dl 2 dl 2 dl 2 dl 2 dl 4 dl] 1 0.3 0 DL } def
/LT8 { PL [2 dl 2 dl 2 dl 2 dl 2 dl 2 dl 2 dl 4 dl] 0.5 0.5 0.5 DL } def
/Pnt { stroke [] 0 setdash
   gsave 1 setlinecap M 0 0 V stroke grestore } def
/Dia { stroke [] 0 setdash 2 copy vpt add M
  hpt neg vpt neg V hpt vpt neg V
  hpt vpt V hpt neg vpt V closepath stroke
  Pnt } def
/Pls { stroke [] 0 setdash vpt sub M 0 vpt2 V
  currentpoint stroke M
  hpt neg vpt neg R hpt2 0 V stroke
  } def
/Box { stroke [] 0 setdash 2 copy exch hpt sub exch vpt add M
  0 vpt2 neg V hpt2 0 V 0 vpt2 V
  hpt2 neg 0 V closepath stroke
  Pnt } def
/Crs { stroke [] 0 setdash exch hpt sub exch vpt add M
  hpt2 vpt2 neg V currentpoint stroke M
  hpt2 neg 0 R hpt2 vpt2 V stroke } def
/TriU { stroke [] 0 setdash 2 copy vpt 1.12 mul add M
  hpt neg vpt -1.62 mul V
  hpt 2 mul 0 V
  hpt neg vpt 1.62 mul V closepath stroke
  Pnt  } def
/Star { 2 copy Pls Crs } def
/BoxF { stroke [] 0 setdash exch hpt sub exch vpt add M
  0 vpt2 neg V  hpt2 0 V  0 vpt2 V
  hpt2 neg 0 V  closepath fill } def
/TriUF { stroke [] 0 setdash vpt 1.12 mul add M
  hpt neg vpt -1.62 mul V
  hpt 2 mul 0 V
  hpt neg vpt 1.62 mul V closepath fill } def
/TriD { stroke [] 0 setdash 2 copy vpt 1.12 mul sub M
  hpt neg vpt 1.62 mul V
  hpt 2 mul 0 V
  hpt neg vpt -1.62 mul V closepath stroke
  Pnt  } def
/TriDF { stroke [] 0 setdash vpt 1.12 mul sub M
  hpt neg vpt 1.62 mul V
  hpt 2 mul 0 V
  hpt neg vpt -1.62 mul V closepath fill} def
/DiaF { stroke [] 0 setdash vpt add M
  hpt neg vpt neg V hpt vpt neg V
  hpt vpt V hpt neg vpt V closepath fill } def
/Pent { stroke [] 0 setdash 2 copy gsave
  translate 0 hpt M 4 {72 rotate 0 hpt L} repeat
  closepath stroke grestore Pnt } def
/PentF { stroke [] 0 setdash gsave
  translate 0 hpt M 4 {72 rotate 0 hpt L} repeat
  closepath fill grestore } def
/Circle { stroke [] 0 setdash 2 copy
  hpt 0 360 arc stroke Pnt } def
/CircleF { stroke [] 0 setdash hpt 0 360 arc fill } def
/C0 { BL [] 0 setdash 2 copy moveto vpt 90 450  arc } bind def
/C1 { BL [] 0 setdash 2 copy        moveto
       2 copy  vpt 0 90 arc closepath fill
               vpt 0 360 arc closepath } bind def
/C2 { BL [] 0 setdash 2 copy moveto
       2 copy  vpt 90 180 arc closepath fill
               vpt 0 360 arc closepath } bind def
/C3 { BL [] 0 setdash 2 copy moveto
       2 copy  vpt 0 180 arc closepath fill
               vpt 0 360 arc closepath } bind def
/C4 { BL [] 0 setdash 2 copy moveto
       2 copy  vpt 180 270 arc closepath fill
               vpt 0 360 arc closepath } bind def
/C5 { BL [] 0 setdash 2 copy moveto
       2 copy  vpt 0 90 arc
       2 copy moveto
       2 copy  vpt 180 270 arc closepath fill
               vpt 0 360 arc } bind def
/C6 { BL [] 0 setdash 2 copy moveto
      2 copy  vpt 90 270 arc closepath fill
              vpt 0 360 arc closepath } bind def
/C7 { BL [] 0 setdash 2 copy moveto
      2 copy  vpt 0 270 arc closepath fill
              vpt 0 360 arc closepath } bind def
/C8 { BL [] 0 setdash 2 copy moveto
      2 copy vpt 270 360 arc closepath fill
              vpt 0 360 arc closepath } bind def
/C9 { BL [] 0 setdash 2 copy moveto
      2 copy  vpt 270 450 arc closepath fill
              vpt 0 360 arc closepath } bind def
/C10 { BL [] 0 setdash 2 copy 2 copy moveto vpt 270 360 arc closepath fill
       2 copy moveto
       2 copy vpt 90 180 arc closepath fill
               vpt 0 360 arc closepath } bind def
/C11 { BL [] 0 setdash 2 copy moveto
       2 copy  vpt 0 180 arc closepath fill
       2 copy moveto
       2 copy  vpt 270 360 arc closepath fill
               vpt 0 360 arc closepath } bind def
/C12 { BL [] 0 setdash 2 copy moveto
       2 copy  vpt 180 360 arc closepath fill
               vpt 0 360 arc closepath } bind def
/C13 { BL [] 0 setdash  2 copy moveto
       2 copy  vpt 0 90 arc closepath fill
       2 copy moveto
       2 copy  vpt 180 360 arc closepath fill
               vpt 0 360 arc closepath } bind def
/C14 { BL [] 0 setdash 2 copy moveto
       2 copy  vpt 90 360 arc closepath fill
               vpt 0 360 arc } bind def
/C15 { BL [] 0 setdash 2 copy vpt 0 360 arc closepath fill
               vpt 0 360 arc closepath } bind def
/Rec   { newpath 4 2 roll moveto 1 index 0 rlineto 0 exch rlineto
       neg 0 rlineto closepath } bind def
/Square { dup Rec } bind def
/Bsquare { vpt sub exch vpt sub exch vpt2 Square } bind def
/S0 { BL [] 0 setdash 2 copy moveto 0 vpt rlineto BL Bsquare } bind def
/S1 { BL [] 0 setdash 2 copy vpt Square fill Bsquare } bind def
/S2 { BL [] 0 setdash 2 copy exch vpt sub exch vpt Square fill Bsquare } bind def
/S3 { BL [] 0 setdash 2 copy exch vpt sub exch vpt2 vpt Rec fill Bsquare } bind def
/S4 { BL [] 0 setdash 2 copy exch vpt sub exch vpt sub vpt Square fill Bsquare } bind def
/S5 { BL [] 0 setdash 2 copy 2 copy vpt Square fill
       exch vpt sub exch vpt sub vpt Square fill Bsquare } bind def
/S6 { BL [] 0 setdash 2 copy exch vpt sub exch vpt sub vpt vpt2 Rec fill Bsquare } bind def
/S7 { BL [] 0 setdash 2 copy exch vpt sub exch vpt sub vpt vpt2 Rec fill
       2 copy vpt Square fill
       Bsquare } bind def
/S8 { BL [] 0 setdash 2 copy vpt sub vpt Square fill Bsquare } bind def
/S9 { BL [] 0 setdash 2 copy vpt sub vpt vpt2 Rec fill Bsquare } bind def
/S10 { BL [] 0 setdash 2 copy vpt sub vpt Square fill 2 copy exch vpt sub exch vpt Square fill
       Bsquare } bind def
/S11 { BL [] 0 setdash 2 copy vpt sub vpt Square fill 2 copy exch vpt sub exch vpt2 vpt Rec fill
       Bsquare } bind def
/S12 { BL [] 0 setdash 2 copy exch vpt sub exch vpt sub vpt2 vpt Rec fill Bsquare } bind def
/S13 { BL [] 0 setdash 2 copy exch vpt sub exch vpt sub vpt2 vpt Rec fill
       2 copy vpt Square fill Bsquare } bind def
/S14 { BL [] 0 setdash 2 copy exch vpt sub exch vpt sub vpt2 vpt Rec fill
       2 copy exch vpt sub exch vpt Square fill Bsquare } bind def
/S15 { BL [] 0 setdash 2 copy Bsquare fill Bsquare } bind def
/D0 { gsave translate 45 rotate 0 0 S0 stroke grestore } bind def
/D1 { gsave translate 45 rotate 0 0 S1 stroke grestore } bind def
/D2 { gsave translate 45 rotate 0 0 S2 stroke grestore } bind def
/D3 { gsave translate 45 rotate 0 0 S3 stroke grestore } bind def
/D4 { gsave translate 45 rotate 0 0 S4 stroke grestore } bind def
/D5 { gsave translate 45 rotate 0 0 S5 stroke grestore } bind def
/D6 { gsave translate 45 rotate 0 0 S6 stroke grestore } bind def
/D7 { gsave translate 45 rotate 0 0 S7 stroke grestore } bind def
/D8 { gsave translate 45 rotate 0 0 S8 stroke grestore } bind def
/D9 { gsave translate 45 rotate 0 0 S9 stroke grestore } bind def
/D10 { gsave translate 45 rotate 0 0 S10 stroke grestore } bind def
/D11 { gsave translate 45 rotate 0 0 S11 stroke grestore } bind def
/D12 { gsave translate 45 rotate 0 0 S12 stroke grestore } bind def
/D13 { gsave translate 45 rotate 0 0 S13 stroke grestore } bind def
/D14 { gsave translate 45 rotate 0 0 S14 stroke grestore } bind def
/D15 { gsave translate 45 rotate 0 0 S15 stroke grestore } bind def
/DiaE { stroke [] 0 setdash vpt add M
  hpt neg vpt neg V hpt vpt neg V
  hpt vpt V hpt neg vpt V closepath stroke } def
/BoxE { stroke [] 0 setdash exch hpt sub exch vpt add M
  0 vpt2 neg V hpt2 0 V 0 vpt2 V
  hpt2 neg 0 V closepath stroke } def
/TriUE { stroke [] 0 setdash vpt 1.12 mul add M
  hpt neg vpt -1.62 mul V
  hpt 2 mul 0 V
  hpt neg vpt 1.62 mul V closepath stroke } def
/TriDE { stroke [] 0 setdash vpt 1.12 mul sub M
  hpt neg vpt 1.62 mul V
  hpt 2 mul 0 V
  hpt neg vpt -1.62 mul V closepath stroke } def
/PentE { stroke [] 0 setdash gsave
  translate 0 hpt M 4 {72 rotate 0 hpt L} repeat
  closepath stroke grestore } def
/CircE { stroke [] 0 setdash 
  hpt 0 360 arc stroke } def
/Opaque { gsave closepath 1 setgray fill grestore 0 setgray closepath } def
/DiaW { stroke [] 0 setdash vpt add M
  hpt neg vpt neg V hpt vpt neg V
  hpt vpt V hpt neg vpt V Opaque stroke } def
/BoxW { stroke [] 0 setdash exch hpt sub exch vpt add M
  0 vpt2 neg V hpt2 0 V 0 vpt2 V
  hpt2 neg 0 V Opaque stroke } def
/TriUW { stroke [] 0 setdash vpt 1.12 mul add M
  hpt neg vpt -1.62 mul V
  hpt 2 mul 0 V
  hpt neg vpt 1.62 mul V Opaque stroke } def
/TriDW { stroke [] 0 setdash vpt 1.12 mul sub M
  hpt neg vpt 1.62 mul V
  hpt 2 mul 0 V
  hpt neg vpt -1.62 mul V Opaque stroke } def
/PentW { stroke [] 0 setdash gsave
  translate 0 hpt M 4 {72 rotate 0 hpt L} repeat
  Opaque stroke grestore } def
/CircW { stroke [] 0 setdash 
  hpt 0 360 arc Opaque stroke } def
/BoxFill { gsave Rec 1 setgray fill grestore } def
/BoxColFill {
  gsave Rec
  /Fillden exch def
  currentrgbcolor
  /ColB exch def /ColG exch def /ColR exch def
  /ColR ColR Fillden mul Fillden sub 1 add def
  /ColG ColG Fillden mul Fillden sub 1 add def
  /ColB ColB Fillden mul Fillden sub 1 add def
  ColR ColG ColB setrgbcolor
  fill grestore } def
%
%
/PatternFill { gsave /PFa [ 9 2 roll ] def
    PFa 0 get PFa 2 get 2 div add PFa 1 get PFa 3 get 2 div add translate
    PFa 2 get -2 div PFa 3 get -2 div PFa 2 get PFa 3 get Rec
    gsave 1 setgray fill grestore clip
    currentlinewidth 0.5 mul setlinewidth
    /PFs PFa 2 get dup mul PFa 3 get dup mul add sqrt def
    0 0 M PFa 5 get rotate PFs -2 div dup translate
	0 1 PFs PFa 4 get div 1 add floor cvi
	{ PFa 4 get mul 0 M 0 PFs V } for
    0 PFa 6 get ne {
	0 1 PFs PFa 4 get div 1 add floor cvi
	{ PFa 4 get mul 0 2 1 roll M PFs 0 V } for
    } if
    stroke grestore } def
/Symbol-Oblique /Symbol findfont [1 0 .167 1 0 0] makefont
dup length dict begin {1 index /FID eq {pop pop} {def} ifelse} forall
currentdict end definefont pop
end
gnudict begin
gsave
0 0 translate
0.100 0.100 scale
0 setgray
newpath
1.000 UL
LTb
550 300 M
63 0 V
2837 0 R
-63 0 V
1.000 UL
LTb
550 593 M
63 0 V
2837 0 R
-63 0 V
1.000 UL
LTb
550 887 M
63 0 V
2837 0 R
-63 0 V
1.000 UL
LTb
550 1180 M
63 0 V
2837 0 R
-63 0 V
1.000 UL
LTb
550 1473 M
63 0 V
2837 0 R
-63 0 V
1.000 UL
LTb
550 1767 M
63 0 V
2837 0 R
-63 0 V
1.000 UL
LTb
550 2060 M
63 0 V
2837 0 R
-63 0 V
1.000 UL
LTb
550 300 M
0 63 V
0 1697 R
0 -63 V
1.000 UL
LTb
1130 300 M
0 63 V
0 1697 R
0 -63 V
1.000 UL
LTb
1710 300 M
0 63 V
0 1697 R
0 -63 V
1.000 UL
LTb
2290 300 M
0 63 V
0 1697 R
0 -63 V
1.000 UL
LTb
2870 300 M
0 63 V
0 1697 R
0 -63 V
1.000 UL
LTb
3450 300 M
0 63 V
0 1697 R
0 -63 V
1.000 UL
LTb
1.000 UL
LTb
550 300 M
2900 0 V
0 1760 V
-2900 0 V
550 300 L
LTb
LTb
1.000 UP
1.000 UP
1.000 UL
LT0
1420 601 M
0 7 V
-31 -7 R
62 0 V
-62 7 R
62 0 V
549 306 R
0 12 V
-31 -12 R
62 0 V
-62 12 R
62 0 V
549 259 R
0 8 V
-31 -8 R
62 0 V
-62 8 R
62 0 V
3363 658 M
0 16 V
-31 -16 R
62 0 V
-62 16 R
62 0 V
1420 604 Pls
2000 920 Pls
2580 1189 Pls
3363 666 Pls
1.000 UP
1.000 UL
LT0
1420 607 M
0 5 V
-31 -5 R
62 0 V
-62 5 R
62 0 V
360 408 R
0 16 V
-31 -16 R
62 0 V
-62 16 R
62 0 V
216 765 R
0 29 V
-31 -29 R
62 0 V
-62 29 R
62 0 V
491 -165 R
0 12 V
-31 -12 R
62 0 V
-62 12 R
62 0 V
3363 694 M
0 15 V
-31 -15 R
62 0 V
-62 15 R
62 0 V
1420 610 Crs
1811 1028 Crs
2058 1815 Crs
2580 1671 Crs
3363 701 Crs
1.000 UL
LT2
550 593 M
29 0 V
30 0 V
29 0 V
29 0 V
29 0 V
30 0 V
29 0 V
29 0 V
30 0 V
29 0 V
29 0 V
30 0 V
29 0 V
29 0 V
29 0 V
30 0 V
29 0 V
29 0 V
30 0 V
29 0 V
29 0 V
29 0 V
30 0 V
29 0 V
29 0 V
30 0 V
29 0 V
29 0 V
29 0 V
30 0 V
29 0 V
29 0 V
30 0 V
29 0 V
29 0 V
30 0 V
29 0 V
29 0 V
29 0 V
30 0 V
29 0 V
29 0 V
30 0 V
29 0 V
29 0 V
29 0 V
30 0 V
29 0 V
29 0 V
30 0 V
29 0 V
29 0 V
30 0 V
29 0 V
29 0 V
29 0 V
30 0 V
29 0 V
29 0 V
30 0 V
29 0 V
29 0 V
29 0 V
30 0 V
29 0 V
29 0 V
30 0 V
29 0 V
29 0 V
30 0 V
29 0 V
29 0 V
29 0 V
30 0 V
29 0 V
29 0 V
30 0 V
29 0 V
29 0 V
29 0 V
30 0 V
29 0 V
29 0 V
30 0 V
29 0 V
29 0 V
29 0 V
30 0 V
29 0 V
29 0 V
30 0 V
29 0 V
29 0 V
30 0 V
29 0 V
29 0 V
29 0 V
30 0 V
29 0 V
1.000 UL
LTb
550 300 M
2900 0 V
0 1760 V
-2900 0 V
550 300 L
1.000 UP
stroke
grestore
end
showpage
}}%
\put(2000,50){\makebox(0,0){$\langle u_p \rangle$}}%
\put(100,1180){%
\special{ps: gsave currentpoint currentpoint translate
270 rotate neg exch neg exch translate}%
\makebox(0,0)[b]{\shortstack{$F_{k^\prime}^{D=1+1}-F_{k^\prime}^{D^\prime}$}}%
\special{ps: currentpoint grestore moveto}%
}%
\put(3450,200){\makebox(0,0){ 1}}%
\put(2870,200){\makebox(0,0){ 0.8}}%
\put(2290,200){\makebox(0,0){ 0.6}}%
\put(1710,200){\makebox(0,0){ 0.4}}%
\put(1130,200){\makebox(0,0){ 0.2}}%
\put(550,200){\makebox(0,0){ 0}}%
\put(500,2060){\makebox(0,0)[r]{ 0.025}}%
\put(500,1767){\makebox(0,0)[r]{ 0.02}}%
\put(500,1473){\makebox(0,0)[r]{ 0.015}}%
\put(500,1180){\makebox(0,0)[r]{ 0.01}}%
\put(500,887){\makebox(0,0)[r]{ 0.005}}%
\put(500,593){\makebox(0,0)[r]{ 0}}%
\put(500,300){\makebox(0,0)[r]{-0.005}}%
\end{picture}%
\endgroup
 

%% file: figures/su3_2x2diffs.tex
\begingroup%
  \makeatletter%
  \newcommand{\GNUPLOTspecial}{%
    \@sanitize\catcode`\%=14\relax\special}%
  \setlength{\unitlength}{0.1bp}%
\begin{picture}(3600,2160)(0,0)%
{\GNUPLOTspecial{"
/gnudict 256 dict def
gnudict begin
/Color false def
/Solid false def
/gnulinewidth 5.000 def
/userlinewidth gnulinewidth def
/vshift -33 def
/dl {10.0 mul} def
/hpt_ 31.5 def
/vpt_ 31.5 def
/hpt hpt_ def
/vpt vpt_ def
/Rounded false def
/M {moveto} bind def
/L {lineto} bind def
/R {rmoveto} bind def
/V {rlineto} bind def
/N {newpath moveto} bind def
/C {setrgbcolor} bind def
/f {rlineto fill} bind def
/vpt2 vpt 2 mul def
/hpt2 hpt 2 mul def
/Lshow { currentpoint stroke M
  0 vshift R show } def
/Rshow { currentpoint stroke M
  dup stringwidth pop neg vshift R show } def
/Cshow { currentpoint stroke M
  dup stringwidth pop -2 div vshift R show } def
/UP { dup vpt_ mul /vpt exch def hpt_ mul /hpt exch def
  /hpt2 hpt 2 mul def /vpt2 vpt 2 mul def } def
/DL { Color {setrgbcolor Solid {pop []} if 0 setdash }
 {pop pop pop 0 setgray Solid {pop []} if 0 setdash} ifelse } def
/BL { stroke userlinewidth 2 mul setlinewidth
      Rounded { 1 setlinejoin 1 setlinecap } if } def
/AL { stroke userlinewidth 2 div setlinewidth
      Rounded { 1 setlinejoin 1 setlinecap } if } def
/UL { dup gnulinewidth mul /userlinewidth exch def
      dup 1 lt {pop 1} if 10 mul /udl exch def } def
/PL { stroke userlinewidth setlinewidth
      Rounded { 1 setlinejoin 1 setlinecap } if } def
/LTw { PL [] 1 setgray } def
/LTb { BL [] 0 0 0 DL } def
/LTa { AL [1 udl mul 2 udl mul] 0 setdash 0 0 0 setrgbcolor } def
/LT0 { PL [] 1 0 0 DL } def
/LT1 { PL [4 dl 2 dl] 0 1 0 DL } def
/LT2 { PL [2 dl 3 dl] 0 0 1 DL } def
/LT3 { PL [1 dl 1.5 dl] 1 0 1 DL } def
/LT4 { PL [5 dl 2 dl 1 dl 2 dl] 0 1 1 DL } def
/LT5 { PL [4 dl 3 dl 1 dl 3 dl] 1 1 0 DL } def
/LT6 { PL [2 dl 2 dl 2 dl 4 dl] 0 0 0 DL } def
/LT7 { PL [2 dl 2 dl 2 dl 2 dl 2 dl 4 dl] 1 0.3 0 DL } def
/LT8 { PL [2 dl 2 dl 2 dl 2 dl 2 dl 2 dl 2 dl 4 dl] 0.5 0.5 0.5 DL } def
/Pnt { stroke [] 0 setdash
   gsave 1 setlinecap M 0 0 V stroke grestore } def
/Dia { stroke [] 0 setdash 2 copy vpt add M
  hpt neg vpt neg V hpt vpt neg V
  hpt vpt V hpt neg vpt V closepath stroke
  Pnt } def
/Pls { stroke [] 0 setdash vpt sub M 0 vpt2 V
  currentpoint stroke M
  hpt neg vpt neg R hpt2 0 V stroke
  } def
/Box { stroke [] 0 setdash 2 copy exch hpt sub exch vpt add M
  0 vpt2 neg V hpt2 0 V 0 vpt2 V
  hpt2 neg 0 V closepath stroke
  Pnt } def
/Crs { stroke [] 0 setdash exch hpt sub exch vpt add M
  hpt2 vpt2 neg V currentpoint stroke M
  hpt2 neg 0 R hpt2 vpt2 V stroke } def
/TriU { stroke [] 0 setdash 2 copy vpt 1.12 mul add M
  hpt neg vpt -1.62 mul V
  hpt 2 mul 0 V
  hpt neg vpt 1.62 mul V closepath stroke
  Pnt  } def
/Star { 2 copy Pls Crs } def
/BoxF { stroke [] 0 setdash exch hpt sub exch vpt add M
  0 vpt2 neg V  hpt2 0 V  0 vpt2 V
  hpt2 neg 0 V  closepath fill } def
/TriUF { stroke [] 0 setdash vpt 1.12 mul add M
  hpt neg vpt -1.62 mul V
  hpt 2 mul 0 V
  hpt neg vpt 1.62 mul V closepath fill } def
/TriD { stroke [] 0 setdash 2 copy vpt 1.12 mul sub M
  hpt neg vpt 1.62 mul V
  hpt 2 mul 0 V
  hpt neg vpt -1.62 mul V closepath stroke
  Pnt  } def
/TriDF { stroke [] 0 setdash vpt 1.12 mul sub M
  hpt neg vpt 1.62 mul V
  hpt 2 mul 0 V
  hpt neg vpt -1.62 mul V closepath fill} def
/DiaF { stroke [] 0 setdash vpt add M
  hpt neg vpt neg V hpt vpt neg V
  hpt vpt V hpt neg vpt V closepath fill } def
/Pent { stroke [] 0 setdash 2 copy gsave
  translate 0 hpt M 4 {72 rotate 0 hpt L} repeat
  closepath stroke grestore Pnt } def
/PentF { stroke [] 0 setdash gsave
  translate 0 hpt M 4 {72 rotate 0 hpt L} repeat
  closepath fill grestore } def
/Circle { stroke [] 0 setdash 2 copy
  hpt 0 360 arc stroke Pnt } def
/CircleF { stroke [] 0 setdash hpt 0 360 arc fill } def
/C0 { BL [] 0 setdash 2 copy moveto vpt 90 450  arc } bind def
/C1 { BL [] 0 setdash 2 copy        moveto
       2 copy  vpt 0 90 arc closepath fill
               vpt 0 360 arc closepath } bind def
/C2 { BL [] 0 setdash 2 copy moveto
       2 copy  vpt 90 180 arc closepath fill
               vpt 0 360 arc closepath } bind def
/C3 { BL [] 0 setdash 2 copy moveto
       2 copy  vpt 0 180 arc closepath fill
               vpt 0 360 arc closepath } bind def
/C4 { BL [] 0 setdash 2 copy moveto
       2 copy  vpt 180 270 arc closepath fill
               vpt 0 360 arc closepath } bind def
/C5 { BL [] 0 setdash 2 copy moveto
       2 copy  vpt 0 90 arc
       2 copy moveto
       2 copy  vpt 180 270 arc closepath fill
               vpt 0 360 arc } bind def
/C6 { BL [] 0 setdash 2 copy moveto
      2 copy  vpt 90 270 arc closepath fill
              vpt 0 360 arc closepath } bind def
/C7 { BL [] 0 setdash 2 copy moveto
      2 copy  vpt 0 270 arc closepath fill
              vpt 0 360 arc closepath } bind def
/C8 { BL [] 0 setdash 2 copy moveto
      2 copy vpt 270 360 arc closepath fill
              vpt 0 360 arc closepath } bind def
/C9 { BL [] 0 setdash 2 copy moveto
      2 copy  vpt 270 450 arc closepath fill
              vpt 0 360 arc closepath } bind def
/C10 { BL [] 0 setdash 2 copy 2 copy moveto vpt 270 360 arc closepath fill
       2 copy moveto
       2 copy vpt 90 180 arc closepath fill
               vpt 0 360 arc closepath } bind def
/C11 { BL [] 0 setdash 2 copy moveto
       2 copy  vpt 0 180 arc closepath fill
       2 copy moveto
       2 copy  vpt 270 360 arc closepath fill
               vpt 0 360 arc closepath } bind def
/C12 { BL [] 0 setdash 2 copy moveto
       2 copy  vpt 180 360 arc closepath fill
               vpt 0 360 arc closepath } bind def
/C13 { BL [] 0 setdash  2 copy moveto
       2 copy  vpt 0 90 arc closepath fill
       2 copy moveto
       2 copy  vpt 180 360 arc closepath fill
               vpt 0 360 arc closepath } bind def
/C14 { BL [] 0 setdash 2 copy moveto
       2 copy  vpt 90 360 arc closepath fill
               vpt 0 360 arc } bind def
/C15 { BL [] 0 setdash 2 copy vpt 0 360 arc closepath fill
               vpt 0 360 arc closepath } bind def
/Rec   { newpath 4 2 roll moveto 1 index 0 rlineto 0 exch rlineto
       neg 0 rlineto closepath } bind def
/Square { dup Rec } bind def
/Bsquare { vpt sub exch vpt sub exch vpt2 Square } bind def
/S0 { BL [] 0 setdash 2 copy moveto 0 vpt rlineto BL Bsquare } bind def
/S1 { BL [] 0 setdash 2 copy vpt Square fill Bsquare } bind def
/S2 { BL [] 0 setdash 2 copy exch vpt sub exch vpt Square fill Bsquare } bind def
/S3 { BL [] 0 setdash 2 copy exch vpt sub exch vpt2 vpt Rec fill Bsquare } bind def
/S4 { BL [] 0 setdash 2 copy exch vpt sub exch vpt sub vpt Square fill Bsquare } bind def
/S5 { BL [] 0 setdash 2 copy 2 copy vpt Square fill
       exch vpt sub exch vpt sub vpt Square fill Bsquare } bind def
/S6 { BL [] 0 setdash 2 copy exch vpt sub exch vpt sub vpt vpt2 Rec fill Bsquare } bind def
/S7 { BL [] 0 setdash 2 copy exch vpt sub exch vpt sub vpt vpt2 Rec fill
       2 copy vpt Square fill
       Bsquare } bind def
/S8 { BL [] 0 setdash 2 copy vpt sub vpt Square fill Bsquare } bind def
/S9 { BL [] 0 setdash 2 copy vpt sub vpt vpt2 Rec fill Bsquare } bind def
/S10 { BL [] 0 setdash 2 copy vpt sub vpt Square fill 2 copy exch vpt sub exch vpt Square fill
       Bsquare } bind def
/S11 { BL [] 0 setdash 2 copy vpt sub vpt Square fill 2 copy exch vpt sub exch vpt2 vpt Rec fill
       Bsquare } bind def
/S12 { BL [] 0 setdash 2 copy exch vpt sub exch vpt sub vpt2 vpt Rec fill Bsquare } bind def
/S13 { BL [] 0 setdash 2 copy exch vpt sub exch vpt sub vpt2 vpt Rec fill
       2 copy vpt Square fill Bsquare } bind def
/S14 { BL [] 0 setdash 2 copy exch vpt sub exch vpt sub vpt2 vpt Rec fill
       2 copy exch vpt sub exch vpt Square fill Bsquare } bind def
/S15 { BL [] 0 setdash 2 copy Bsquare fill Bsquare } bind def
/D0 { gsave translate 45 rotate 0 0 S0 stroke grestore } bind def
/D1 { gsave translate 45 rotate 0 0 S1 stroke grestore } bind def
/D2 { gsave translate 45 rotate 0 0 S2 stroke grestore } bind def
/D3 { gsave translate 45 rotate 0 0 S3 stroke grestore } bind def
/D4 { gsave translate 45 rotate 0 0 S4 stroke grestore } bind def
/D5 { gsave translate 45 rotate 0 0 S5 stroke grestore } bind def
/D6 { gsave translate 45 rotate 0 0 S6 stroke grestore } bind def
/D7 { gsave translate 45 rotate 0 0 S7 stroke grestore } bind def
/D8 { gsave translate 45 rotate 0 0 S8 stroke grestore } bind def
/D9 { gsave translate 45 rotate 0 0 S9 stroke grestore } bind def
/D10 { gsave translate 45 rotate 0 0 S10 stroke grestore } bind def
/D11 { gsave translate 45 rotate 0 0 S11 stroke grestore } bind def
/D12 { gsave translate 45 rotate 0 0 S12 stroke grestore } bind def
/D13 { gsave translate 45 rotate 0 0 S13 stroke grestore } bind def
/D14 { gsave translate 45 rotate 0 0 S14 stroke grestore } bind def
/D15 { gsave translate 45 rotate 0 0 S15 stroke grestore } bind def
/DiaE { stroke [] 0 setdash vpt add M
  hpt neg vpt neg V hpt vpt neg V
  hpt vpt V hpt neg vpt V closepath stroke } def
/BoxE { stroke [] 0 setdash exch hpt sub exch vpt add M
  0 vpt2 neg V hpt2 0 V 0 vpt2 V
  hpt2 neg 0 V closepath stroke } def
/TriUE { stroke [] 0 setdash vpt 1.12 mul add M
  hpt neg vpt -1.62 mul V
  hpt 2 mul 0 V
  hpt neg vpt 1.62 mul V closepath stroke } def
/TriDE { stroke [] 0 setdash vpt 1.12 mul sub M
  hpt neg vpt 1.62 mul V
  hpt 2 mul 0 V
  hpt neg vpt -1.62 mul V closepath stroke } def
/PentE { stroke [] 0 setdash gsave
  translate 0 hpt M 4 {72 rotate 0 hpt L} repeat
  closepath stroke grestore } def
/CircE { stroke [] 0 setdash 
  hpt 0 360 arc stroke } def
/Opaque { gsave closepath 1 setgray fill grestore 0 setgray closepath } def
/DiaW { stroke [] 0 setdash vpt add M
  hpt neg vpt neg V hpt vpt neg V
  hpt vpt V hpt neg vpt V Opaque stroke } def
/BoxW { stroke [] 0 setdash exch hpt sub exch vpt add M
  0 vpt2 neg V hpt2 0 V 0 vpt2 V
  hpt2 neg 0 V Opaque stroke } def
/TriUW { stroke [] 0 setdash vpt 1.12 mul add M
  hpt neg vpt -1.62 mul V
  hpt 2 mul 0 V
  hpt neg vpt 1.62 mul V Opaque stroke } def
/TriDW { stroke [] 0 setdash vpt 1.12 mul sub M
  hpt neg vpt 1.62 mul V
  hpt 2 mul 0 V
  hpt neg vpt -1.62 mul V Opaque stroke } def
/PentW { stroke [] 0 setdash gsave
  translate 0 hpt M 4 {72 rotate 0 hpt L} repeat
  Opaque stroke grestore } def
/CircW { stroke [] 0 setdash 
  hpt 0 360 arc Opaque stroke } def
/BoxFill { gsave Rec 1 setgray fill grestore } def
/BoxColFill {
  gsave Rec
  /Fillden exch def
  currentrgbcolor
  /ColB exch def /ColG exch def /ColR exch def
  /ColR ColR Fillden mul Fillden sub 1 add def
  /ColG ColG Fillden mul Fillden sub 1 add def
  /ColB ColB Fillden mul Fillden sub 1 add def
  ColR ColG ColB setrgbcolor
  fill grestore } def
%
%
/PatternFill { gsave /PFa [ 9 2 roll ] def
    PFa 0 get PFa 2 get 2 div add PFa 1 get PFa 3 get 2 div add translate
    PFa 2 get -2 div PFa 3 get -2 div PFa 2 get PFa 3 get Rec
    gsave 1 setgray fill grestore clip
    currentlinewidth 0.5 mul setlinewidth
    /PFs PFa 2 get dup mul PFa 3 get dup mul add sqrt def
    0 0 M PFa 5 get rotate PFs -2 div dup translate
	0 1 PFs PFa 4 get div 1 add floor cvi
	{ PFa 4 get mul 0 M 0 PFs V } for
    0 PFa 6 get ne {
	0 1 PFs PFa 4 get div 1 add floor cvi
	{ PFa 4 get mul 0 2 1 roll M PFs 0 V } for
    } if
    stroke grestore } def
/Symbol-Oblique /Symbol findfont [1 0 .167 1 0 0] makefont
dup length dict begin {1 index /FID eq {pop pop} {def} ifelse} forall
currentdict end definefont pop
end
gnudict begin
gsave
0 0 translate
0.100 0.100 scale
0 setgray
newpath
1.000 UL
LTb
600 300 M
63 0 V
2787 0 R
-63 0 V
1.000 UL
LTb
600 551 M
63 0 V
2787 0 R
-63 0 V
1.000 UL
LTb
600 803 M
63 0 V
2787 0 R
-63 0 V
1.000 UL
LTb
600 1054 M
63 0 V
2787 0 R
-63 0 V
1.000 UL
LTb
600 1306 M
63 0 V
2787 0 R
-63 0 V
1.000 UL
LTb
600 1557 M
63 0 V
2787 0 R
-63 0 V
1.000 UL
LTb
600 1809 M
63 0 V
2787 0 R
-63 0 V
1.000 UL
LTb
600 2060 M
63 0 V
2787 0 R
-63 0 V
1.000 UL
LTb
600 300 M
0 63 V
0 1697 R
0 -63 V
1.000 UL
LTb
1170 300 M
0 63 V
0 1697 R
0 -63 V
1.000 UL
LTb
1740 300 M
0 63 V
0 1697 R
0 -63 V
1.000 UL
LTb
2310 300 M
0 63 V
0 1697 R
0 -63 V
1.000 UL
LTb
2880 300 M
0 63 V
0 1697 R
0 -63 V
1.000 UL
LTb
3450 300 M
0 63 V
0 1697 R
0 -63 V
1.000 UL
LTb
1.000 UL
LTb
600 300 M
2850 0 V
0 1760 V
-2850 0 V
600 300 L
LTb
LTb
1.000 UP
1.000 UP
1.000 UL
LT0
714 380 M
0 232 V
683 380 M
62 0 V
683 612 M
62 0 V
1027 311 M
0 129 V
996 311 M
62 0 V
996 440 M
62 0 V
967 539 R
0 377 V
1994 979 M
62 0 V
-62 377 R
62 0 V
3165 613 M
0 153 V
3134 613 M
62 0 V
-62 153 R
62 0 V
714 496 Pls
1027 375 Pls
2025 1167 Pls
3165 690 Pls
1.000 UP
1.000 UL
LT0
714 347 M
0 346 V
683 347 M
62 0 V
683 693 M
62 0 V
282 194 R
0 151 V
996 887 M
62 0 V
-62 151 R
62 0 V
967 595 R
0 377 V
-31 -377 R
62 0 V
-62 377 R
62 0 V
3165 717 M
0 169 V
3134 717 M
62 0 V
-62 169 R
62 0 V
714 520 Crs
1027 963 Crs
2025 1821 Crs
3165 802 Crs
1.000 UL
LT2
600 551 M
29 0 V
29 0 V
28 0 V
29 0 V
29 0 V
29 0 V
29 0 V
28 0 V
29 0 V
29 0 V
29 0 V
28 0 V
29 0 V
29 0 V
29 0 V
29 0 V
28 0 V
29 0 V
29 0 V
29 0 V
29 0 V
28 0 V
29 0 V
29 0 V
29 0 V
28 0 V
29 0 V
29 0 V
29 0 V
29 0 V
28 0 V
29 0 V
29 0 V
29 0 V
29 0 V
28 0 V
29 0 V
29 0 V
29 0 V
29 0 V
28 0 V
29 0 V
29 0 V
29 0 V
28 0 V
29 0 V
29 0 V
29 0 V
29 0 V
28 0 V
29 0 V
29 0 V
29 0 V
29 0 V
28 0 V
29 0 V
29 0 V
29 0 V
28 0 V
29 0 V
29 0 V
29 0 V
29 0 V
28 0 V
29 0 V
29 0 V
29 0 V
29 0 V
28 0 V
29 0 V
29 0 V
29 0 V
29 0 V
28 0 V
29 0 V
29 0 V
29 0 V
28 0 V
29 0 V
29 0 V
29 0 V
29 0 V
28 0 V
29 0 V
29 0 V
29 0 V
29 0 V
28 0 V
29 0 V
29 0 V
29 0 V
28 0 V
29 0 V
29 0 V
29 0 V
29 0 V
28 0 V
29 0 V
29 0 V
1.000 UL
LTb
600 300 M
2850 0 V
0 1760 V
-2850 0 V
600 300 L
1.000 UP
stroke
grestore
end
showpage
}}%
\put(2025,50){\makebox(0,0){$\langle u_w \rangle$}}%
\put(100,1180){%
\special{ps: gsave currentpoint currentpoint translate
270 rotate neg exch neg exch translate}%
\makebox(0,0)[b]{\shortstack{$F_{k^\prime}^{D=1+1}-F_{k^\prime}^{D^\prime}$}}%
\special{ps: currentpoint grestore moveto}%
}%
\put(3450,200){\makebox(0,0){ 1}}%
\put(2880,200){\makebox(0,0){ 0.8}}%
\put(2310,200){\makebox(0,0){ 0.6}}%
\put(1740,200){\makebox(0,0){ 0.4}}%
\put(1170,200){\makebox(0,0){ 0.2}}%
\put(600,200){\makebox(0,0){ 0}}%
\put(550,2060){\makebox(0,0)[r]{ 0.0012}}%
\put(550,1809){\makebox(0,0)[r]{ 0.001}}%
\put(550,1557){\makebox(0,0)[r]{ 0.0008}}%
\put(550,1306){\makebox(0,0)[r]{ 0.0006}}%
\put(550,1054){\makebox(0,0)[r]{ 0.0004}}%
\put(550,803){\makebox(0,0)[r]{ 0.0002}}%
\put(550,551){\makebox(0,0)[r]{ 0}}%
\put(550,300){\makebox(0,0)[r]{-0.0002}}%
\end{picture}%
\endgroup
 

%% file: figures/su6_2x2diffs.tex
\begingroup%
  \makeatletter%
  \newcommand{\GNUPLOTspecial}{%
    \@sanitize\catcode`\%=14\relax\special}%
  \setlength{\unitlength}{0.1bp}%
\begin{picture}(3600,2160)(0,0)%
{\GNUPLOTspecial{"
/gnudict 256 dict def
gnudict begin
/Color false def
/Solid false def
/gnulinewidth 5.000 def
/userlinewidth gnulinewidth def
/vshift -33 def
/dl {10.0 mul} def
/hpt_ 31.5 def
/vpt_ 31.5 def
/hpt hpt_ def
/vpt vpt_ def
/Rounded false def
/M {moveto} bind def
/L {lineto} bind def
/R {rmoveto} bind def
/V {rlineto} bind def
/N {newpath moveto} bind def
/C {setrgbcolor} bind def
/f {rlineto fill} bind def
/vpt2 vpt 2 mul def
/hpt2 hpt 2 mul def
/Lshow { currentpoint stroke M
  0 vshift R show } def
/Rshow { currentpoint stroke M
  dup stringwidth pop neg vshift R show } def
/Cshow { currentpoint stroke M
  dup stringwidth pop -2 div vshift R show } def
/UP { dup vpt_ mul /vpt exch def hpt_ mul /hpt exch def
  /hpt2 hpt 2 mul def /vpt2 vpt 2 mul def } def
/DL { Color {setrgbcolor Solid {pop []} if 0 setdash }
 {pop pop pop 0 setgray Solid {pop []} if 0 setdash} ifelse } def
/BL { stroke userlinewidth 2 mul setlinewidth
      Rounded { 1 setlinejoin 1 setlinecap } if } def
/AL { stroke userlinewidth 2 div setlinewidth
      Rounded { 1 setlinejoin 1 setlinecap } if } def
/UL { dup gnulinewidth mul /userlinewidth exch def
      dup 1 lt {pop 1} if 10 mul /udl exch def } def
/PL { stroke userlinewidth setlinewidth
      Rounded { 1 setlinejoin 1 setlinecap } if } def
/LTw { PL [] 1 setgray } def
/LTb { BL [] 0 0 0 DL } def
/LTa { AL [1 udl mul 2 udl mul] 0 setdash 0 0 0 setrgbcolor } def
/LT0 { PL [] 1 0 0 DL } def
/LT1 { PL [4 dl 2 dl] 0 1 0 DL } def
/LT2 { PL [2 dl 3 dl] 0 0 1 DL } def
/LT3 { PL [1 dl 1.5 dl] 1 0 1 DL } def
/LT4 { PL [5 dl 2 dl 1 dl 2 dl] 0 1 1 DL } def
/LT5 { PL [4 dl 3 dl 1 dl 3 dl] 1 1 0 DL } def
/LT6 { PL [2 dl 2 dl 2 dl 4 dl] 0 0 0 DL } def
/LT7 { PL [2 dl 2 dl 2 dl 2 dl 2 dl 4 dl] 1 0.3 0 DL } def
/LT8 { PL [2 dl 2 dl 2 dl 2 dl 2 dl 2 dl 2 dl 4 dl] 0.5 0.5 0.5 DL } def
/Pnt { stroke [] 0 setdash
   gsave 1 setlinecap M 0 0 V stroke grestore } def
/Dia { stroke [] 0 setdash 2 copy vpt add M
  hpt neg vpt neg V hpt vpt neg V
  hpt vpt V hpt neg vpt V closepath stroke
  Pnt } def
/Pls { stroke [] 0 setdash vpt sub M 0 vpt2 V
  currentpoint stroke M
  hpt neg vpt neg R hpt2 0 V stroke
  } def
/Box { stroke [] 0 setdash 2 copy exch hpt sub exch vpt add M
  0 vpt2 neg V hpt2 0 V 0 vpt2 V
  hpt2 neg 0 V closepath stroke
  Pnt } def
/Crs { stroke [] 0 setdash exch hpt sub exch vpt add M
  hpt2 vpt2 neg V currentpoint stroke M
  hpt2 neg 0 R hpt2 vpt2 V stroke } def
/TriU { stroke [] 0 setdash 2 copy vpt 1.12 mul add M
  hpt neg vpt -1.62 mul V
  hpt 2 mul 0 V
  hpt neg vpt 1.62 mul V closepath stroke
  Pnt  } def
/Star { 2 copy Pls Crs } def
/BoxF { stroke [] 0 setdash exch hpt sub exch vpt add M
  0 vpt2 neg V  hpt2 0 V  0 vpt2 V
  hpt2 neg 0 V  closepath fill } def
/TriUF { stroke [] 0 setdash vpt 1.12 mul add M
  hpt neg vpt -1.62 mul V
  hpt 2 mul 0 V
  hpt neg vpt 1.62 mul V closepath fill } def
/TriD { stroke [] 0 setdash 2 copy vpt 1.12 mul sub M
  hpt neg vpt 1.62 mul V
  hpt 2 mul 0 V
  hpt neg vpt -1.62 mul V closepath stroke
  Pnt  } def
/TriDF { stroke [] 0 setdash vpt 1.12 mul sub M
  hpt neg vpt 1.62 mul V
  hpt 2 mul 0 V
  hpt neg vpt -1.62 mul V closepath fill} def
/DiaF { stroke [] 0 setdash vpt add M
  hpt neg vpt neg V hpt vpt neg V
  hpt vpt V hpt neg vpt V closepath fill } def
/Pent { stroke [] 0 setdash 2 copy gsave
  translate 0 hpt M 4 {72 rotate 0 hpt L} repeat
  closepath stroke grestore Pnt } def
/PentF { stroke [] 0 setdash gsave
  translate 0 hpt M 4 {72 rotate 0 hpt L} repeat
  closepath fill grestore } def
/Circle { stroke [] 0 setdash 2 copy
  hpt 0 360 arc stroke Pnt } def
/CircleF { stroke [] 0 setdash hpt 0 360 arc fill } def
/C0 { BL [] 0 setdash 2 copy moveto vpt 90 450  arc } bind def
/C1 { BL [] 0 setdash 2 copy        moveto
       2 copy  vpt 0 90 arc closepath fill
               vpt 0 360 arc closepath } bind def
/C2 { BL [] 0 setdash 2 copy moveto
       2 copy  vpt 90 180 arc closepath fill
               vpt 0 360 arc closepath } bind def
/C3 { BL [] 0 setdash 2 copy moveto
       2 copy  vpt 0 180 arc closepath fill
               vpt 0 360 arc closepath } bind def
/C4 { BL [] 0 setdash 2 copy moveto
       2 copy  vpt 180 270 arc closepath fill
               vpt 0 360 arc closepath } bind def
/C5 { BL [] 0 setdash 2 copy moveto
       2 copy  vpt 0 90 arc
       2 copy moveto
       2 copy  vpt 180 270 arc closepath fill
               vpt 0 360 arc } bind def
/C6 { BL [] 0 setdash 2 copy moveto
      2 copy  vpt 90 270 arc closepath fill
              vpt 0 360 arc closepath } bind def
/C7 { BL [] 0 setdash 2 copy moveto
      2 copy  vpt 0 270 arc closepath fill
              vpt 0 360 arc closepath } bind def
/C8 { BL [] 0 setdash 2 copy moveto
      2 copy vpt 270 360 arc closepath fill
              vpt 0 360 arc closepath } bind def
/C9 { BL [] 0 setdash 2 copy moveto
      2 copy  vpt 270 450 arc closepath fill
              vpt 0 360 arc closepath } bind def
/C10 { BL [] 0 setdash 2 copy 2 copy moveto vpt 270 360 arc closepath fill
       2 copy moveto
       2 copy vpt 90 180 arc closepath fill
               vpt 0 360 arc closepath } bind def
/C11 { BL [] 0 setdash 2 copy moveto
       2 copy  vpt 0 180 arc closepath fill
       2 copy moveto
       2 copy  vpt 270 360 arc closepath fill
               vpt 0 360 arc closepath } bind def
/C12 { BL [] 0 setdash 2 copy moveto
       2 copy  vpt 180 360 arc closepath fill
               vpt 0 360 arc closepath } bind def
/C13 { BL [] 0 setdash  2 copy moveto
       2 copy  vpt 0 90 arc closepath fill
       2 copy moveto
       2 copy  vpt 180 360 arc closepath fill
               vpt 0 360 arc closepath } bind def
/C14 { BL [] 0 setdash 2 copy moveto
       2 copy  vpt 90 360 arc closepath fill
               vpt 0 360 arc } bind def
/C15 { BL [] 0 setdash 2 copy vpt 0 360 arc closepath fill
               vpt 0 360 arc closepath } bind def
/Rec   { newpath 4 2 roll moveto 1 index 0 rlineto 0 exch rlineto
       neg 0 rlineto closepath } bind def
/Square { dup Rec } bind def
/Bsquare { vpt sub exch vpt sub exch vpt2 Square } bind def
/S0 { BL [] 0 setdash 2 copy moveto 0 vpt rlineto BL Bsquare } bind def
/S1 { BL [] 0 setdash 2 copy vpt Square fill Bsquare } bind def
/S2 { BL [] 0 setdash 2 copy exch vpt sub exch vpt Square fill Bsquare } bind def
/S3 { BL [] 0 setdash 2 copy exch vpt sub exch vpt2 vpt Rec fill Bsquare } bind def
/S4 { BL [] 0 setdash 2 copy exch vpt sub exch vpt sub vpt Square fill Bsquare } bind def
/S5 { BL [] 0 setdash 2 copy 2 copy vpt Square fill
       exch vpt sub exch vpt sub vpt Square fill Bsquare } bind def
/S6 { BL [] 0 setdash 2 copy exch vpt sub exch vpt sub vpt vpt2 Rec fill Bsquare } bind def
/S7 { BL [] 0 setdash 2 copy exch vpt sub exch vpt sub vpt vpt2 Rec fill
       2 copy vpt Square fill
       Bsquare } bind def
/S8 { BL [] 0 setdash 2 copy vpt sub vpt Square fill Bsquare } bind def
/S9 { BL [] 0 setdash 2 copy vpt sub vpt vpt2 Rec fill Bsquare } bind def
/S10 { BL [] 0 setdash 2 copy vpt sub vpt Square fill 2 copy exch vpt sub exch vpt Square fill
       Bsquare } bind def
/S11 { BL [] 0 setdash 2 copy vpt sub vpt Square fill 2 copy exch vpt sub exch vpt2 vpt Rec fill
       Bsquare } bind def
/S12 { BL [] 0 setdash 2 copy exch vpt sub exch vpt sub vpt2 vpt Rec fill Bsquare } bind def
/S13 { BL [] 0 setdash 2 copy exch vpt sub exch vpt sub vpt2 vpt Rec fill
       2 copy vpt Square fill Bsquare } bind def
/S14 { BL [] 0 setdash 2 copy exch vpt sub exch vpt sub vpt2 vpt Rec fill
       2 copy exch vpt sub exch vpt Square fill Bsquare } bind def
/S15 { BL [] 0 setdash 2 copy Bsquare fill Bsquare } bind def
/D0 { gsave translate 45 rotate 0 0 S0 stroke grestore } bind def
/D1 { gsave translate 45 rotate 0 0 S1 stroke grestore } bind def
/D2 { gsave translate 45 rotate 0 0 S2 stroke grestore } bind def
/D3 { gsave translate 45 rotate 0 0 S3 stroke grestore } bind def
/D4 { gsave translate 45 rotate 0 0 S4 stroke grestore } bind def
/D5 { gsave translate 45 rotate 0 0 S5 stroke grestore } bind def
/D6 { gsave translate 45 rotate 0 0 S6 stroke grestore } bind def
/D7 { gsave translate 45 rotate 0 0 S7 stroke grestore } bind def
/D8 { gsave translate 45 rotate 0 0 S8 stroke grestore } bind def
/D9 { gsave translate 45 rotate 0 0 S9 stroke grestore } bind def
/D10 { gsave translate 45 rotate 0 0 S10 stroke grestore } bind def
/D11 { gsave translate 45 rotate 0 0 S11 stroke grestore } bind def
/D12 { gsave translate 45 rotate 0 0 S12 stroke grestore } bind def
/D13 { gsave translate 45 rotate 0 0 S13 stroke grestore } bind def
/D14 { gsave translate 45 rotate 0 0 S14 stroke grestore } bind def
/D15 { gsave translate 45 rotate 0 0 S15 stroke grestore } bind def
/DiaE { stroke [] 0 setdash vpt add M
  hpt neg vpt neg V hpt vpt neg V
  hpt vpt V hpt neg vpt V closepath stroke } def
/BoxE { stroke [] 0 setdash exch hpt sub exch vpt add M
  0 vpt2 neg V hpt2 0 V 0 vpt2 V
  hpt2 neg 0 V closepath stroke } def
/TriUE { stroke [] 0 setdash vpt 1.12 mul add M
  hpt neg vpt -1.62 mul V
  hpt 2 mul 0 V
  hpt neg vpt 1.62 mul V closepath stroke } def
/TriDE { stroke [] 0 setdash vpt 1.12 mul sub M
  hpt neg vpt 1.62 mul V
  hpt 2 mul 0 V
  hpt neg vpt -1.62 mul V closepath stroke } def
/PentE { stroke [] 0 setdash gsave
  translate 0 hpt M 4 {72 rotate 0 hpt L} repeat
  closepath stroke grestore } def
/CircE { stroke [] 0 setdash 
  hpt 0 360 arc stroke } def
/Opaque { gsave closepath 1 setgray fill grestore 0 setgray closepath } def
/DiaW { stroke [] 0 setdash vpt add M
  hpt neg vpt neg V hpt vpt neg V
  hpt vpt V hpt neg vpt V Opaque stroke } def
/BoxW { stroke [] 0 setdash exch hpt sub exch vpt add M
  0 vpt2 neg V hpt2 0 V 0 vpt2 V
  hpt2 neg 0 V Opaque stroke } def
/TriUW { stroke [] 0 setdash vpt 1.12 mul add M
  hpt neg vpt -1.62 mul V
  hpt 2 mul 0 V
  hpt neg vpt 1.62 mul V Opaque stroke } def
/TriDW { stroke [] 0 setdash vpt 1.12 mul sub M
  hpt neg vpt 1.62 mul V
  hpt 2 mul 0 V
  hpt neg vpt -1.62 mul V Opaque stroke } def
/PentW { stroke [] 0 setdash gsave
  translate 0 hpt M 4 {72 rotate 0 hpt L} repeat
  Opaque stroke grestore } def
/CircW { stroke [] 0 setdash 
  hpt 0 360 arc Opaque stroke } def
/BoxFill { gsave Rec 1 setgray fill grestore } def
/BoxColFill {
  gsave Rec
  /Fillden exch def
  currentrgbcolor
  /ColB exch def /ColG exch def /ColR exch def
  /ColR ColR Fillden mul Fillden sub 1 add def
  /ColG ColG Fillden mul Fillden sub 1 add def
  /ColB ColB Fillden mul Fillden sub 1 add def
  ColR ColG ColB setrgbcolor
  fill grestore } def
%
%
/PatternFill { gsave /PFa [ 9 2 roll ] def
    PFa 0 get PFa 2 get 2 div add PFa 1 get PFa 3 get 2 div add translate
    PFa 2 get -2 div PFa 3 get -2 div PFa 2 get PFa 3 get Rec
    gsave 1 setgray fill grestore clip
    currentlinewidth 0.5 mul setlinewidth
    /PFs PFa 2 get dup mul PFa 3 get dup mul add sqrt def
    0 0 M PFa 5 get rotate PFs -2 div dup translate
	0 1 PFs PFa 4 get div 1 add floor cvi
	{ PFa 4 get mul 0 M 0 PFs V } for
    0 PFa 6 get ne {
	0 1 PFs PFa 4 get div 1 add floor cvi
	{ PFa 4 get mul 0 2 1 roll M PFs 0 V } for
    } if
    stroke grestore } def
/Symbol-Oblique /Symbol findfont [1 0 .167 1 0 0] makefont
dup length dict begin {1 index /FID eq {pop pop} {def} ifelse} forall
currentdict end definefont pop
end
gnudict begin
gsave
0 0 translate
0.100 0.100 scale
0 setgray
newpath
1.000 UL
LTb
600 300 M
63 0 V
2787 0 R
-63 0 V
1.000 UL
LTb
600 496 M
63 0 V
2787 0 R
-63 0 V
1.000 UL
LTb
600 691 M
63 0 V
2787 0 R
-63 0 V
1.000 UL
LTb
600 887 M
63 0 V
2787 0 R
-63 0 V
1.000 UL
LTb
600 1082 M
63 0 V
2787 0 R
-63 0 V
1.000 UL
LTb
600 1278 M
63 0 V
2787 0 R
-63 0 V
1.000 UL
LTb
600 1473 M
63 0 V
2787 0 R
-63 0 V
1.000 UL
LTb
600 1669 M
63 0 V
2787 0 R
-63 0 V
1.000 UL
LTb
600 1864 M
63 0 V
2787 0 R
-63 0 V
1.000 UL
LTb
600 2060 M
63 0 V
2787 0 R
-63 0 V
1.000 UL
LTb
600 300 M
0 63 V
0 1697 R
0 -63 V
1.000 UL
LTb
1170 300 M
0 63 V
0 1697 R
0 -63 V
1.000 UL
LTb
1740 300 M
0 63 V
0 1697 R
0 -63 V
1.000 UL
LTb
2310 300 M
0 63 V
0 1697 R
0 -63 V
1.000 UL
LTb
2880 300 M
0 63 V
0 1697 R
0 -63 V
1.000 UL
LTb
3450 300 M
0 63 V
0 1697 R
0 -63 V
1.000 UL
LTb
1.000 UL
LTb
600 300 M
2850 0 V
0 1760 V
-2850 0 V
600 300 L
LTb
LTb
1.000 UP
1.000 UP
1.000 UL
LT0
714 498 M
0 36 V
683 498 M
62 0 V
-62 36 R
62 0 V
282 397 R
0 46 V
996 931 M
62 0 V
-62 46 R
62 0 V
2025 867 M
0 125 V
1994 867 M
62 0 V
-62 125 R
62 0 V
3165 628 M
0 59 V
-31 -59 R
62 0 V
-62 59 R
62 0 V
714 516 Pls
1027 954 Pls
2025 930 Pls
3165 657 Pls
1.000 UP
1.000 UL
LT0
714 556 M
0 48 V
683 556 M
62 0 V
-62 48 R
62 0 V
282 1207 R
0 56 V
-31 -56 R
62 0 V
-62 56 R
62 0 V
967 -679 R
0 117 V
-31 -117 R
62 0 V
-62 117 R
62 0 V
3165 652 M
0 63 V
-31 -63 R
62 0 V
-62 63 R
62 0 V
714 580 Crs
1027 1839 Crs
2025 1246 Crs
3165 683 Crs
1.000 UL
LT2
600 496 M
29 0 V
29 0 V
28 0 V
29 0 V
29 0 V
29 0 V
29 0 V
28 0 V
29 0 V
29 0 V
29 0 V
28 0 V
29 0 V
29 0 V
29 0 V
29 0 V
28 0 V
29 0 V
29 0 V
29 0 V
29 0 V
28 0 V
29 0 V
29 0 V
29 0 V
28 0 V
29 0 V
29 0 V
29 0 V
29 0 V
28 0 V
29 0 V
29 0 V
29 0 V
29 0 V
28 0 V
29 0 V
29 0 V
29 0 V
29 0 V
28 0 V
29 0 V
29 0 V
29 0 V
28 0 V
29 0 V
29 0 V
29 0 V
29 0 V
28 0 V
29 0 V
29 0 V
29 0 V
29 0 V
28 0 V
29 0 V
29 0 V
29 0 V
28 0 V
29 0 V
29 0 V
29 0 V
29 0 V
28 0 V
29 0 V
29 0 V
29 0 V
29 0 V
28 0 V
29 0 V
29 0 V
29 0 V
29 0 V
28 0 V
29 0 V
29 0 V
29 0 V
28 0 V
29 0 V
29 0 V
29 0 V
29 0 V
28 0 V
29 0 V
29 0 V
29 0 V
29 0 V
28 0 V
29 0 V
29 0 V
29 0 V
28 0 V
29 0 V
29 0 V
29 0 V
29 0 V
28 0 V
29 0 V
29 0 V
1.000 UL
LTb
600 300 M
2850 0 V
0 1760 V
-2850 0 V
600 300 L
1.000 UP
stroke
grestore
end
showpage
}}%
\put(2025,50){\makebox(0,0){$\langle u_w \rangle$}}%
\put(100,1180){%
\special{ps: gsave currentpoint currentpoint translate
270 rotate neg exch neg exch translate}%
\makebox(0,0)[b]{\shortstack{$F_{k^\prime}^{D=1+1}-F_{k^\prime}^{D^\prime}$}}%
\special{ps: currentpoint grestore moveto}%
}%
\put(3450,200){\makebox(0,0){ 1}}%
\put(2880,200){\makebox(0,0){ 0.8}}%
\put(2310,200){\makebox(0,0){ 0.6}}%
\put(1740,200){\makebox(0,0){ 0.4}}%
\put(1170,200){\makebox(0,0){ 0.2}}%
\put(600,200){\makebox(0,0){ 0}}%
\put(550,2060){\makebox(0,0)[r]{ 0.004}}%
\put(550,1864){\makebox(0,0)[r]{ 0.0035}}%
\put(550,1669){\makebox(0,0)[r]{ 0.003}}%
\put(550,1473){\makebox(0,0)[r]{ 0.0025}}%
\put(550,1278){\makebox(0,0)[r]{ 0.002}}%
\put(550,1082){\makebox(0,0)[r]{ 0.0015}}%
\put(550,887){\makebox(0,0)[r]{ 0.001}}%
\put(550,691){\makebox(0,0)[r]{ 0.0005}}%
\put(550,496){\makebox(0,0)[r]{ 0}}%
\put(550,300){\makebox(0,0)[r]{-0.0005}}%
\end{picture}%
\endgroup
 

%% file: version4.bbl
\vspace*{0.20in}

\begin{thebibliography}{99}

\bibitem{BurTep06}
F. Bursa and M. Teper,
Phys. Rev. D74 (2006) 125010 [hep-lat/0511081].

\bibitem{Belova83}
T. Belova, Y. Makeenko, M. Polikarpov and A. Veselov,
Nucl. Phys. B230[FS10] (1984) 473.

\bibitem{NN2006}
R. Narayanan and H. Neuberger,
JHEP 0603 (2006) 064 [hep-th/0601210].

\bibitem{Brzoska04}
A. Brzoska, F. Lenz, J. Negele and M.Thies,
Phys. Rev. D71 (2005) 034008 [hep-th/0412003].

\bibitem{GW}
D. Gross and E. Witten,
Phys. Rev. D 21 (1980) 446.

\bibitem{Campo98}
M. Campostrini,
Nucl. Phys. Proc. Suppl. 73 (1999) 724
[hep-lat/9809072].

\bibitem{OxG01}
B. Lucini and M. Teper,
JHEP  0106 (2001) 050 [hep-lat/0103027].

\bibitem{OxT05}
B. Lucini, M. Teper and U. Wenger,
JHEP 0502 (2005) 033 [hep-lat/0502003].

\bibitem{DurOle}
B. Durhuus and P. Olesen,
Nucl. Phys. B184 (1981) 461.

\bibitem{BasGriVian}
A. Bassetto, L. Griguolo and F. Vian,
Nucl. Phys. B559 (1999) 563 [hep-th/9906125].

\bibitem{Teper:2004pk}
M. Teper, hep-th/0412005.

\bibitem{Jaffe:2004}
R. Jaffe, Invited talk at `QCD and Strings', ECT, July 2004.

\bibitem{Narayanan:2005en}
R. Narayanan and H. Neuberger,
PoS Lat2005 (2006) 005 [hep-lat/0509014].

\bibitem{APO84}
J. Ambj\o rn, P. Olesen and C. Peterson,
Nucl. Phys. B240 (1984) 533.

\bibitem{Deldar00}
S. Deldar,
Phys. Rev. D62 (2000) 034509 [hep-lat/9911008].

\bibitem{Bali00}
G. Bali,
Phys. Rev. D62 (2000) 114503 [hep-lat/0006022].

\bibitem{LucTep01}
B. Lucini and M. Teper,
Phys. Rev. D64 (2001) 105019 [hep-lat/0107007].



\end{thebibliography}
